\documentclass[11pt]{article}
\usepackage{amsthm,amssymb,amsfonts,amsmath,amscd}
\usepackage{nccmath}
\usepackage{graphicx}
\usepackage[pdfa]{hyperref}

\makeatletter \@addtoreset{equation}{section} \makeatother

\newtheorem{theorem}{Theorem}[section]
\newtheorem{lemma}{Lemma}[section]

\newtheorem{definition}{Definition}[section]
\newtheorem{proposition}{Proposition}[section]

\newcommand{\mdet}{\mathrm{det}}

\newcommand{\tr}{\mathrm{tr}\,}
\newcommand{\Tr}{\mathrm{Tr}\,}

\begin{document}

\title{Universality for 1 d random band matrices}

\author{ Mariya Shcherbina
\thanks{Institute for Low Temperature Physics, Kharkiv, Ukraine\& Karazin Kharkiv National University, Kharkiv, Ukraine,  e-mail: shcherbi@ilt.kharkov.ua} \and
 Tatyana Shcherbina
\thanks{ Department of Mathematics, Princeton University, Princeton, USA, e-mail: tshcherbyna@princeton.edu. Supported in part by NSF grant DMS-1700009.}}

\date{}
\maketitle

\begin{abstract}
 We consider 1d random Hermitian $N\times N$ block band matrices consisting of $W\times W$ random Gaussian blocks (parametrized by  $j,k \in\Lambda=[1,n]\cap \mathbb{Z}$, $N=nW$) with a
fixed entry's variance $J_{jk}=W^{-1}(\delta_{j,k}+\beta\Delta_{j,k})$ in each block.  Considering the limit $W, n\to\infty$, we prove that the behaviour  of the second correlation function of such matrices in the bulk of the spectrum, as $W\gg \sqrt{N}$,  is determined by the Wigner -- Dyson statistics. 
The method of the proof is based on the rigorous application of supersymmetric transfer matrix approach developed in \cite{SS:sigma}.

\end{abstract}

\section{Introduction}
Random band matrices (RBM) provide
a natural and important model to study eigenvalue statistic and quantum transport
in disordered systems as they interpolate between classical Wigner matrices, i.e. Hermitian
random matrices with all independent identically distributed elements, and random
Schr$\ddot{\hbox{o}}$dinger operators, where only a random on-site potential is present in addition to the
deterministic Laplacian on a regular box in $d$-dimension lattice.
Such matrices have various application in physics: the
eigenvalue statistics of RBM is in relevance in quantum chaos, the quantum dynamics
associated with RBM can be used to model conductance in thick wires, etc. 
 
 The density of states $\rho$ of a general class of RBM with $W\gg 1$ is given by the well-known Wigner semicircle law (see
\cite{BMP:91, MPK:92}):
\begin{equation}\label{rho_sc}
\rho(E)=(2\pi)^{-1}\sqrt{4-E^2},\quad E\in[-2,2].
\end{equation}
 The main feature of RBM is that they can be used
to model the celebrated Anderson metal-insulator phase transition in $d\ge 3$ (see the review (\cite{Sp:12}) for the details). Moreover,
the crossover for RBM can be investigated even in $d = 1$ by varying the bandwidth $W$.

More precisely, the key physical parameter of RBM  is the localization length,
which describes the length scale of the eigenvector corresponding to the energy $E\in (-2,2)$.
The system is called delocalized if for all $E$ in the bulk of spectrum  the localization length is
comparable with the system size, and it is called localized otherwise.
Physically, delocalized systems correspond to electric
conductors, and localized systems are insulators.

The questions of the localization length  are closely related to the universality conjecture
of the bulk local regime of the random matrix theory. The bulk local regime deals with the behaviour of eigenvalues of $N\times N$
random matrices on the intervals whose length is of the order $O(N^{-1})$. The main objects of the local regime are $k$-point correlation functions
$R_k$ ($k=1,2,\ldots$), which can be defined by the equalities:
\begin{multline} \label{R}
\mathbf{E}\left\{ \sum_{j_{1}\neq ...\neq j_{k}}\varphi_k
(\lambda_{j_{1}}^{(N)},\dots,\lambda_{j_{k}}^{(N)})\right\}\\ =\int_{\mathbb{R}^{k}} \varphi_{k}
(\lambda_{1}^{(N)},\ldots,\lambda_{k}^{(N)})R_{k}(\lambda_{1}^{(N)},\ldots,\lambda_{k}^{(N)})
d\lambda_{1}^{(N)}\ldots d\lambda_{k}^{(N)},
\end{multline}
where $\varphi_{k}: \mathbb{R}^{k}\rightarrow \mathbb{C}$ is
bounded, continuous and symmetric in its arguments and the
summation is over all $k$-tuples of distinct integers $
j_{1},\dots,j_{k}\in\{1,\ldots,N\}$.
According to the Wigner -- Dyson universality conjecture, the local behaviour
of the eigenvalues does not depend on the matrix probability
law (ensemble) and is determined only by the symmetry type of matrices (real
symmetric, Hermitian, or quaternion real in the case of real eigenvalues and orthogonal,
unitary or symplectic in the case of eigenvalues on the unit circle).
For example, the conjecture states that for Hermitian random matrices in the bulk of the spectrum
and in the range of parameters for which the eigenvectors are
delocalized
\begin{equation}\label{Un}
\displaystyle\frac{1}{(N\rho(E))^k}
R_k\left(E+\displaystyle\frac{\xi_1}{\rho(E)\,N},
\ldots,E+\displaystyle\frac{\xi_k}{\rho(E)\,N}\right)
\stackrel{w}{\longrightarrow}\det \Big\{\dfrac{\sin \pi(\xi_i-\xi_j)}
{\pi(\xi_i-\xi_j)}\Big\}_{i,j=1}^k,\quad N\to\infty
\end{equation}
for any fixed $k$. This means that the limit coincides with that for GUE.

 One of the main long standing problem in the field is to prove a  fundamental  physical conjecture formulated in late 80th (see \cite{Ca-Co:90}, \cite{FM:91}). 
The conjecture states that  the eigenvectors of $N\times N$ RBM are completely delocalized and the local  spectral  statistics  
governed  by  the Wigner-Dyson statistics  
for large bandwidth $W$, and by Poisson statistics for a small $W$ (with exponentially localized eigenvectors). The transition is conjectured to be sharp and for RBM 
in  one  spatial  dimension  occurs around  the  critical  value $W=\sqrt{N}$. This  is  an  analogue  of  the  celebrated  Anderson  
metal-insulator  transition  for random Schr$\ddot{\hbox{o}}$dinger operators.

The conjecture on the crossover in RBM with $W\sim\sqrt N$ is supported by physical derivation due to Fyodorov and Mirlin (see \cite{FM:91}) based on supersymmetric formalism, and also by the so-called Thouless scaling. However, so far there were only a few partial results on the mathematical level of rigour.
Localization of eigenvectors in the bulk of the spectrum was first shown for $W\ll N^{1/8}$ \cite{S:09}, and then the bound was improved to $N^{1/7}$  \cite{wegb:16} .
On the other side, 
by a development of the Erd\H{o}s-Schlein-Yau approach to Wigner matrices (see \cite{EYY:10}),  there were obtained some results where the weaker form of delocalization was 
proved for  $W\gg N^{6/7}$ in \cite{EK:11} , $W\gg N^{4/5}$ in \cite{Yau:12}, $W\gg N^{7/9}$ in \cite{HM:18}.
The combination of this approach with the new ideas based on quantum unique ergodicity gave first GUE/GOE gap distributions for RBM with $W\sim N$ \cite{BEYY:16}, 
and then were developed in \cite{BYY:18}  to obtain bulk universality and complete delocalization in the  range  $W\gg N^{3/4}$ (see review \cite{B:rev} for the details).
We mention also that at the edge of the spectrum, the transition for 1d band matrices (with critical exponent
$N^{5/6}$) was understood in \cite{S:band} by the method of moments.

The main aim of this paper is to prove bulk universality  in the range $W\gg \sqrt{N}$  for the Gaussian Hermitian block RBM, which are RBM with some specific covariance profile. More precisely, we consider Hermitian matrices $H_N$, $N=n W$ with elements $H_{jk,\alpha\beta}$,
where $j,k \in 1,\ldots, n$ (they parametrize the lattice sites) and $\alpha, \beta= 1,\ldots, W$ (they
parametrize the orbitals on each site). The entries $H_{jk,\alpha\beta}$ are random
Gaussian variables with mean zero such that
\begin{equation}\label{H}
\langle H_{j_1k_1,\alpha_1\beta_1}H_{j_2k_2,\alpha_2\beta_2}\rangle=\delta_{j_1k_2}\delta_{j_2k_1}
\delta_{\alpha_1\beta_2}\delta_{\beta_1\alpha_2} J_{j_1k_1}.
\end{equation}
Here $J_{jk}\ge 0$ are matrix elements of the positive-definite symmetric $n\times n$ matrix $J$, such that
\[
\sum\limits_{j=1}^nJ_{jk}=1/W.
\]
Such models were first introduced and studied by Wegner (see \cite{S-W:80}, \cite{We:79}). 

We restricted ourself to the case
\begin{equation}\label{J}
J=1/W+\beta\Delta/W, \quad \beta<1/4,
\end{equation}
where $W\gg 1$ and $\Delta$ is the discrete Laplacian on $[1,n]\cap \mathbb{Z}$ with Neumann boundary conditions.
This model  is one of the possible realizations of the Gaussian random band matrices with the band width $2W+1$
(note that the model can be defined similarly in any dimensions $d>1$   taking  $j,k\in [1,n]^d\cap \mathbb{Z}^d$
in (\ref{H})).

The main result of the paper is the following theorem
\begin{theorem}\label{thm:1}
In the dimension $d=1$ the behaviour of the second order correlation function (\ref{R}) of (\ref{H}) -- (\ref{J}), as $W\ge C n\log^5n $,
in the bulk of the spectrum coincides with those for the GUE. More precisely, if $\Lambda=[1,n]\cap \mathbb{Z}$ and $H_N$, $N=Wn$ are  matrices (\ref{H}) 
with $J$ of (\ref{J}), then 
for any $E\in (-2,2)$
\begin{equation}\label{Un_2}
(N\rho(E))^{-2}
R_2\left(E+\displaystyle\frac{\xi_1}{\rho(E)\,N},
E+\displaystyle\frac{\xi_2}{\rho(E)\,N}\right)\longrightarrow
1-\dfrac{\sin^2 (\pi(\xi_1-\xi_2))}
{\pi^2(\xi_1-\xi_2)^2},
\end{equation}
in the limit $n, W\to\infty$, $W\ge Cn \log^5 n$.
\end{theorem}
In order to prove Theorem \ref{thm:1}, we applied a rigorous form of the supersymmetric (SUSY) transfer matrix approach.
 The approach based on the fact that the main spectral characteristics of RBM  (such as density of states, second correlation functions, or
the average of an elements of the resolvent) can be written as the averages of certain observables in some SUSY statistical mechanics models containing both
complex and Grassmann variables
(so-called \textit{dual} representation in terms of SUSY). The rigorous analysis of such integral representation usually is very complicated and requires  powerful 
analytic and statistical mechanics tools. In our case the specific form of the covariance (\ref{J}) allows to combine the SUSY techniques with a transfer matrix approach. The supersymmetric transfer matrix formalism in this context 
 was first suggested by Efetov (see \cite{Ef}) and on a heuristic level it was adapted specifically for RBM
in \cite{FM:94} (see also references therein). However the rigorous application of the method to the main spectral characteristics of RBM is quite difficult due to
the complicated structure and non self-adjointness of the corresponding transfer operator. During the last years,  the techniques were developed step by step (see \cite{SS:rev} for details). First we applied it
 in \cite{SS:den} to obtain the precise estimate for the density of state. Then the method was elaborated in \cite{SS:ChP} to study the localized regime of the second correlation function
 of characteristic polynomials, which together with the result of \cite{TSh:14} finished the proof of the transition around $W\sim N^{1/2}$ on the level of characteristic 
 polynomials. The next crucial step was done in \cite{SS:sigma}, where we applied the techniques to the so-called sigma-model approximation, 
which is often used by physicists to study complicated statistical mechanics systems.  In such approximation spins take values in
some symmetric space ($\pm 1$ for Ising model, $S^1$ for the rotator, $S^2$ for the classical
Heisenberg model, etc.).  It is expected that sigma-models have all the qualitative physics
of more complicated models with the same symmetry (for more detailes see, e.g., \cite{Sp:12}). The sigma-model approximation for RBM was introduced by Efetov (see \cite{Ef}),
and the spins there are $4\times 4$ matrices with both complex and Grassmann entries.
As it was shown in \cite{SS:sigma}, the mechanism of the crossover for the sigma-model is essentially the same as for the correlation functions of characteristic polynomials (see \cite{SS:ChP}), 
but  the structure of the transfer
operator for the sigma-model is more complicated:  it is a $6\times 6$ matrix kernel whose entries are kernels depending on two unitary $2\times 2$ matrices $U,U'$ and two hyperbolic  $2\times 2$ 
matrices $S, S'$. As it will be shown below, 
in the case of the second correlation function of (\ref{H}) -- (\ref{J}) which is the main point of interest in this paper, the transfer operator $\mathcal{K}$ becomes
$70\times 70$ matrix whose elements are kernels defined on $L_2(U(2))\otimes L_2(\mathcal{H}_2^+L)$, where $U(2)$ is $2\times 2$ unitary group, 
$\mathcal{H}_2^+$ is a space of $2\times 2$ positive hermitian matrices, and $L=\hbox{diag}\{1,-1\}$, and so the spectral analysis of $\mathcal{K}$
provides serious structural problems. The key idea of this analysis is to prove that the main part of $\mathcal{K}$ is still $6\times 6$ matrix kernel appeared 
in the transfer operator corresponding to the sigma-model approximation.

We would like to mention also that the model (\ref{H}) -- (\ref{J}) in any dimension but with a finite number of blocks was analysed in \cite{TSh:14_1} via
SUSY techniques combined with a delicate steepest descent method. Combining the approach of \cite{TSh:14_1} with the Green's function comparison strategy, the delocalization for $W\gg N^{6/7}$  has been proved  in \cite{EB:15} for the block band matrices (\ref{H}) with a rather general non-Gaussian element's distribution. 

Notice that according to the properties of the Stieljes transform, to prove Theorem \ref{thm:1}, it suffices to show that
\begin{equation}\label{main}
\lim\limits_{\varepsilon\to 0}\lim\limits_{n\to\infty}F_2(z_1,z_2)\\
=1-\dfrac{\sin^2 (\pi(\xi_1-\xi_2))}
{\pi^2(\xi_1-\xi_2)^2},
\end{equation}
where 
\begin{align}\label{F_2}
& F_2(z_1,z_2):=\\ \notag
&(2\pi i N\rho(E))^{-2}\mathbf{E}\Big\{\Tr
\Big((H-z_1)^{-1}-(H-\overline{z}_1)^{-1}\Big)\cdot \Tr
\Big((H-z_2)^{-1}-(H-\overline{z}_2)^{-1}\Big)\Big\}
\end{align}
with
 \begin{align}\label{z}
 z_1&=E+i\varepsilon/N+\xi_1/N\rho(E),\quad z_2=
E+i\varepsilon/N+\xi_2/N\rho(E),\\ \notag
z_1^\prime&=E+i\varepsilon/N+\xi_1^\prime/N\rho(E),\quad z_2^\prime=
E+i\varepsilon/N+\xi_2^\prime/N\rho(E),
\end{align}
$\varepsilon>0$, and $\xi_1,\xi_2,\xi_1^\prime,\xi_2^\prime\in [-C,C]\subset
\mathbb{R}$.

Since
\begin{align*}\notag
& (2\pi i N\rho(E))^{2}F_2(z_1,z_2)\\ \notag
&=\mathbf{E}\Big\{\tr
(H-z_1)^{-1}\cdot \tr
(H-z_2)^{-1}\Big\}+\mathbf{E}\Big\{\overline{\tr
(H-z_1)^{-1}\cdot \tr
(H-z_2)^{-1}}\Big\}\\ \notag
&-\mathbf{E}\Big\{\tr
(H-z_1)^{-1}\cdot \tr
(H-\overline{z}_2)^{-1}\Big\}-\mathbf{E}\Big\{\overline{\tr
(H-z_1)^{-1}\cdot \tr
(H-\overline{z}_2)^{-1}}\Big\},
\end{align*}
we get
\begin{multline}\label{F_expr}
F_2(z_1,z_2)=(2\pi)^{-2}\dfrac{\partial^2}{\partial \xi_1^\prime \partial \xi_2^\prime}\Big(
\mathcal{R}_{Wn}^{+-}(E,\varepsilon,\bar\xi)+\overline{\mathcal{R}_{Wn}^{+-}}(E,\varepsilon,\bar\xi)\\-\mathcal{R}_{Wn}^{++}(E,\varepsilon,\bar\xi)-\overline{\mathcal{R}_{Wn}^{++}}(E,\varepsilon,\bar\xi)\Big)
\Big|_{\xi^\prime=\xi},
\end{multline}
where $\xi^\prime=\xi$ means $\xi_1^\prime=\xi_1$, $\xi_2^\prime=\xi_2$, and 
\begin{align}\label{G_2}
 \mathcal{R}_{Wn}^{+-}(E,\varepsilon,\bar\xi)&=\mathbf{E}\bigg\{\dfrac{\mdet(H-z_1)\mdet(H-\overline{z}_2)}
{\mdet(H-z_1^\prime)\mdet(H-\overline{z}_2^\prime)}\bigg\},\\ \notag
\mathcal{R}_{Wn}^{++}(E,\varepsilon,\bar\xi)&=\mathbf{E}\bigg\{\dfrac{\mdet(H-z_1)\mdet(H-z_2)}
{\mdet(H-z_1^\prime)\mdet(H-z_2^\prime)}\bigg\}
\end{align}
are generalized correlation functions with $\bar \xi=(\xi_1,\xi_2,\xi_1^\prime,\xi_2^\prime)$.

Introduce also
\begin{align}\label{a_pm}
a_{\pm}=\frac{iE\pm\sqrt{4-E^2}}{2},\quad a_+=:e^{ i\varphi_0}.
\end{align}
Now similarly to \cite{SS:sigma}, the main result of Theorem \ref{thm:1} follows from two theorems dealing with the behaviour of
generalized correlation functions (\ref{G_2}):
\begin{theorem}\label{t:2}
Given $\mathcal{R}_{Wn\beta}^{++}$ of (\ref{G_2}) ,(\ref{H}) and  (\ref{J}),
with  any fixed $\beta$, $\varepsilon>0$, and 
 $\xi=(\xi_1,\xi_2, \xi_1',\xi_2')\in\mathbb{C}^4$ 
 ($|\Im\xi_j|<\varepsilon\cdot\rho(E)/2$) we have,   as $n,W\to\infty$, $W\ge Cn\log^5 n$:
\begin{align}\label{G++_lim}
&\mathcal{R}_{Wn}^{++}(E,\varepsilon,\bar \xi)\to e^{{ia_+}(\xi_1'+\xi_2'-\xi_1-\xi_2)/{\rho(E)}},\\
&\frac{\partial^2\mathcal{R}_{Wn}^{++}}{\partial\xi_1'\partial\xi_2'}(E,\varepsilon,\xi)\Big|_{\xi'=\xi}\to-a_+^2/\rho^2(E).
\notag\end{align}
\end{theorem}
\begin{theorem}\label{t:1} Given $\mathcal{R}_{Wn\beta}^{+-}$ of (\ref{G_2}) ,(\ref{H}) and  (\ref{J}),
with  any fixed $\beta$, $\varepsilon>0$, and 
 $\xi=(\xi_1,\xi_2, \xi_1',\xi_2')\in\mathbb{C}^4$ 
 ($|\Im\xi_j|<\varepsilon\cdot\rho(E)/2$) we have,   as $n,W\to\infty$, $W\ge Cn\log^5 n$:
\begin{align}\label{t1.1}
&\mathcal{R}_{nW}^{+-}(E,\varepsilon,\xi)\to e^{iE(\sigma_1-\sigma_2)}e^{-2c\alpha_2}\Big(\Big(\frac{\alpha_1}{2\alpha _2}+\frac{\alpha_2}{2\alpha_1}\Big)
\sinh(2c\alpha_1)+\cosh(2c\alpha_1)\\&\hskip6cm-(\sigma_1-\sigma_2)^2\frac{\sinh(2c\alpha_1)}{2\alpha _1 \alpha_2}\Big) 
\notag\end{align}
where 
\begin{align}
\label{alp}
\sigma_1=\frac{\xi_1+\xi_2}{2i\rho(E)},\quad \sigma_2=\frac{\xi_1'+\xi_2'}{2i\rho(E)},\quad\alpha_1=\varepsilon+\frac{\xi_1-\xi_2}{2i\rho(E)},\quad
\alpha_2=\varepsilon+\frac{\xi_1'-\xi_2'}{2i\rho(E)},
\end{align}
and $c=\pi\rho(E)$.
%\begin{equation*}%\label{c_xi}
%C_{E,\xi}=e^{E(\xi_1+\xi_2-\xi_1'-\xi_2')/2\rho(E)}.
%\end{equation*}
In addition,
\begin{align}\label{t1.2}
\dfrac{\partial^2}{\partial \xi_1'\partial\xi_2'}\mathcal{R}_{nW}^{+-}(E,\varepsilon,\xi)\Big|_{\xi'=\xi}\to&\dfrac{1}{\rho^2(E)}+\dfrac{1-e^{-4c\alpha_1}}{4\alpha_1^2\rho^2(E)}.
\end{align}
\end{theorem}
Notice that Theorems \ref{t:2} -- \ref{t:1} and (\ref{F_2}) imply
\begin{align*}
\lim\limits_{\varepsilon\to 0}\lim\limits_{n,W\to\infty}F_2(z_1,z_2)&=\lim\limits_{\varepsilon\to 0}\Big(\dfrac{2+a_+^2+a_-^2}{4\pi^2\rho^2(E)}+\dfrac{2-e^{-4c\alpha_1}-e^{-4c\bar\alpha_1}}{4\pi^2(2\alpha_1\rho(E))^2}\Big)\\
&=\dfrac{(a_+-a_-)^2}{4\pi^2\rho^2(E)}-\dfrac{2-e^{i2\pi(\xi_1-\xi_2) }-e^{-i2\pi(\xi_1-\xi_2) }}{4\pi^2(\xi_1-\xi_2)^2}\\
&=1+\dfrac{(e^{i\pi(\xi_1-\xi_2) }-e^{-i\pi(\xi_1-\xi_2) })^2}{4\pi^2(\xi_1-\xi_2)^2}=1-\dfrac{\sin^2(\pi(\xi_1-\xi_2))}{\pi^2(\xi_1-\xi_2)^2},
\end{align*}
and so Theorem \ref{thm:1} indeed follows from Theorems \ref{t:2} -- \ref{t:1}. Here we used $a_+a_-=-1$, $a_+-a_-=2\pi\rho(E).$

\section{Representation of $\mathcal{R}_{Wn}^{+-}$ and $\mathcal{R}_{Wn}^{++}$ in the operator form}
We will use the integral representations for $\mathcal{R}_{Wn}^{+-}(E,\varepsilon,\bar\xi)$, $\mathcal{R}_{Wn}^{++}(E,\varepsilon,\bar\xi)$ obtained in \cite{TSh:14_1} (see Sec. 2). First  set
\begin{align}\label{L_cal}
&\mathcal{L}_\pm=L_\pm^{(1)}\cup L_\pm^{(2)},\quad L_\pm ^{(1)}=\{te^{\pm i\varphi_0}, \,t\in[0,1]\},\\
&L_\pm^{(2)}=\{e^{\pm i\varphi_0}+te^{\pm i\psi}, \,t\in[0,\infty]\},\quad \psi=\left\{\begin{array}{ll}
\varphi_0,&\varphi_0\le \pi/4\\
\max\{\varphi_0-\pi/4;\pi/2-\varphi_0\},&\varphi_0>\pi/4.
\end{array}\right.
\notag
\end{align}
\subsection{Operator expression for $\mathcal{R}_{Wn}^{+-}$}
\begin{proposition}[\textbf{\cite{TSh:14_1}}]\label{p:int_rep1} 
For $E\in (-2,2)$, $\hat{\xi}=\hbox{diag}\,\{\xi_1,\xi_2\}$, $\hat{\xi}'=\hbox{diag}\,\{\xi_1',\xi_2'\}$, and $\varepsilon>0$ we have
\begin{align}\label{int_rep_pm}
&\mathcal{R}_{Wn}^{+-}(E,\varepsilon,\bar\xi):=W^{4n}\displaystyle\int \prod\limits_{j=1}^n \dfrac{dX_jdY_j}{(-\pi^2)}
\prod\limits_{j=1}^n d\rho_jd\tau_j\\ \notag
&\times \exp\Big\{\dfrac{\beta W}{2}\sum\limits_{j=1}^{n-1}\Tr (X_j-X_{j+1})^2-
\dfrac{\beta W}{2}\sum\limits_{j=1}^{n-1}\Tr (Y_j-Y_{j+1})^2\Big\}\\ \notag
&\times  \exp\Big\{\dfrac{W}{2}\sum\limits_{j=1}^n \Big(\Tr X_j^2-\Tr Y_j^2-iW\Tr X_j\big(\Lambda_\varepsilon+\dfrac{\hat{\xi}}{N\rho(E)}\big)+iW\Tr Y_j\big(\Lambda_\varepsilon+\dfrac{\hat{\xi}'}{N\rho(E)}\big)\Big)\Big\}\\ \notag
&\times \exp\Big\{\beta\sum\limits_{j=1}^{n-1}\Tr (\rho_j-\rho_{j+1})(\tau_j-\tau_{j+1})-\sum\limits_{j=1}^{n}\Tr \rho_j\tau_j\Big\}
\prod\limits_{j=1}^n\dfrac{\mdet^W Y_j}{\mdet^W (X_j+W^{-1}\rho_jY_j^{-1}\tau_j)}
\end{align}
where  $\Lambda_\varepsilon=E\cdot I_2+i\varepsilon L/N$,
\begin{align*}
 L=\left(\begin{array}{ll}
1&0\\
0& -1
\end{array}
\right),
\end{align*}
$X_j$ is $2\times 2$ unitary matrix, $\rho_j$, $\tau_j$ are $2\times 2$ matrices whose entries are independent Grassmann variables, and
$Y_j=T_j^{-1}\hat{B}_jT_j$, $\,T_j\in \mathring{U}(1,1)$, 
\[
\hat{B}_j=\left(\begin{array}{ll}
b_{j,1}&0\\
0& b_{j,2}
\end{array}
\right),\quad b_{j,1}\in \mathcal{L}_+,\quad b_{j,2}\in \mathcal{L}_-. 
\]
 Here
\begin{align*}
%&d\overline{F}=\prod\limits_{j=1}^nd F_j,\quad d F_j= dX_j\,d Y_j\,d\rho_j\,d\tau_j,\\
&dY_j=\dfrac{\pi}{2}(b_{j,1}-b_{j,2})^2db_{j,1}\,db_{j,2}\, d\mu(T_j),\quad d\rho_jd\tau_j=\prod\limits_{l,s=1}^2d\rho_{j, ls}\,d\tau_{j, ls},
\end{align*}
and $dX_j$, $d\mu(T_j)$ are  Haar measures over $U(2)$ and $\mathring{U}(1,1)$ respectively.

%Then we can represent
%\begin{align*}
%&\mathcal{R}_{Wn}^{+-}(E,\varepsilon,\bar\xi)=\int \,dM_\pm(E,\hat{\xi}';\bar F),\\
%%&\dfrac{\partial}{\partial \xi_1^\prime } G_2^{+-}(z,\xi')\bigg|_{\xi'=\xi}=\int O_1(\bar{X}) \,dM_\pm(E,\hat{\xi};\bar F),\\
%%&\dfrac{\partial}{ \partial \xi_2^\prime} G_2^{+-}(z,\xi')\bigg|_{\xi'=\xi}=\int O_2(\bar{Y}) \,dM_\pm(E,\hat{\xi};\bar F),\\
%%&\dfrac{\partial^2}{\partial \xi_1^\prime \partial \xi_2^\prime} G_2^{+-}(z,\xi')\bigg|_{\xi'=\xi}=\int O_1(\bar{X}) \cdot O_2(\bar{Y}) \,dM_\pm(E,\hat{\xi};\bar F),
%\end{align*}
%where $ \hat{\xi}'=\hbox{diag}\,\{\xi, \xi_2', -\xi'_1,\xi\}$, $ \hat{\xi}=\xi\mathcal{L}$,
%\begin{align*}
%&O_1(\bar{X})=\dfrac{i}{n\rho(E)}\sum\limits_{j=1}^n\Big(X_{j, 11}+iE/2-\varepsilon/N+i\xi/N\rho(E)\Big),\\
%&O_2(\bar{Y})=-\dfrac{i}{n\rho(E)}\sum\limits_{j=1}^n\Big(Y_{j, 22}+iE/2+\varepsilon/N-i\xi/N\rho(E)\Big).
%\end{align*}

\end{proposition}

Let $\mathcal{F}_1:U(2)\to U(2)$,  $\mathcal{F}_2:\mathcal{H}_2^+L\to \mathcal{H}_2^+L$ be the operators of multiplication by
\begin{align}\label{F_cal}
\mathcal{F}_1 (X)&=\exp\Big\{
\frac{W}{4}\, \Tr X^2-\frac{iW}{2}\Tr X\big(\Lambda_\varepsilon+\dfrac{\hat{\xi}}{N\rho(E)}\big)-\frac{W}{2}\, \log
\mdet X\Big\},\\ \notag
\mathcal{F}_2 (Y)&=\exp\Big\{-
\frac{W}{4}\,\Tr Y^2+\frac{iW}{2}\Tr Y\big(\Lambda_\varepsilon+\dfrac{\hat{\xi}}{N\rho(E)}\big)+\frac{W}{2}\,\log
 \mdet Y\Big\}, 
\end{align}
Introduce compact integral operators $K_1$ and $K_2$ in $L_2[U(2)]$ and $L_2[\mathcal{H}_2^+L]$ with the kernels
\begin{align}\label{K_X,Y}
K_1(X,X')&=W^2\,\mathcal{F}_1 (X)\exp\Big\{-\dfrac{\beta W}{2}\Tr (X-X')^2\Big\}\mathcal{F}_1 (X'),\\
K_2(Y,Y')&=W^2\,\mathcal{F}_2 (Y)\exp\Big\{\dfrac{\beta W}{2}\Tr (Y-Y')^2\Big\}\mathcal{F}_2 (Y'),\notag\\
 \hat Q(\rho,\tau;\rho',\tau')&=\exp\Big\{\beta \,\Tr (\rho-\rho')(\tau-\tau')-\Tr\rho'\tau'\Big\} \notag \\
& \mathrm{det}^{-W}(I+(X')^{-1}\rho'(Y')^{-1}\tau'/W)
\label{Q} \end{align}
%\end{align}
and let
\begin{align}\label{K}
\mathcal{K}=K_1\otimes K_2\cdot \hat Q.
\end{align}
Then Proposition \ref{p:int_rep1} yield
\begin{align}\label{trans}
\mathcal{R}_{n\beta}^{+-}(E,\varepsilon,\bar\xi)=&W^4
\int  \mathcal{F}_1(X)\mathcal{F}_2(Y)\mathcal{K}^{n-1}(X,Y,\rho,\tau;X',Y',\rho',\tau')\mathcal{F}_1(X')\mathcal{F}_2(Y')\\ \notag
&\times  \mathrm{det}^{-W}(I+(X)^{-1}\rho'(Y)^{-1}\tau'/W)\dfrac{dX dY d\rho d\tau}{(-\pi^2)}\dfrac{dX' dY' d\rho' d\tau'}{(-\pi^2)}
\end{align}
Notice that $\hat{Q}$ and the operator of multiplication by $\mathrm{det}^{-W}(I+(X')^{-1}\rho'(Y')^{-1}\tau'/W)$  can be considered as operators acting on the space 
$\mathfrak{Q}_{256}\cong (L_2(U(2))\otimes L_2[\mathcal{H}_2^+L])^{256}$ of polynomials 
of Grassmann variables $\rho_{ls}'$, $\tau_{ls}'$,
$l,s=1,2$ with coefficients from $L_2(U(2))\otimes L_2[\mathcal{H}_2^+L]$. Hence, in the natural basis of monomials in Grassmann space, all our operators 
can be considered as $256\times 256$ matrices whose entries are operators on $L_2(U(2))\otimes L_2[\mathcal{H}_2^+L]$.

Thus, introducing the resolvent $ \mathcal{G}(z)=(\mathcal{K}-z)^{-1}$,
one can write
\begin{align}\label{rep_0}
\mathcal{R}_{Wn}^{+-}(E,\varepsilon,\bar\xi)&=W^4(\mathcal{K}^{n-1} f,g)=-\frac{W^4}{2\pi i}\oint\limits_{\omega_A} z^{n-1}(\mathcal{G}(z)f,g)dz,
\\ \notag
 f(X,Y,\rho,\tau)&=\mathcal{F}_1 (X)\mathcal{F}_2 (Y) e^{(0)},\\ \notag
  g(X,Y,\rho,\tau)&=\mathcal{F}_1 (X)\mathcal{F}_2 (Y) \cdot \big(\mathrm{det}^{-W}(I+(X)^{-1}\rho(Y)^{-1}\tau/W)\big)^t e^{(c)}
\notag\end{align}
since  from the consideration  below it follows  that all eigenvalues of $\mathcal{K}$ are
inside the circle $\omega_A=\{z:|z|=1+A/n\}$ with sufficiently large $A$.
Vectors  $e^{(0)}$ and $e^{(c)}$ here are the vectors in the space of Grassmann variables $\mathfrak{Q}_{256}$ corresponding
to 1 and to $\prod_{l,s=1}^2 \rho_{ls}\tau_{ls}$ respectively, and $\big(\mathrm{det}^{-W}(I+(X)^{-1}\rho(Y)^{-1}\tau/W)\big)^t $ is the transposed
 operator to the operator multiplication by $\mathrm{det}^{-W}(I+(X)^{-1}\rho(Y)^{-1}\tau/W)$ in the space $\mathfrak{Q}_{256}$.
Appearing of  $e^{(c)}$ in the inner product in the r.h.s. reflect the fact that, by definition, the integral over the Grassmann variables 
of some polynomial from $\mathfrak{Q}_{256}$ gives
the coefficient under $\prod_{l,s=1}^2 \rho_{ls}\tau_{ls}$.

\subsection{Operator expression for $\mathcal{R}_{Wn}^{++}$}
\begin{proposition}[\textbf{\cite{TSh:14_1}}]\label{p:int_rep2}
For $E\in (-2,2)$, $\hat{\xi}=\hbox{diag}\,\{\xi_1,\xi_2\}$, $\hat{\xi}'=\mathrm{diag}\,\{\xi_1',\xi_2'\}$, and $\varepsilon>0$ we have
\begin{align}\label{int_rep_+}
&\mathcal{R}_{Wn}^{++}(E,\varepsilon,\bar\xi)=W^{4n}\int \exp\Big\{\sum\nolimits_{n,W}\Big\}
%\prod\limits_{j=1}^n 
\prod\limits_{j=1}^n\dfrac{\mdet^W Y_j\,dX_jdY_j^+d\rho_jd\tau_j}{(-\pi^2)\mdet^W (X_j+W^{-1}\rho_j(Y_j^+)^{-1}\tau_j)}
%}\prod\limits_{j=1}^n
\end{align}
%\\ \notag
where
\begin{align*}
\sum\nolimits_{n,W}=&\dfrac{\beta W}{2}\sum\limits_{j=1}^{n-1}\Tr (X_j-X_{j+1})^2-
\dfrac{\beta W}{2}\sum\limits_{j=1}^{n-1}\Tr (Y_j^+-Y_{j+1}^+)^2
%\Big\}
%\\ \notag
%&\times  \exp\Big\{
+\dfrac{W}{2}\sum\limits_{j=1}^n \Big(\Tr X_j^2-\Tr (Y_j^+)^2\Big)\\ \notag
&+\sum\limits_{j=1}^n \Big(-iW\Tr X_j\big((iE+\varepsilon/N)I_2+\dfrac{\hat{\xi}}{N\rho(E)}\big)+iW\Tr Y_j^+\big((iE+\varepsilon/N)I_2+\dfrac{\hat{\xi}'}{N\rho(E)}\big)\Big)\\ \notag
&+\beta\sum\limits_{j=1}^{n-1}\Tr (\rho_j-\rho_{j+1})(\tau_j-\tau_{j+1})-\sum\limits_{j=1}^{n}\Tr \rho_j\tau_j,
\end{align*}
with  $Y_j^+\in \mathcal{H}_2^+$,
\[
dY_j^+=\mathbf{1}_{Y_j^+>0}\cdot d\Re Y^+_{12,j}\,
d\Im Y^+_{12,j}\,d Y^+_{11,j}\,d Y^+_{22,j},
\]
and $X_j$, $\rho_j$, $\tau_j$ are the same as in Proposition \ref{p:int_rep1}.
%Then we can represent
%\begin{align*}
%&\mathcal{R}_{Wn}^{++}(E,\varepsilon,\bar\xi)=\int \,dM_+(E,\hat{\xi}';\bar F^+),\\
%%&\dfrac{\partial}{\partial \xi_1^\prime } G_2^{++}(z,\xi')\bigg|_{\xi'=\xi}=\int O_1(\bar{X}) \,dM_+(E,\hat{\xi};\bar F^+),\\
%%&\dfrac{\partial}{ \partial \xi_2^\prime} G_2^{++}(z,\xi')\bigg|_{\xi'=\xi}=\int O_2(\bar{Y}^+) \,dM_+(E,\hat{\xi};\bar F^+),\\
%%&\dfrac{\partial^2}{\partial \xi_1^\prime \partial \xi_2^\prime} G_2^{++}(z,\xi')\bigg|_{\xi'=\xi}=\int O_1(\bar{X}) \cdot O_2(\bar{Y}^+) \,dM_+(E,\hat{\xi};\bar F^+),
%%
%\end{align*}
\end{proposition}

Similarly to (\ref{F_cal}) we define $\mathcal{F}_1^+:U(2)\to U(2)$,  $\mathcal{F}_2^+:\mathcal{H}_2^+\to \mathcal{H}_2^+$ as the operators of multiplication by
\begin{align}\label{F+_cal}
\mathcal{F}_1^+ (X)&=\exp\Big\{
\frac{W}{4}\, \Tr X^2-\frac{iW}{2}\Tr X\big((iE+\varepsilon/N)I_2+\dfrac{\hat{\xi}}{N\rho(E)}\big)-\frac{W}{2}\, \log
 \mdet X\Big\},\\ \notag
\mathcal{F}_2^+ (Y)&=\exp\Big\{-
\frac{W}{4}\, \Tr Y^2+\frac{iW}{2}\Tr Y\big((iE+\varepsilon/N)I_2+\dfrac{\hat{\xi}}{N\rho(E)}\big)+\frac{W}{2}\, \log
\mdet Y\Big\},
\end{align}
and introduce compact integral operators $K_1^+$ and $K_2^+$ in $L_2[U(2)]$ and $L_2[\mathcal{H}_2^+]$ with the kernels
\begin{align}\label{K+_X,Y}
K_1^+(X,X')&=W^2\,\mathcal{F}_1^+ (X)\exp\Big\{-\dfrac{\beta W}{2}\Tr (X-X')^2\Big\}\mathcal{F}_1^+ (X'),\\
K_2^+(Y,Y')&=W^2\,\mathcal{F}_2^+ (Y)\exp\Big\{\dfrac{\beta W}{2}\Tr (Y-Y')^2\Big\}\mathcal{F}_2^+ (Y').\notag
 \end{align}
%\end{align}
Set
\begin{align}\label{K+}
\mathcal{K}^+=K_1^+\otimes K_2^+\cdot Q.
\end{align}
Similarly to Section 2.1, introducing the resolvent $ \mathcal{G}^+(z)=(\mathcal{K}^+-z)^{-1}$,
one can write
\begin{align}\label{rep+_0}
\mathcal{R}_{Wn}^{++}(E,\varepsilon,\bar\xi)&=W^4((\mathcal{K}^+)^{n-1} f,g)=-\frac{W^4}{2\pi i}\oint_{\omega_A} z^{n-1}(\mathcal{G}^+(z)f,g)dz,
\\ \notag
  f^+(X,Y,\rho,\tau)&=\mathcal{F}_1^+ (X)\mathcal{F}_2^+ (Y) e^{(0)},\\ \notag
  g^+(X,Y,\rho,\tau)&=\mathcal{F}_1^+ (X)\mathcal{F}_2^+ (Y)\cdot \big(\mathrm{det}^{-W}(I+(X)^{-1}\rho(Y)^{-1}\tau/W)\big)^t e^{(c)},
\notag\end{align}
since  from the consideration  below it follows  that all eigenvalues of $\mathcal{K}^+$ are
inside the circle $\omega_A=\{z: |z|=1+A/n\}$ with sufficiently large $A$.

\section{Proof of Theorem \ref{t:1}}
In this section we are going to prove Theorem \ref{t:1} using (\ref{int_rep_pm}). 

Note that below we assume that $\alpha_1,\alpha_2$ of (\ref{alp}) are real and $\alpha_1>\varepsilon/2$, $\alpha_2>\varepsilon/2$, since
  it suffices to prove Theorem \ref{t:1} only for $\xi$ such that 
\begin{align}\label{cond_xi}
&\Re\xi_1=\Re\xi_2,\quad\Re\xi_1'=\Re\xi_2',\quad \xi_1,\xi_2,\xi_1',\xi_2'\in\Omega_{c\varepsilon}\\
&\Omega_{c\varepsilon}=\{\xi:\Im\xi>-c\varepsilon\},\quad (0<c<1).
\notag\end{align}
 Indeed, assume that  $\{\mathcal{R}_{Wn}^{+-}(E,\varepsilon,\xi)\}$ are uniformly bounded in  $n,W$  
for $\xi_1,\xi_2,\xi_1',\xi_2'\in\Omega_{c\varepsilon}$. Consider
$\{\mathcal{R}_{Wn}^{+-}(E,\varepsilon,\xi)\}$  as  functions on $\xi_1$ with fixed $\xi_2,\xi_1',\xi_2'$ such that $\Re\xi_1'=\Re\xi_2'$.
Since these functions are analytic in $\Omega_{c\varepsilon}$,  the standard complex analysis argument yields that (\ref{t1.1})  on the segment
$\Re\xi_1=\Re\xi_2$ implies (\ref{t1.1})  for any $\xi_1\in\Omega_{c\varepsilon}$, hence for any $\xi_1,\xi_2\in \Omega_{c\varepsilon}$.
Then, fixing any $\xi_1,\xi_2,\xi_2'$, we can consider $\{\mathcal{R}_{n\beta}^{+-}(E,\varepsilon,\xi)\}$
as a sequence of analytic functions on $\xi_1'$. Since, by the above argument, (\ref{t1.1}) is valid on the segment $\Re\xi_1'=\Re\xi_2'$, 
the same argument yields that (\ref{t1.1}) is valid for any $\xi_1',\xi_2'$. 

%Therefore, it is enough to prove Theorem \ref{t:1} for real $\alpha_1>\varepsilon/2$, 
%$\alpha_2>\varepsilon/2$.

To check that $\{\mathcal{R}_{Wn}^{+-}(E,\varepsilon,\xi)\}$ are uniformly bounded in  $n,W$  for $\xi_1,\xi_2,\xi_1',\xi_2'\in\Omega_{c\varepsilon}$, we apply the  Cauchy-Schwartz inequality to $\mathcal{R}_{Wn}^{+-}(E,\varepsilon,\xi)$ in the form (\ref{G_2}). Then we get
\begin{align*}
&|\mathcal{R}_{Wn}^{+-}(E,\varepsilon,\xi)|^2\le |\mathcal{R}_{Wn}^{+-}(E,\varepsilon,\xi_1)|\,|\mathcal{R}_{Wn}^{+-}(E,\varepsilon,\xi_2)|
%\Rightarrow &\mathcal{R}_{Wn}^{+-}(E,\varepsilon,\xi)|^2\le |\mathcal{R}_{Wn}^{+-}(E,\varepsilon,\xi_1)|\,|\mathcal{R}_{Wn}^{+-}(E,\varepsilon,\xi_2)|
\end{align*}
where $\xi_1=(\xi_1,\xi_1,\xi_1',\xi_1')$, $\xi_2=(\xi_2,\xi_2,\xi_2',\xi_2')$. Since $\xi_1,\xi_2$ satisfy (\ref{cond_xi}), the uniform boundedness of
the r.h.s. follows from the uniform convergence (in $\xi$ satisfying (\ref{cond_xi})) of (\ref{t1.1}).

\subsection{Preliminary transformation of $\mathcal{K}$}\label{s:K}
%\subsection{Transformation of $\mathcal{K}$}
 Consider (\ref{int_rep_pm}). Diagonalizing $X_i$, $Y_i$ by matrices $U_i\in \mathring{U}(2)$,  $S_i\in \mathring{U}(1,1)$ such that
\begin{align}\label{U,V}
X_i=U_i ^*D_{ai} U_i,\quad D_{ai}=\mathrm{diag }\{a_{i1},a_{i2}\},\quad Y_i=S_i^{-1}D_{bi}S_i, \quad
D_{bi}=\mathrm{diag }\{b_{i1},b_{i2}\},
\end{align}
we change the measure $dX_i dY_i/(-\pi^2)$ from (\ref{int_rep_pm}) to
\[
(2\pi)^{-2}d \bar a_i \,d \bar b_i  \,d U_i \,d S_i\, (a_{i1}-a_{i2})^2(b_{i1}-b_{i2})^2,
\]
where $d\bar a_i=d a_{i1} d a_{i2}$, $a_{i1}, a_{i2}$ belong to the unit circle $\mathbb{T}=\{z: |z|=1\}$,  $d\bar b_i=d b_{i1} d b_{i2}$ with 
$b_{i1}\in \mathcal{L}_+$, $b_{i2}\in\mathcal{L}_-$, and $dU_i$, $dS_i$ are the Haar measures on
$\mathring{U}(2)$, $ \mathring{U}(1,1)$ correspondingly.

Consider also the change of Grasmann variables
\begin{align}\label{r',t'}
(U_i\rho_i S^{-1}_i)_{\alpha\beta}\to \rho_{i,\alpha\beta},\quad(S_i\tau_i U^{*}_i)_{\alpha\beta} \to\tau_{i,\alpha\beta},
\end{align}
and let $\widehat C'(U_i,S_i)$ be the matrix corresponding to the above change in the space $\mathfrak{Q}_{256}$ of all polynomials of
 Grassmann variables $\rho_{ls}$, $\tau_{ls}$, $l,s=1,2$ with coefficients on $L_2[\mathring{U}(2)]\otimes L_2[\mathring{U}(1,1)]$.
 
One can see that
\begin{align}\label{I_1}
 &\mathcal{K}^{n-1}=C_{\bar\xi,\varepsilon}^{(n-1)/n}\int\prod_{j=2}^{n-1} dU_jdS_jd \bar a_jd\bar b_j d\rho_jd\tau_j\prod_{i=1}^{n-1}\widehat F(\bar a_{i},\bar b_{i},U_{i},S_{i})\widehat F(\bar a_{i+1},\bar b_{i+1}U_{i+1},S_{i+1})\\
&\times A_{ab}(\bar a_{i+1},\bar a_{i};\bar b_{i+1},\bar b_{i})K_{US}(U_{i}^*U_{i+1},S_i^{-1}S_{i+1})(\widehat C'(U_i,S_i))^{-1} \beta^{-2} \hat Q(\rho,\tau;\rho',\tau')\widehat C'(U_i,S_i)\notag\\
&C_{\bar\xi,\varepsilon}=e^{iE(\sigma_1-\sigma_2)+2\pi\rho(E)(\alpha_1-\alpha_2)}\notag
\notag\end{align}
where $\widehat Q$ is defined by (\ref{Q}) with $X'$ and $Y'$ replaced by diagonal matrices $\mathrm{diag}\{a'_1,a'_2\}$ and $\mathrm{diag}\{b_1',b_2'\}$, $\bar a_i=\{a_{i1},a_{i2}\}$, $\bar b_i=\{b_{i1},b_{i2}\}$, and
\begin{align}\label{K_U}
&K_U(U^*U'):=(t_a W)e^{-t_aW|(U^*U')_{12}|^2},\quad t_a=\beta(a_1-a_2)(a_1'-a_2');\\
&K_{S}(S^{-1}S'):=(t_b W)e^{-t_bW|(S^{-1}S')_{12}|^2},\quad t_b=\beta(b_1-b_2)(b_1'-b_2');\notag\\
&K_{US}(U^*U',S^{-1}S')=K_U(U^*U')K_S(S^{-1}S').
\notag\end{align}
Constant $C_{\bar\xi,\varepsilon}$  is a kind of "normalization constant" here, which appear because of our definition of $F_U(\bar a,U)$ and
$F_S(\bar b,U)$ (see (\ref{F_a,b})).

We also defined
\begin{align}\label{K_ab}
&A_a(a,a')= \Big(\dfrac{W}{2\pi}\Big)^{1/2}e^{- W\Lambda(a,a')};\quad
A_b(b,b')= \Big(\dfrac{W}{2\pi}\Big)^{1/2}e^{W\Lambda(b,b')};\\
&\Lambda(x,y)=\frac{\beta }{2}(x-y)^2-\frac{1}{2}\phi_0(x)-\frac{1}{2}\phi_0(y)+\Re\phi_0(a_+);\notag\\
&\phi_0(x)={x^2}/{2}-ixE-\log x;\label{phi_0} \\
&A_{ab}(\bar a,\bar b; \bar a', \bar b')=A_a(a_1,a_1') A_a(a_2,a_2')A_b(b_1,b_1') A_b(b_2,b_2');\notag
\end{align}
and put
\begin{align}\label{F_a,b}
&F_U(\bar a,U)=\exp\Big\{\frac{\sigma_1(a_1+a_2)}{n}+\frac{\alpha_1(a_1-a_2)}{2n}\varphi_U-\frac{\eta_1}{n}\Big\},\,\quad \varphi_U=\Tr ULU^*L\\
&F_S(\bar b,S)=\exp\Big\{-\frac{\sigma_2(b_1+b_2)}{n}-\frac{\alpha_2(b_1-b_2)}{2n}\varphi_S+\frac{\eta_2}{n}\Big\},\,\quad  \varphi_S=\Tr SLS^{-1}L\notag\\
&\eta_{1,2}=\sigma_{1,2}(a_++a_-)+\alpha_{1,2}(a_+-a_-),\quad \widehat F(\bar a,\bar b, U, S)=F_U^{1/2}(\bar a,U)F_S^{1/2}(\bar b,S)\notag
\end{align}
where $\alpha_{1,2}$ and $\sigma_{1,2}$ are defined in (\ref{alp}).

  Notice that  $\widehat Q$ of (\ref{Q}) and the operator multiplication by $\mathrm{det}^{-W}(I+(X)^{-1}\rho(Y)^{-1}\tau/W)$ keep the difference
  between numbers of $\rho$ and  $\tau$.
 Thus for all Grassmann operators below we can consider the restriction of these operators  to the subspace $\mathfrak{Q}_{70}\subset \mathfrak{Q}_{256}$ corresponding to the vectors with equal numbers of $\rho$ and $\tau$ (it is easy to see that there are $70$ of such monomials).
 To simplify notation, all such restriction will be denoted by the same symbols.

% By  construction,
% \[ C\widehat e_1=\widehat e_1, \quad C\widehat e_2=\det U\det S\widehat e_2=\widehat e_2,\quad (C^{-1})^*\widehat e_2=\widehat e_2, \]
 Since $K_{US},A_{a,b}$ do not contain Grassmann variables, we can move $(\widehat C'(U_i,S_i))^{-1} $ in each multiplier of (\ref{I_1}) to the left.
 Moreover, since   $\widehat C'(U,S)$ corresponds to the change of variables (\ref{r',t'}), we have
 \begin{equation}\label{C_i}
\widehat C'(U_{i+1},S_{i+1})(\widehat C'(U_i,S_i))^{-1}=\widehat C'(U_{i+1}U_i^*,S_{i+1}S_i^{-1}).
 \end{equation} 
 Hence (\ref{rep_0}) can be rewritten as
 \begin{align}\label{repr} 
&\mathcal{R}_{Wn}^{+-}(E,\varepsilon,\bar\xi)=
W^4C_{\bar\xi,\varepsilon}(\widetilde{\mathcal{K}}^{n-1} f,g)=-\frac{C_{\bar\xi,\varepsilon}W^4}{2\pi i}\oint_L z^{n-1}(\mathcal{G}(z)f,g) dz,\quad\\
&\mathcal{G}(z)=(\widetilde{\mathcal{K}}-z)^{-1},\notag\quad
 f=v(\bar a,\bar b,U,S)(a_1-a_2)(b_1-b_2) e^{(0)},\notag\\
& \hskip3.5cm g= v(\bar a,\bar b, U,S) (a_1-a_2)(b_1-b_2) (\widehat D(\bar a,\bar b))^{(t)}e^{(c)}\notag\\
& v(\bar a,\bar b, U,S):=(2\pi)^{-1}e^{W\phi_0(a_1)/2+W\phi_0(a_2)/2}e^{-W\phi_0(b_1)/2-W\phi_0(b_2)/2}\widehat F(\bar a,\bar b,U,S) \notag\\
&\widetilde{\mathcal{K}}(\bar a,\bar b,\rho,\tau, U,S;\bar a',\bar b',\rho', \tau',U',S' )
\notag\\
&\quad \quad=A_{ab}(\bar a,\bar a';\bar b,\bar b')\widehat F(\bar a,\bar b,U,S)\widehat Q(\rho,\tau;\rho',\tau')\widehat C(\tilde U,\tilde S)\widehat F(\bar a',\bar b',U',S') 
\label{repr'}\end{align}
where  the operator $ (\widehat D(\bar a,\bar b))^{(t)}$ being the transposed
 operator to $\widehat D(\bar a,\bar b)$ which corresponds to the multiplication in the Grassmann space by 
 $\big(\det(1+W^{-1}D_a^{-1}\rho D_b^{-1}\tau)\big)^{-W}$ and we set
 \begin{equation}\label{hat_C}
\widehat C(\tilde U,\tilde S)=K_{US}(\tilde U,\tilde S)\widehat C'(\tilde U,\tilde S),\quad \tilde U=U(U')^*,\quad \tilde S=S(S')^{-1}
\end{equation} 
To study the entries of $\widehat C$, it is convenient to introduce "difference" operators.
\begin{definition}\label{def_diff}
 Given function $v$ defined on the space of $2\times 2$ matrices, we denote by $\big( v(U)\big)_U$  the integral operator with the kernel 
$v(U(U')^*)K_{U}(U(U')^*)$ and by
$\big( v(S)\big)_S$  the integral operator with the kernel $v(S(S')^{-1})K_{S}(S(S')^{-1})$, where
$K_{U}$ and $K_{S}$ were defined in (\ref{K_U}). 
 \end{definition}

Recall that $\mathcal{G}(z)$ acts in the Grassmann space $\mathfrak{Q}_{70}$ and it can be considered as a $70\times 70$ matrix whose entries are operators on $L_2(U(2))\otimes L_2[\mathcal{H}_2^+L]$. Our strategy of the proof of Theorem \ref{t:1} is to replace the resolvent $\mathcal{G}(z)$ by some $6\times 6$ block matrix
whose entries are operators in $L_2(\mathring{U}_2)\otimes L_2(\mathring{U}(1,1))$. 
To this aim,  we will use multiple times the following simple  proposition 
\begin{proposition}\label{p:res} Let the matrix $H(z)$ has the block  form
\begin{equation*}
H(z)=\left(\begin{array}{cc}H_{11}&H_{12}\\H_{21}&H_{22}\end{array}\right).
\end{equation*}
Then 
\begin{align}\label{inv}
G(z)&:=H^{-1}(z)
% \left(\begin{array}{cc}H_{11}&H_{12}\\H_{21}&H_{22}\end{array}\right)^{-1}&
 =\left(\begin{array}{cc}G_{11}&-G_{11}H_{12}H_{22}^{-1}\\-H_{22}^{-1}H_{21}G_{11}&H_{22}^{-1}+H_{22}^{-1}H_{21}G_{11}H_{12}H_{22}^{-1}\end{array}\right)\\
G_{11}&=M_1^{-1},\quad M_1=H_{11}-H_{12}H_{22}^{-1}H_{21},\
\notag\end{align}
If  $H_{22}^{-1}$ is an analytic function for $|z|>1-\delta$, and $\|H_{22}^{-1}\|\le C$, then
\begin{align}\label{lres}
&\oint_{\omega_A}z^{n-1} (G(z)f, g)dz=\oint_{\omega_A}z^{n-1}(G_{11} f^{(1)}(z), g^{(1)}(z)) dz+ O(e^{-nc})\\
& f^{(1)}(z)=f_0-H_{12}H_{22}^{-1}f_1,\quad  g^{(1)}(z)=g_0-H_{21}^T(H_{22}^T)^{-1}g_1
\notag\end{align}
where $\omega_A=\{z:|z|=1+A/n\}$, $f=(f_0,f_1)$, $g=(g_0,g_1)$ where $f_0$ and $g_0$ are the projection of $f$ and $g$ on the subspace, corresponding to
$H_{11}$, while  $f_1$ and $g_1$ are the projection of $f$ and $g$ on the subspace, corresponding to
$H_{22}$.
\end{proposition}
\textit{Proof.} Formula (\ref{inv})  is the well-known block matrix inversion formula. Now apply the formula (\ref{inv})  and write
\begin{align*}
&\oint_{\omega_A}z^{n-1} (G(z)f, g)dz=\oint_{\omega_A}z^{n-1}(G_{11} f^{(1)}(z), g^{(1)}(z)) dz+ \oint_{\omega_A} z^{n-1}(H_{22}^{-1}f_1,g_1) dz.
\end{align*}
For the second integral change the integration contour from ${\omega_A}$ to $|z|=1-\delta$. Then the inequality
\[|z|^{n-1}\le (1-\delta)^{n-1}\le Ce^{-nc}\]
yields (\ref{lres}). 

$\square$

It is easy to check that  points $a_\pm$ (see (\ref{a_pm})) are the stationary points of the function $\phi_0$ of (\ref{K_ab}).
We start the proof from the restriction of the integration with respect to $\bar a_i,\bar b_i$ by
the neighbourhood of  $a_\pm$.
Set
\begin{align*}
\Omega_{+}=&\{x:|x-a_+|\le\log W/W^{1/2}\}, \quad \Omega_{-}=\{x:|x-a_-|\le\log W/W^{1/2}\}, \\
\widetilde \Omega_{\pm}=&\{a_1,a_1',b_1,b_1'\in\Omega_+, a_2,a_2',b_2,b_2'\in\Omega_-\}, \\
\widetilde \Omega_{+}=&\{a_1, a_2,a_1',a_2'b_1,b_1'\in\Omega_+ ,\,b_2,b_2'\in\Omega_-\},\\
\widetilde \Omega_{-}=&\{a_1, a_2,a_1',a_2'b_2,b_2'\in\Omega_- ,\,b_2,b_2'\in\Omega_+\}
\end{align*}
and let $\mathbf{1}_{\widetilde \Omega_{\pm}}$, $\mathbf{1}_{\widetilde\Omega_{+}}$ $\mathbf{1}_{\widetilde\Omega_{-}}$ be indicator functions of the above domains.

\begin{lemma}\label{l:b_K_ab}
Let $\mathcal{L}_\pm$ be as defined in  (\ref{L_cal}). Then
\begin{align}\label{b_K.1}
&\Big(\int_{\mathcal{L}_+\setminus\Omega_+}+\int_{\mathcal{L}_-\setminus\Omega_-}\Big)|A_b(b,b')||db'|\le Ce^{-c\log^2W},\\
&\int_{\mathbb{T}\setminus(\Omega_+\cup\Omega_-)}|A_a(a,a')||da'|\le Ce^{-c\log^2W}.
\label{b_K.1'}\end{align}

\end{lemma}
The proof of the lemma is given in Section \ref{s:b_K_ab}.

Lemma \ref{l:b_K_ab} yields that 
\begin{align}\label{b_out}
\int dU'dS'd\bar a'd\bar b'(1-\mathbf{1}_{\Omega_{\pm}}-\mathbf{1}_{\widetilde\Omega_{+}}-\mathbf{1}_{\widetilde\Omega_{-}})
\|\widetilde{\mathcal{K}}\|\le e^{-c\log^2W}
\end{align}
Set 
\[H_{11}=(\mathbf{1}_{\Omega_{\pm}}\widetilde{\mathcal{K}}\,\mathbf{1}_{\Omega_{\pm}})\oplus(\mathbf{1}_{\widetilde\Omega_{+}}\widetilde{\mathcal{K}}\,\mathbf{1}_{\widetilde\Omega_{+}})
\oplus(\mathbf{1}_{\widetilde\Omega_{-}}\widetilde{\mathcal{K}}\,\mathbf{1}_{\widetilde\Omega_{-}})=\mathcal{K}_\pm\oplus \mathcal{K}_{+}\oplus \mathcal{K}_-.\]
Then (\ref{b_out}) yields
\[\|H_{22}\|+\|H_{12}\|+\|H_{21}\|\le Ce^{-c\log^2W}.\]
Therefore for any $|z|>\frac{1}{2}$
\[\|H_{12}(H_{22}-z)^{-1}H_{21}\|\le  Ce^{-c\log^2W}.\]
Moreover, it will be proven below that 
\begin{equation*}
\|(H_{11}-z)^{-1}\|\le Cn,
\end{equation*}
and so for $G_{11}$ of (\ref{inv}) we have
\[
\|G_{11}-(H_{11}-z)^{-1}\|\le e^{-c\log^2W/2}
\]
Thus
we obtain by Proposition \ref{p:res}
\begin{align}
 \mathcal{R}_{Wn}^{+-}(E,\varepsilon,\bar\xi)=-\dfrac{W^4C_{\bar\xi,\varepsilon}}{2\pi i}\oint z^{n-1}((H_{11}-z)^{-1}f,g)dz+O(e^{-c\log^2W/2})+O(e^{-nc_1}).
\label{repr1} \end{align}
In view of  the block structure of $H_{11}$, its resolvent  also has a block structure, hence 
\begin{align}\notag
\mathcal{R}_{Wn}^{+-}(E,\varepsilon,\bar\xi)=&-\frac{W^4C_{\bar\xi,\varepsilon}}{2\pi i}\oint_{\omega_A} z^{n-1}
(\mathcal{G}_{\pm}(z)f_{\pm},g_\pm)dz-\frac{W^4C_{\bar\xi,\varepsilon}}{2\pi i}\oint_{\omega_A}  z^{n-1}(\mathcal{G}_{+}(z)f_{+},g_+)dz
\\&-\frac{W^4C_{\bar\xi,\varepsilon}}{2\pi i}\oint_{\omega_A}  z^{n-1}(\mathcal{G}_{-}(z)f_{-},g_-)dz
=I_{\pm}+I_{+}+I_{-},
\label{repr2}\end{align}
where 
\begin{align*}
\mathcal{G}_{\pm}=&(\mathcal{K}_{\pm}-z)^{-1},\quad 
 \mathcal{G}_{+}(z)=(\mathcal{K}_{+}-z)^{-1},\quad
\mathcal{G}_{-}(z)=(\mathcal{K}_{-}-z)^{-1} 
\end{align*}
and $f_{\pm},f_+,f_-$, $g_{\pm},g_+,g_-$ are projections of $f$ and $g$ onto the subspaces corresponding to
$\mathcal{K}_{\pm},\,\mathcal{K}_{+},\,\mathcal{K}_{-}$

\subsection{Analysis of $I_\pm$}
We start from the  analysis of  $\widehat Q$ of (\ref{Q}).
  We shall call the product of Grassmann variables "good", if it is composed only from the multipliers of the form 
 \begin{align}\label{n_mu,nu}
 n_{\mu\nu}=\rho_{\mu\nu}\tau_{\nu\mu}.
 \end{align}
 We shall  call "semi-good" the expressions
 \begin{align}\label{eta}
 \eta_{1,2}\, p(n_{11},n_{22}),\quad \eta_1=\rho_{12}\tau_{12},\quad \eta_2=\rho_{21}\tau_{21}
 \end{align}
 with a polynomial $p$.
  All the rest Grassmann expressions we call "non-good". By (\ref{Q}) (recall that now $X'$ and $Y'$ are diagonal matrices $D_a$ and $D_b$ of (\ref{U,V}))
  \begin{align}\notag
  \beta^{-2}\widehat Q(\rho,\tau;\rho',\tau')=&\beta^{-2} \Big(\prod_{\mu,\nu=1,2}e^{\beta(\rho_{\mu\nu}-\rho'_{\mu\nu})(\tau_{\nu\mu}-\tau'_{\nu\mu})}
  e^{-(1+a^{-1}_\mu b^{-1}_\nu)\rho'_{\mu\nu}\tau'_{\nu\mu}}\Big) (1+W^{-1}\mathcal{X}+W^{-2}\mathcal{Y})\\
  =&\widehat\Pi \, (1+W^{-1}\mathcal{X}+W^{-2}\mathcal{Y})), \label{hat_Q} 
  \end{align}
  where
  \begin{align}\label{dec}
\widehat \Pi \,=\left(\begin{array}{ccc}\Pi&0&0\\0&\Pi'&0\\0&0&\widetilde \Pi\end{array}\right).
 \end{align}
 
%  \begin{align}\label{hat_Pi}
%  \widehat\Pi= \Pi\oplus \Pi'\oplus\widetilde \Pi
% \end{align}
 with $\Pi$ corresponding to "good" vectors, $\Pi'$ to "semi-good" ones, and $\widetilde \Pi$ to "non-good", i.e.
 \begin{align}\label{Q(x)}
& \Pi:=\beta^{-2} (Q_{12}\otimes Q_{21})\otimes Q_{11}\otimes Q_{22},\quad \Pi'= I_2\otimes Q_{11}\otimes Q_{22},\\
& Q_{\mu\nu}:=Q(c_{\mu\nu}),\quad Q(c)=\left(\begin{array}{ll}\beta-c&1\\-\beta c &\beta\end{array}\right),
 \quad c_{\mu\nu}:=1+a^{-1}_{\mu}b^{-1}_\nu,\,\,\mu,\nu=1,2.
\notag \end{align}
Matrices $\mathcal{X},\mathcal{Y}$ correspond to the multiplication  by
 \begin{align}\label{X,Y}
\mathcal{X}= &-\frac{1}{2}\Tr(D_{a'}^{-1}\rho' D_{b'}^{-1}\tau')^2
\\
\mathcal{Y}=&\frac{1}{8} \Big((\Tr D_{a'}^{-1}\rho' D_{b'}^{-1}\tau')^2\Big)^2-\frac{1}{3}\Tr(D_{a'}^{-1}\rho' D_{b'}^{-1}\tau')^3+O(W^{-1})
 \notag\end{align}
with $D_{a'}$, $D_{b'}$ of (\ref{U,V}).

Let us make a change of variables
 \begin{align}\label{ch_ab}
 &a_{1i}=a_+(1+i\theta_+\widetilde{a}_{1i}/\sqrt{W}),\quad a_{2i}=a_-(1+i\theta_-\widetilde{a}_{2i}/\sqrt{W}),\\
 &b_{1i}=a_+(1+\theta_+\widetilde{b}_{1i}/\sqrt{W}),\quad b_{2i}=a_-(1+\theta_-\widetilde{b}_{2i}/\sqrt{W}),
\notag \end{align}
where $a_\pm$ is defined in (\ref{a_pm}), and $\theta_\pm$ will be chosen later (see (\ref{kappa})).
Notice that this change of variables replaces the factor $W^4$ in front of the first integral in the r.h.s. of (\ref{repr2}) to $W^2$.

 Then we have
 \begin{align}\label{exp_Pi}
 &\Pi=C_*\Pi_0
(1+O(W^{-1/2}\log W)),\quad C_*=\lambda_0^+\lambda_0^-\\
 &\Pi'=C_*\Pi_0'(1+O(W^{-1/2}\log W)),\notag\\
 & \widetilde \Pi=C_*\widetilde \Pi_0(1+O(W^{-1/2}\log W))
\notag 
\end{align}
where
\begin{align*}
&\Pi_0=C_*^{-1}\Pi\Big|_{\widetilde{a}_{1}=\widetilde{a}_{2}=\widetilde{b}_{1}=\widetilde{b}_{2}=0}, \quad
 \Pi_0'=C_*^{-1}\Pi'\Big|_{\widetilde{a}_{1}=\widetilde{a}_{2}=\widetilde{b}_{1}=\widetilde{b}_{2}=0},  \quad
\widetilde \Pi_0=C_*^{-1}\widetilde \Pi\Big|_{\widetilde{a}_{1}=\widetilde{a}_{2}=\widetilde{b}_{1}=\widetilde{b}_{2}=0}.
\notag\end{align*}
In addition, notice that for $\mathcal{X}_0=\mathcal{X}\Big|_{\widetilde{a}_{1}=\widetilde{a}_{2}=\widetilde{b}_{1}=\widetilde{b}_{2}=0}$ we have
\begin{align}\label{X_0}
\mathcal{X}_0=-\Big((a_+^{-2}n'_{11}-a_-^{-2}n'_{22})(n'_{21}-n'_{12})+(\rho'_{11}\tau'_{12}\rho'_{22}\tau'_{21}+\rho'_{12}\tau'_{22}\rho'_{21}\tau'_{11})\Big).
\end{align} 
 Evidently
 \begin{align}\Pi_0=&\big(Q_{0}\otimes Q_0\otimes Q_+\otimes Q_-\big),\quad
 Q_{0}:=\beta^{-1}Q(0),\quad Q_{\pm}:=Q(c_\pm)/\lambda_0^\pm,
\label{Q_pm} 
\end{align}
where
\begin{align}
c_\pm=&1+a_{\pm}^{-2},
\label{c_pm}\end{align}
 and $\lambda_0^+:=\lambda_0(Q(c_+))$ and $\lambda_0^-:=\lambda_0(Q(c_-))$ are the biggest eigenvalues of $Q(c_\pm)$. It is easy to see that 
 eigenvalues of $Q(c_\pm)$ are solutions of the equations
\begin{align}
&\lambda^2-(2\beta-c_\pm)\lambda+\beta^2=0\notag\\
&\Rightarrow \lambda_0^\pm=\beta-c_\pm/2-\sqrt{c_\pm^2/4-\beta c_\pm},\quad |\lambda_0^\pm|>\beta.
 \label{la_pm} \end{align}
Let us study the structure of $\Pi_0$.
One can  see that the vectors $\{e_0^\pm,e_1^\pm\}$
\begin{align}\label{e^pm}
&e^{+}_{0}=e^{c^+_{0}n_{11}},\quad e^{+}_{1}=e^{c^+_{1}n_{11}},\\ \notag
&e^{-}_{0}=e^{c^-_{0}n_{22}},\quad e^{-}_{1}=e^{c^-_{1}n_{22}},\\ \notag
&c^\pm_{0}=c_\pm/2+\sqrt{c_\pm^2/4-\beta c_\pm},\quad c^\pm_{1}=c_\pm/2-\sqrt{c_\pm^2/4-\beta c_\pm}
\end{align}
are eigenvectors for $Q_\pm$, such that  $\lambda_{0\pm}=1$, $\lambda_{1\pm}<1$. Evidently they make a basis in which
$$Q_\pm=\mathrm{diag}\{1,\lambda_{1\pm}\}.$$ 
Thus
\begin{align}\label{P_ij}
\Pi_0=&Q_{0}\otimes Q_0\otimes\Big(P_{0}\oplus(\lambda_{1+}P_{10})\oplus(\lambda_{1-}P_{01})\oplus(\lambda_{1+}\lambda_{1-}P_{11})\Big),
\end{align}
where $P_{\alpha\beta}$ means the projection corresponding to the vector $e^+_\alpha\otimes e^-_\beta$ ($\alpha,\beta=0,1$) and 
we set $P_{0}=P_{00}$. More precisely, if we consider any 
 $ p(n_{11},n_{22})$ then
\begin{align}\label{P_00}
&P_{0} p=(p,s^+\otimes s^-)\,e^+_0\otimes e^-_0,
\quad
\mathrm{where}\quad s^\pm=\left(\begin{array}{c}-c_1^\pm\\1\end{array}\right)(c_0^\pm-c_1^\pm)^{-1},
%\quad s^\pm \in \mathbb{C}^2,
\end{align}
Other projectors of (\ref{P_ij}) can be defined similarly.
\begin{lemma}\label{l:R}
 Given $\widetilde \Pi_0$ and $\Pi'_0$ of (\ref{exp_Pi}), we have
 \begin{align}\label{norm_R}
\lambda_0( \widetilde \Pi_0)<1-\delta  \,\,\,(\delta>0), \quad \quad \lambda_0( \Pi'_0)= 1,
 \end{align}
 and there are only two eigenvectors of $ \Pi'_0$ which correspond to $1$:
$\eta_1e^{+}_{0}e^{-}_{0}$ and $\eta_2e^{+}_{0}e^{-}_{0}$ with $\eta_1$ and $\eta_2$ of (\ref{eta}).
\end{lemma}
\textit{Proof.}
Any "non-good" product can be represented as an "absolutely non-good" part $\tilde\eta$ and  a "good" part $p(n_{11},n_{22},n_{12},n_{21})$.  Here "absolutely non-good" are  the products which do not contain any $n_{\alpha,\beta}$. It is easy to see that  the exponential part of $\widehat Q$ (see (\ref{hat_Q}) ) transforms
$\tilde \eta\to \beta^m\tilde \eta$, where $m\ge 2$ is the degree $\tilde\eta$.  Since a good part cannot be  $e^{+}_{0}e^-_0$ (we exclude these two vectors by
the condition of the lemma), the corresponding eigenvalue is less than $\lambda^+_0\lambda^-_0$. Thus, after multiplication by $\beta^{-2}C_*^{-1}$ it becomes less than $1$.

$\square$

It will be shown below that if $\bar a,\bar b\in\Omega_{\pm}$, then the matrix $\widehat C$ of (\ref{hat_C}) becomes  diagonal in the main order
(see Lemma \ref{l:b_C}). Moreover, it will be shown also (see Proposition  \ref{p:A_ab}) that the main part of $A_{ab}$ is the product of
operators whose kernel can be  obtained from $A_a$ and $A_b$ by leaving only quadratic terms in $\Lambda(x,y)$ (see (\ref{K_ab})).
Hence in the main order
\begin{align}
\mathcal{K}_\pm\sim A_*^+\otimes A_*^+\otimes A_*^-\otimes A_*^-\otimes (\Pi_0\oplus \Pi'_0\oplus\widetilde \Pi_0),
\end{align}
where $A_*^\pm$ is defined in (\ref{K_a*}). 
By  Proposition  \ref{p:A_ab} $\lambda_0(A^+_*)=\lambda_0(A^-_*)=1$, and all other eigenvalues are less than $1$.
Hence, it is naturally to expect that, as $n\to\infty$, only the projection onto the corresponding eigenvector 
$A^\pm_*$ will give non-zero 
contribution.  Similarly, by our preliminary analysis of $\Pi_0$, the largest eigenvalue of its main part is $1$ and the 
corresponding root subspace is
\begin{align}\label{frak_L}
\mathfrak{L}=\mathfrak{L}^{(0)}\otimes (e_0^+\otimes e_0^- ),\quad \mathfrak{L}^{(0)}=\mathrm{Lin}\{1, n_{12},n_{21},n_{21}n_{12},\eta_1,\eta_2\}
\end{align}
(see (\ref{eta}), (\ref{e^pm}) and (\ref{P_00})),  and the
part of $\Pi_0$, corresponding to this eigenvalue, is \\
$((Q_0\otimes Q_0)\oplus I_2)\otimes P_{0}$ (it is a  matrix of rank 6). However, 
one cannot leave only the main part of $\mathcal{K}_\pm$, because of the multiplier $W^4$  in front of the integral in $I_\pm$ (see (\ref{repr2})).
As was mentioned above, this factor is reduced 
to $W^2$ after the change of variables (\ref{ch_ab}), which shows that we need to take into account not only
the main part of  $\mathcal{K}_\pm$, but also the terms of order $W^{-1}$ and $W^{-2}$.

It is convenient to use the transformation  $B\to TBT$ in the space of $6\times 6$ matrices  with  a matrix $T$
\begin{align}\label{T}
T=\left(\begin{array}{cc}T_0\otimes T_0&0\\0&I_2\end{array}\right)\quad  
 T_0=\left(\begin{array}{cc}0&W^{1/2}\\W^{-1/2}&0\end{array}\right), \quad T_0^2=I_2,\quad T^2=I_6.
\end{align}
Let  $P_{\mathfrak{L}}$ be the projection onto the subspace (\ref{frak_L}) such that  transforms all "non-good" vectors into $0$ and
for $x\otimes p(n_{11},n_{22})$ with $x\in \mathfrak{L}^{(0)}$
\begin{align}\label{P_frak_L}
P_{\mathfrak{L}}(x\otimes p)=x\otimes P_0p,
%\quad x_1\in\mathfrak{L}_0,\,\,x_2\in \mathrm{Lin}\{1,n_{11},n_{22},n_{11}n_{22}\},
\end{align}
with $P_0$ of (\ref{P_00}).

We apply the transformation $T$ to $P_{\mathfrak{L}}\mathcal{G}(z)P_{\mathfrak{L}}$.
Recall that  $e^{(0)}$ and $e^{(c)}$ in the definition of $f$ and $g$  (see (\ref{rep_0})) are the vectors in the space of Grassmann variables $\mathfrak{Q}_{70}$ corresponding
to 1 and to $\prod_{l,s=1}^2 \rho_{ls}\tau_{sl}$ respectively.
Since $$TP_{\mathfrak{L}} e^{(0)}=c_1W^{-1}P_{\mathfrak{L}} e^{(c)},\quad T^*P_{\mathfrak{L}} e^{(c)}=c_2W^{-1}P_{\mathfrak{L}} e^{(0)},$$ the transformation via (\ref{T}) will "kill" the multiplier $W^2$ in front of the integral in $I_\pm$,
but it will multiply some entries by $W$ and even $W^2$. Our aim is to
prove that entries  of $TP_{\mathfrak{L}}\mathcal{G}(z)P_{\mathfrak{L}}T$ are bounded. 

We  shall use also the following definition.
 \begin{definition}\label{d:type} Let $p(U)$ (or $p(S)$) be  products of matrix entries of $U$ (or $S$). We say that
 the operator $\big( p\big)_U$ or $\big( p\big)_S$ (see (\ref{def_diff}) for the definition of $\big( p\big)_U$) is of the type $m$, if the  number of non-diagonal entries of $U$ (or $S$) in
 $p$  is $m$. 
 We say that the operator $\big(\prod_{i=1}^p F_0r_ir_{i}'F_0B_i\big)$ is of the joint type $m$, if the operators
 $r_i$ (of the form $\big( p\big)_U$) and $r_i'$ (of the form $\big( p\big)_S$) are of the type $m_i,m_i'$ respectively, $m_1+m_1'+\dots +m_{p}+m_{p}'=m$, and $B_i=\varphi_i(F_0K_iF_0)$ where $K_i$ are of the type $0$ and $\varphi_i$ are analytic
 functions in $\mathbb{B}_{1+\delta}=\{z:|z|\le 1+\delta\}$. We denote  $O_*(\mathfrak{r}^{(m)})$ the linear combinations of the operators of the joint type
 at least $m$.  \end{definition}
 Here and below
 \begin{align}\label{K_0}
 K_F=F_0K_0F_0, \quad K_{0}=K_{0U}K_{0S},\quad
 F_0=F_{0U}F_{0S}=e^{-c(\alpha_1|U_{12}|^2+\alpha_2|S_{12}|^2)/n},\,
 \end{align}
where $c=(a_+-a_-)/2$,  $K_{0U}$, $K_{0S}$ are defined by (\ref{K_U}) for $t_a=t_b=t_*:=\beta(a_+-a_-)^2$, and $F_{0U}$, $F_{0S}$ are defined in (\ref{F_a,b}) with $\bar a=\bar b=(a_+,a_-)$. It is important for us that for $\bar\xi,\bar\xi'\in\Omega_{c\varepsilon}$ (see (\ref{cond_xi})) $\alpha_1,\alpha_2>0$, hence
the multiplication operator $F_0$ satisfies the bound $$0\le F_0\le 1.$$

Notice also that by  (\ref{r',t'}) and (\ref{hat_C}) we obtain   that the non-diagonal entries of $\widehat C$
are  linear combinations of operators of non-zero type.
\begin{theorem}\label{t:M_0} There exist matrices $\mathfrak{M}'(z)$, $\mathfrak{M}(z)$ such that for
  $T$ of (\ref{T}) and $P_{\mathfrak{L}}$ of (\ref{P_frak_L}),  we have
\begin{align}\label{P_0GP_0}
P_{\mathfrak{L}}\mathcal{G}(z)P_{\mathfrak{L}}=T(\mathfrak{M}'(z)-z)^{-1}T,\,\,\mathfrak{M}'(z)=\mathfrak{M}(z)+O(W^{-1}),\,\,
 \mathfrak{M}(z):=\left(\begin{array}{cc} \mathcal{M}& \mathcal{M}'\\ \mathcal{M}''& \mathcal{M}'''\end{array}\right)     
\end{align}
where 
\begin{align}
\mathcal{M}&=\left(\begin{array}{cccc}
K_F&u(\frac{K_F}{z})+K_FF_1+\widetilde K_1& -u(\frac{K_F}{z})+K_FF_2+\widetilde K_2&u_0(\frac{K_F}{z})+\widetilde K_5 \\
0&K_F& 0&-u(\frac{K_F}{z})+K_FF_2+\widetilde K_3\\
0&0&K_F&u(\frac{K_F}{z}) +K_FF_1+\widetilde K_4\\
0&0& 0&K_F
 \end{array}\right)\label{cal_M}\\
\mathcal{M}'&=\left(\begin{array}{cccc}
 \tilde K_6&0&0&0 \\
\tilde K_7&0& 0&0
 \end{array}\right)^{(t)},\,
  \mathcal{M}''=\left(\begin{array}{cccc}
0&0&0&  \tilde K_8\\
0&0& 0&\tilde K_9
 \end{array}\right) ,\,    \mathcal{M}'''=\left(\begin{array}{cc}F_0K_1F_0&0\\0&F_0K_2F_0 \end{array}\right) 
 \label{cal_M'}   \end{align}
 Here $K_F$ and $F_0$ are defined in (\ref{K_0}) and  $K_{1},K_2$ are any operators of the type $0$. Operators $F_1$ and $F_2$ have the form
 \begin{align}
 & F_1= n^{-1}\Big(y_1(K_F/z)(2\sigma_1+\alpha_1\varphi_{U})+y_2(K_F/z)(2\sigma_2-\alpha_2\varphi_{S})\Big),\notag\\
& F_2=n^{-1}\Big(y_2(K_F/z)(2\sigma_1-\alpha_1\varphi_{U})+y_1(K_F/z)(2\sigma_2+\alpha_2\varphi_{S})\Big),\label{M_0.1}
\end{align}
with $\varphi_U,\varphi_S,\alpha_{1,2},\sigma_{1,2}$  defined in (\ref{F_a,b}). Functions $u$ and $u_0$ have the form
\begin{align}\label{g=0}
u(\zeta)=(\zeta-1)u_1(\zeta),\quad u_0(\zeta)=(\zeta-1)u_2(\zeta),
\end{align}
and    $u_1(\zeta), u_2(\zeta), y_1(\zeta),y_2(\zeta)$ are analytic in $\mathbb{B}_{1+\delta}=\{\zeta:|\zeta|<1+\delta,\, \delta>0\}$.
Moreover, 
\begin{align}\label{r_tiK}
\widetilde K_i=WO_*(\mathfrak{r}^{(1)})O_*(\mathfrak{r}^{(1)}),\, (i\not=5),\quad \widetilde K_5=WO_*(\mathfrak{r}^{(1)})O_*(\mathfrak{r}^{(1)})+W^2O_*(\mathfrak{r}^{(\mu)})O_*(\mathfrak{r}^{(\nu)}),
\end{align}
%and each of operators contains at least two terms of non-zero type.
where $\mu,\nu>0$, $\mu+\nu\ge 4$.
In addition,
\begin{align}\label{repr.fin}
&I_{\pm}=-\frac{C_{\bar\xi,\varepsilon}}{2\pi i}\oint z^{n-1}\big(\big(\mathfrak{M}'-z\big)^{-1}\widehat f,\widehat g\big)dz+ o(1),\quad\widehat f= \widehat f_0+\widetilde f,
\quad\widehat g= \widehat g_0+\widetilde g,\\
&  \widehat f_0=( f_{01}, f_{02}, -f_{02}, d_{1*},0,0)F_0,
%\label{hat_f}\\
 \quad \widehat g_0=( d_{2*}, g_{02}, -g_{02}, g_{04},0, 0)F_0,\notag\\
&\widetilde f=WO_*(\mathfrak{r}^{(1)})O_*(\mathfrak{r}^{(1)})\varphi_{1}(K_F)F_0f'+O(W^{-1})F_0f'',\notag\\
&\widetilde g=WO_*(\mathfrak{r}^{(1)})O_*(\mathfrak{r}^{(1)})\varphi_{2}(K_F)F_0g'
+O(W^{-1})F_0g'',
\notag\end{align}
where  $ f_{0i}, g_{0i}$ are 
analytic  in $\mathbb{B}_{1+\delta}$ functions of $z^{-1}K_F$,
$f_{0i}F_0$ is a result of the application of $f_{0i}(K_F/z)$ to $F_0$, $O(W^{-1})F_0$ is a result of the application of the operator with $O(W^{-1})$
norm to $F_0$, and $O_*(\mathfrak{r}^{(2)})\varphi_{1,2}(K_F)F_0$ is a result of the application of the operator $O_*(\mathfrak{r}^{(2)})\varphi_{1,2}(K_F)$
 to $F_0$, and $f',g',f'',g''\in \mathbb{C}^6$.

%$\widetilde K=W\widetilde K'$, where $\widetilde K_i'$ are linear combinations of operators of the joint type at least 2, containing at least $2$ operators of the type at least $1$.
%%\begin{align}\label{ti_K}
%%\widetilde  K=W\widetilde B_1F_0\widetilde r_1F_0\widetilde B_2F_0\widetilde r_2F_0.
%%\end{align}
%Here $\widetilde r_1$ and $\widetilde r_2$ are some non diagonal entries of $\widehat C$, and $\widetilde B_1,\widetilde B_2$ are some analytic 
%functions of the diagonal entries of $\widehat C$.
\end{theorem}
We postpone the proof of Theorem \ref{t:M_0} to Section \ref{s:M_0}. Now we prove Theorem \ref{t:1} on the basis of Theorem \ref{t:M_0}.
We need also two following lemmas, whose proofs can be found in  Section \ref{s:bG}.
\begin{lemma}\label{l:bG} Given  $\mathfrak{M}',\mathfrak{M}$ of (\ref{P_0GP_0}), we have
 for any $z\in\omega_A=\{z: |z|=1+A/n\}$ 
\begin{align}\label{b_1}
%&\|(\mathfrak{M}-\mathfrak{M}_0)\widehat G_0\widehat f\|^2\le C(n/W)^2,\quad \|(\mathfrak{M}-\mathfrak{M}_0)\widehat G_0\widehat g\|^2\le C(n/W)^2\\
%&|(\widehat G_0(\mathfrak{M}-\mathfrak{M}_0)\widehat G_0\widehat f,\widehat g)|\le nW^{-1}/|z-1|,\quad
%% \|\big((\mathfrak{M}-\mathfrak{M}_0)\widehat G_0\big)^{(t)}\widehat g\|^2\le Cn/W,\\
 \|(\mathfrak{M}'-z)^{-1}\|\le C/|z-1|, \quad  \|\mathfrak{G}(z)\|\le C/|z-1|, \quad\Big(\mathfrak{G}(z) =(\mathfrak{M}-z)^{-1}\Big).
\end{align} 
%The same bound is valid for $(\mathfrak{M}'-z)^{-1}$.
%and the same bounds are valid for $\widehat g$.
\end{lemma}
Denote by $\mathcal{M}_0 $ an upper triangular matrix which 
is  obtained from $\mathcal{M}$ of (\ref{cal_M}) by replacing all $\widetilde K_i$ with zeros
and  $K_F$ with $F^2_0$. Set
\begin{align}\label{M_0}
&\mathfrak{M}_0=\left(\begin{array}{cc} \mathcal{M}_0&0\\ 0& \mathcal{M}'''\end{array}\right)\Rightarrow
\mathfrak{G}_0(z):=(\mathfrak{M}_0-z)^{-1}=\left(\begin{array}{cc}\widehat{G}_0(z)&0\\ 0& (\mathcal{M}'''-z)^{-1}\end{array}\right),\\
&\mathrm{where}\quad
\widehat{G}_0(z):=(\mathcal{M}_0-z)^{-1}.
\notag\end{align}
Introduce also
\begin{align}\label{P_L}
L=\log^2 n,\quad \mathcal{P}_L= \mathcal{P}_{LU}\otimes  \mathcal{P}_{LS},
\end{align}
where $\mathcal{P}_{LU}$   is an orthogonal projection of $L_2(\mathring{U}_2)$ onto the subspace $\oplus_{l<L}\mathcal{L}^{(lU)}$ and $\mathcal{P}_{LS}$ is a similar orthogonal
projection in $L_2(\mathring{U}(1,1))$ (see (\ref{LU})).
\begin{lemma}\label{l:bGP} Let  $\mathfrak{M}$ and $\mathfrak{M}_0$ be defined by (\ref{P_0GP_0}) and  of (\ref{M_0}), 
and $\mathcal{E}_0$ be  projection in $\mathcal{L}_2=L_2(\mathring{U}_2)\otimes L_2(\mathring{U}(1,1))$ onto the subspace of function  on $|U_{12}|^2$ and $|S_{12}|^2$. Then
\begin{align}\label{PM}
\|(\mathfrak{M}_0-\mathfrak{M})\mathcal{P}_L\mathcal{E}_0\|\le L^2/W,
\quad \|(\mathfrak{M}_0-\mathfrak{M})^{(t)}\mathcal{P}_L\mathcal{E}_0\|\le L^2/W
\end{align}
and
\begin{align}\label{f-Pf}
&\|(1-\mathcal{P}_L)\widehat f_0\|\le e^{-c\log^{4/3}n}, \quad\|(1-\mathcal{P}_L) \mathfrak{G}_0\widehat f_0\|\le e^{-c\log^{4/3}n},\quad
 \| \mathfrak{G}_0\widehat f_0\|^2\le n/|z-1|,\\
&  \|\widehat f_0\|^2\le Cn,\qquad\qquad \qquad\quad\|\widetilde f\|^2\le Cn L^4/W^2+O(e^{-c\log^{4/3}n}).
\label{b_ti_f}\end{align}
where $\widehat f_0$, $\widetilde f$ are defined in (\ref{repr.fin}). Similar bounds hold for $\widehat g_0$, $\widetilde g$.
  \end{lemma}
 \textit{Proof of Theorem \ref{t:1}}
By Lemma \ref{l:bG} and (\ref{b_ti_f})
\begin{align}\label{diff2}
|((\mathfrak{M}'-z)^{-1}\widehat f,\widehat g)-&((\mathfrak{M}'-z)^{-1}\widehat f_0,\widehat g_0)|\\
&\le C |z-1|^{-1}\Big( \|\widetilde f\|\|\widehat g_0\|+\|\widetilde g\|\|\widehat f_0\|
+\|\widetilde f\|\|\widetilde g\|\Big)
+O(e^{-c\log^{4/3}n})\notag\\
&\le L^2n/W|z-1|+O(e^{-c\log^{4/3}n}).
\notag\end{align}
Then, using that $|z^n|\le C$ for $z\in \omega_A$,
\begin{align}\label{int}
\oint_{\omega_A}\frac{|dz|}{|z-1|}\le C\log n,
\end{align}
and $L^2n\log n/W\to 0$ (since $W\ge Cn \log^5 n$ by assumptions of Theorem \ref{t:1}), we get from (\ref{repr.fin}) and (\ref{diff2}) 
\begin{align}\label{repr.fin1}
I_\pm=-\frac{C_{\bar\xi,\varepsilon}}{2\pi i}\oint_{\omega_A}z^{n-1}((\mathfrak{M}'-z)^{-1}\widehat f_0,\widehat g_0)dz+o(1)
\end{align}
To analyse the r.h.s. we use  the resolvent identity
\begin{align}\label{res_id}
A_1^{-1}-A_2^{-1}=-A_1^{-1}(A_1-A_2)A_2^{-1}=-A_2^{-1}(A_1-A_2)A_1^{-1}
\end{align}
with $A_1=\mathfrak{M}'-z$, $A_2=\mathfrak{M}_0-z$.
Since $\mathfrak{M}'=\mathfrak{M}+O(W^{-1})$, applying twice (\ref{res_id})  we obtain
\begin{align*}
&\Big|((\mathfrak{M}'-z)^{-1}-\mathfrak{G}_0)\widehat f_0,\widehat g_0)\Big|\le
\Big|(\mathfrak{G}_0(O(W^{-1})+\mathfrak{M}-\mathfrak{M}_0)\mathfrak{G}_0\widehat f_0,\widehat g_0)\Big|\\
&+ \|(\mathfrak{M}'-z)^{-1}\|\cdot \|(O(W^{-1})+\mathfrak{M}-\mathfrak{M}_0)\mathfrak{G}_0(z)\widehat f_0\|\cdot
\|(O(W^{-1})+\mathfrak{M}-\mathfrak{M}_0)^{(t)}\mathfrak{G}_0(\bar z)\widehat g_0\|\\&+O(e^{-c\log^{4/3}n})=
\mathcal{T}_1+ \mathcal{T}_2+O(e^{-c\log^{4/3}n})
%\\
%&\le C(n/W)\oint_{\omega_A}\frac{|dz|}{|z-1|}\le Cn\log n/W\to 0,
\end{align*}
Relations (\ref{PM}), (\ref{f-Pf}), and the fact that $\mathfrak{G}_0(z)\widehat f_0\in\mathcal{E}_0\mathcal{L}_2$  yield for $z\in\omega_A$
\begin{align*}
&\|(O(W^{-1})+\mathfrak{M}-\mathfrak{M}_0)\mathfrak{G}_0(z)\widehat f_0||=
\|(O(W^{-1})+\mathfrak{M}-\mathfrak{M}_0)\mathcal{P}_L\mathcal{E}_0\mathfrak{G}_0(z)\widehat f_0\|+O(e^{-c\log^{4/3}n})
\\
&\le
 \dfrac{CL^2}{W}\|\mathfrak{G}_0\widehat f_0\|+O(e^{-c\log^{4/3}n})
 \end{align*}
 Hence, using   Lemma 4.2 for $\|(\mathfrak{M}'-z)^{-1}\|$ and the last bound of (\ref{f-Pf}), we get
 \begin{align*}
 &\mathcal{T}_1\le
\dfrac{CL^2}{W}\|\mathfrak{G}_0\widehat f_0\| \|\mathfrak{G}_0\widehat g_0\|+O(e^{-c\log^{4/3}n})\le \dfrac{CnL^2}{W\cdot |z-1|}
 +O(e^{-c\log^{4/3}n}),\\
 &\mathcal{T}_2\le C|z-1|^{-1} \dfrac{L^4}{W^4}\|\mathfrak{G}_0\widehat f_0\| \|\mathfrak{G}_0\widehat g_0\|+O(e^{-c\log^{4/3}n})
% ||\mathfrak{G}(z) \|\cdot \|(\mathfrak{M}-\widetilde{\mathfrak{M}}_0)\widetilde{\mathfrak{G}}_0(z)\widehat f_0\|\cdot
%\|(\mathfrak{M}-\widetilde{\mathfrak{M}}_0)^{(t)}\widetilde{\mathfrak{G}}_0(\bar z)\widehat g_0\|
\le \dfrac{Cn^2L^4}{W^2\cdot |z-1|}+O(e^{-c\log^{4/3}n}).
\end{align*}
These bounds and (\ref{int}) yield
\begin{align}\label{G-G_0}
&\Big|\oint_{\omega_A}z^{n-1}((\mathfrak{M}'-z)^{-1}-\mathfrak{G}_0)\widehat f_0,\widehat g_0)dz\Big|\\
&\le C\Big(\frac{L^2n}{W}+\frac{L^4n^2}{W^2}\Big)\oint_{\omega_A}\frac{|dz|}{|z-1|}+O(e^{-c\log^{4/3}n})\le C\frac{nL^2\log n}{W}+O(e^{-c\log^{4/3}n})=o(1).
\notag\end{align}
Thus in view of (\ref{repr.fin1}), (\ref{G-G_0}),  and (\ref{M_0}) we get
\begin{align*}
I_{\pm}&=-\frac{C_{\bar\xi,\varepsilon}}{2\pi i}\oint_{\omega_A}z^{n-1}(\mathfrak{G}_0\widehat f_0,\widehat g_0)dz
+o(1)=-\frac{C_{\bar\xi,\varepsilon}}{2\pi i}\oint_{\omega_A}z^{n-1}
(\widehat G_0(z)\widehat f_0,\widehat g_0)dz+o(1)
%\\
%\widehat f_0&=(f_{01},f_{02},-f_{02},f_{04}),\,\widehat g_0=(g_{01},g_{02},-g_{02},g_{04})
\end{align*}

Using the representation (\ref{repr2}) for $\mathcal{R}_{nW}^{+-}$, the  inverse matrix formula for $\widehat G_0(z)$ (see (\ref{A^-1}) below), the form of $\widehat f_0$, $\widehat g_0$ (see (\ref{repr.fin})), and taking into account the bound (\ref{b_I_+}) for $I_+$ and similar bound
for $I_-$ we obtain
\begin{align}\notag
&C_{\bar\xi,\varepsilon}^{-1}\mathcal{R}_{nW}^{+-}(E,\varepsilon,\xi)=-\frac{1}{2\pi i}\oint_{\omega_A} dz\, z^{n-1}\Big(G_F(z)\big(f_{01}d_{2*}+2f_{02}g_{02}+d_{1*}g_{04}\big)F_0^2\\
\label{int_z}&-2G_F^2(z)u(F_0^2/z)\big( f_{02}d_{2*}
-d_{1*}g_{02}\big) F_0^2
-G_F^2(z)\big(F_1-F_2\big)(f_{02}d_{2*}-d_{1*}g_{02}) F_0^2\\
&-G_F^2(z)u_0(F_0^2/z)d_{1*}d_{2*} F_0^2+2G_F^3(z)(F_1+u(F_0^2/z))(F_2-u(F_0^2/z))d_{1*}d_{2*} F_0^2\Big)+o(1),
\notag
\end{align}
where $G_F(z)=(F_0^2-z)^{-1}$. 

Taking the integral with respect to $z$ and using (\ref{g=0}) we get
\begin{align*}
\mathcal{R}_{nW}^{+-}(E,\varepsilon,\xi)=C_{\bar\xi,\varepsilon}\int \big(k_1 \widetilde F_1\widetilde F_2+k_2(\widetilde F_1-\widetilde F_2)+k_3) F_0^{2n}dUdS+o(1),
\end{align*}
where
\begin{align*}
&\widetilde  F_1= y_1(1)(2\sigma_1+\alpha_1\varphi_{U})+y_2(1)(2\sigma_2-\alpha_2\varphi_{S}),\notag\\
& \widetilde F_2=y_2(1)(2\sigma_1-\alpha_1\varphi_{U})+y_1(1)(2\sigma_2+\alpha_2\varphi_{S})\\
&F_0^{2n}=e^{iE(\sigma_1-\sigma_2)}e^{c\alpha_1\varphi_U-c\alpha_2\varphi_S},\quad c=(a_+-a_-)/2=\pi\rho(E) 
\end{align*}
and $k_1,k_2,k_3$ are some constants which we find using the fact that for $\alpha_1=\alpha_2=\alpha$ and $\sigma_1=\sigma_2=\sigma$
the above expression is 1 (see the definition of $\mathcal{R}_{nW}^{+-}(E,\varepsilon,\xi)$ in (\ref{G_2})). It implies immediately that the coefficient at $\sigma^2$ is 0, hence
\begin{align*}
&y_1(1)=-y_2(1)=y\Rightarrow \widetilde  F_1\widetilde  F_2=y^2(\alpha_1\varphi_U+\alpha_2\varphi_S)^2-4y^2(\sigma_1-\sigma_2)^2,
\end{align*}
Performing the integration with respect to $dU$, $dS$ we obtain 
\begin{align*}
\int dUdS e^{c\alpha_1\varphi_U-c\alpha_2\varphi_S}&(\alpha_1\varphi_U+\alpha_2\varphi_S)^2\\=&\frac{e^{-2c\alpha_2}}{c^2}\Big(\Big(
\frac{\alpha_1}{2\alpha _2}+\frac{\alpha_2}{2\alpha _1}\Big)
\sinh(2c\alpha_1)+\cosh(2c\alpha_1)+\frac{\sinh(2c\alpha_1)}{4c^2\alpha _1 \alpha_2}\Big)\\
%\int dUdS e^{c\alpha_1\varphi_U-c\alpha_2\varphi_S}&(\alpha_1\varphi_U+\alpha_2\varphi_S)=
%\frac{e^{-2c\alpha_2}}{c}\Big(\frac{\cosh(2c\alpha_1)}%{4c\alpha_2}+\frac{\sinh(2c\alpha_1)}{4c\alpha_1}\Big)\\
\int dUdS e^{c\alpha_1\varphi_U-c\alpha_2\varphi_S}&=\frac{e^{-2c\alpha_2}\sinh(2c\alpha_1)}{8c^2\alpha _1 \alpha_2}
\end{align*}
Now putting $\alpha_1=\alpha_2=\alpha$ and $\sigma_1=\sigma_2=\sigma$, we can conclude that $k_1y^2c^{-2}=1$ and $k_3=-2$. 
Hence
\begin{align*}
\mathcal{R}_{nW}^{+-}(E,\varepsilon,\xi)=&e^{iE(\sigma_1-\sigma_2)}e^{-2c\alpha_2}\Big(\Big(\frac{\alpha_1}{2\alpha _2}+\frac{\alpha_2}{2\alpha_1}\Big)
\sinh(2c\alpha_1)+\cosh(2c\alpha_1)\\&\hskip2cm-\Big(k_2y(\sigma_1-\sigma_2)+(\sigma_1-\sigma_2)^2\Big)\frac{\sinh(2c\alpha_1)}{2\alpha _1 \alpha_2}\Big). 
\end{align*}
Taking the derivative  with respect to $\xi_1'$ and then putting $\bar \xi=\bar \xi'=0$, we get
\begin{multline*}
(N\rho(E))^{-1}\mathbb E\{\Tr (H_N-E-i\varepsilon/N)^{-1}\}=\dfrac{\partial}{\partial \xi_1'}\mathcal{R}_{nW}^{+-}(E,\varepsilon,\xi)\Big|_{\bar \xi=\bar\xi'=0}\\
=\dfrac{ia_+}{\rho(E)} +k_2y\frac{1-e^{-4c\varepsilon}}{8\varepsilon^2\rho(E) i}+o(1) .
\end{multline*}
But it follows from Theorem \ref{t:2} that
\[\mathbb E\{N^{-1}\Tr (H_N-E-i\varepsilon/N)^{-1}\}=\dfrac{ia_+}{\rho(E)} +o(1)\]
Therefore  we conclude that $k_2=0$ and (\ref{t1.1}) holds.  As a corollary, we obtain (\ref{t1.2}).
%\begin{align*}
%\dfrac{\partial^2}{\partial \xi_1'\partial\xi_2'}\mathcal{R}_{nW}^{+-}(E,\varepsilon,\xi)\Big|_{\xi'=\xi}=&\dfrac{1}{\rho^2(E)}+\dfrac{1-e^{-4c\alpha}}{4\alpha^2\rho^2(E)}.
%\end{align*}
% $\square$
%\begin{align*}
%\mathcal{R}_{nW}^{+-}(E,\varepsilon,\xi)=e^{-c_0(\alpha_1+\alpha_2)}\Big(\delta_1\delta_2(e^{2c_0\alpha_1}-1)/4\alpha_1\alpha_2
%-(\delta_1+\delta_2)e^{2c_0\alpha_1}/2\beta _2+e^{2c_0\alpha_1}\alpha_1\alpha_2\Big)
%\end{align*}\begin{align}
%\end{align}

\subsection{Analysis of $I_+$ and $I_-$}\label{s:I_+}
Analysis of $I_+$ is much simpler than that for $I_{\pm}$. This time it is more convenient to consider $X$ just like 
unitary matrix which is close to the $a_+I_2$. So let us still diagonalize $Y$ according to (\ref{U,V}), and use (\ref{ch_ab}) for $b_1, b_2$,
but let $X$ be parametrized as (cf  (\ref{ch_ab}))
\begin{align*}
X_{11}=a_+(1+i\theta_+\widetilde a_1/\sqrt{W}),\,X_{22}=a_+(1+i\theta_+\widetilde a_2/\sqrt{W}),\, X_{12}=\dfrac{i\theta_+ a_+(x+iy)}{\sqrt{2W}}=-\overline{X_{12}}.
\end{align*}
This change transforms the measure $dX_i dY_i/(-\pi^2)$ from (\ref{int_rep_pm})  to
\[
(a_-\theta_-)(a_+\theta_+)^5(2\pi)^{-3}d \bar a_i \,d \bar b_i  \,d \bar x_i \, d S_i,
\]
and it reduces the factor $W^4$ in front of $I_+$ in (\ref{repr2}) to $W$.

In addition let $\widehat C''(S_i)$ be the matrix corresponding to the change of Grassmann variables 
\begin{align}\label{r',t'.1}
(\rho_i S^{-1}_i)_{\alpha\beta}\to \rho_{i,\alpha\beta},\quad(S_i\tau_i)_{\alpha\beta} \to\tau_{i,\alpha\beta},
\end{align}
 in the space $\mathfrak{Q}_{70}$. Set also
\begin{align*}
\hat F_+(X,\bar b, S)=F_S^{1/2}(\bar b, S)\cdot \exp\Big\{-\dfrac{i}{2n}\Tr X(\varepsilon L+\hat\xi/\rho(E))\Big\}.
\end{align*}
In this parametrization the operator $\mathcal{K}_+$ has the form
\begin{align*}
\mathcal{K}_+(\bar a,\bar b&,x,y,\rho,\tau, S;\bar a',\bar b',x',y',\rho', \tau', S' ) \\
=&  (-\lambda^+_0)^{-5/2}(-\lambda^-_0)^{-1/2}A_{ab}^+(\bar a,\bar a',x,x',y,y';\bar b,\bar b')\\
&\times \hat F_+(X,\bar b, S)\beta^{-1}\widehat Q(\rho,\tau;\rho',\tau', X', D_b)\widehat C''(S(S')^{-1})K_{S}(S(S')^{-1})\hat F_+(X',\bar b', S')
\end{align*}
where
\begin{align*}
A_{ab}^+(\bar a,\bar a',x,x',y,y';\bar b,\bar b')=A_{a}^+(\widetilde a_1,\widetilde a_1')A_{a}^+(\widetilde a_2,\widetilde a_2')
A_{b}^+(\widetilde b_1,\widetilde b_1')A_{b}^-(\widetilde b_2,\widetilde b_2')A_x^+(x,x')A_x^+(y,y')
\end{align*}
and (cf (\ref{kappa}) and (\ref{K_a*}))
\begin{align}\label{A_x}
&A_x^+(x,x')\\ \notag
&=(-\lambda^+_0)^{1/2}\frac{a_+ \theta_+}{(2\pi)^{1/2}}\exp\Big\{(a_+\theta_+)^2\big(\beta (x-x')^2/2-\frac{c_+}{4}(x^2+(x')^2)\big)\Big\}(1+o(W^{-1})).
\end{align}
The main order of this operator has  the form
$A^+_*(\widetilde x,\widetilde x')$, hence the largest eigenvalue of this operator is
$1+O(W^{-1})$. It is easy to see that similarly to consideration below (see (\ref{hat_Q}) -- (\ref{Q_pm})) we have
\[
\hat Q(\rho,\tau;\rho',\tau', X', D_b)=\beta^{2} \Big(\big(Q_0
\otimes Q_+\otimes Q_+\otimes Q_0)\oplus R\Big) (1+o(1)),
\]
and so
\begin{align*}
\mathcal{K}_+=&\beta (-\lambda^+_0)^{-1/2}(-\lambda^+_0)^{-1/2}  A^+_*\otimes A^+_*\otimes A^+_*\otimes A^-_*\otimes A^+_*\otimes A^+_*\\
&\otimes \Big(\big(Q_0
\otimes Q_+\otimes Q_+\otimes Q_0)\oplus R\Big)\otimes (K_S \hat C'')(1+o(1)).
\end{align*}
Hence using that the maximum eigenvalues of $A^+_*, A^-_*,Q_0,Q_+$ are $1$, the largest eigenvalue of $K_S \hat C''$ is $1+O(W^{-1})$,  and $|\lambda_0(R)|<1$ (similarly to Lemma \ref{l:R}),
we obtain 
\begin{align*}
\lambda_0(\mathcal{K}_+)=\beta (-\lambda^+_0)^{-1/2}(-\lambda^+_0)^{-1/2}(1+o(1))=(q_+q_-)^{1/2}(1+o(1))\Longrightarrow
|\lambda_0(\mathcal{K}_+)|<1-c.
\end{align*}
Here we used the definition  of $q_\pm$ (see (\ref{la_k})).

Thus, choosing the circle of radius $1-c$ as the integration contour in (\ref{repr2}) , we get that
\begin{align}\label{b_I_+}
I_+=O(e^{-cn})
\end{align}
The bound for $I_-$ can be obtained similarly.

\section{Proof of Theorem \ref{t:M_0} }\label{s:M_0}

In what follows it will be convenient for us to consider  decomposition of the Grassmann space $\mathfrak{Q}_{70}$ into "good", "semi-good" and "non-good" subspaces (see (\ref{hat_Q})-(\ref{dec})) and write
$\widehat Q$   and $\widehat C$ (see (\ref{hat_C})) as $3\times 3$ block matrices corresponding to this decomposition. 
\begin{lemma}\label{l:dec_C}
The matrix $\widehat C$  (see (\ref{hat_C})) in the block representation (\ref{dec})  has the form
\begin{align}\label{cal_R}
&\widehat C=\widehat C_d+\mathcal{C},\quad \mathcal{C}=\left(\begin{array}{ccc}O(W^{-1})&O_*(\mathfrak{r}^{(1)})O_*(\mathfrak{r}^{(1)})&O_*(\mathfrak{r}^{(1)})\\O_*(\mathfrak{r}^{(1)})O_*(\mathfrak{r}^{(1)})&O(W^{-1})&
O_*(\mathfrak{r}^{(1)})\\
O_*(\mathfrak{r}^{(1)})&O_*(\mathfrak{r}^{(1)})&O(W^{-1})\end{array}\right),\quad 
\end{align}
where  $\widehat C_d$ is a diagonal matrix whose entries are $K_{US}$ in the first block and some operators of  the type zero 
(see  Definition (\ref{d:type})) for other blocks
(their norm is not more
than $1+C/W$ by Lemma \ref{l:U,V} below), and $O_*(\mathfrak{r}^{(1)})$ is defined in (\ref{d:type}).

Moreover,
for $\widehat C_0=\widehat C\Big|_{t_a=t_b=t_*}$ we have
\begin{align}\label{41}
&\widehat C^{(11)}_0(e_0^+e_0^-)=e_0^+e_0^--(t_*W)^{-1}(n_{12}-n_{21})(c_0^+e_0^--c_0^-e_0^+)+O(W^{-2}),\\
&\frac{\partial^2\widehat C^{(11)}_0(n_{12}e_0^+e_0^-)}{\partial n_{12}\partial n_{21}}=(t_*W)^{-1}(c_0^+e_0^--c_0^-e_0^+)+O(W^{-2})
=-\frac{\partial^2\widehat C^{(11)}_0(n_{21}e_0^+e_0^-)}{\partial n_{12}\partial n_{21}}+O(W^{-2}).
\notag\end{align}
\end{lemma}
\begin{definition} \label{d:2} We  write  $M=O_*(\mathcal{C})$ for some $70\times 70$ matrix $M$, if 
 the entries of $M$ in the  block representation as in (\ref{dec}) satisfy the same bounds as the entries of the corresponding block of $\mathcal{C}$
 (see (\ref{cal_R})).
\end{definition}

The proof  is based on the following lemma  proven in Section \ref{s:U,V}.
\begin{lemma}\label{l:U,V}  Operators of the form $\big( v)_U$ and $\big( v\big)_S$ are reduced by
the $l$-th subspace of irreducible
representation of the shift operator on $U(2)$ and  $U(1,1)$ respectively. If $p(U)$ is some product of  matrix entries of $U$ of non-zero
type, then we have for the reduced operator $\big(p)^{(l)}_U$
\begin{align}\label{isometr}
\big(p)^{(l)}_U=\lambda_{ij}^{(l)}(p)\mathcal{E}^{(l)U}_{ij}, \quad   \lambda_{ij}^{(l)}(p)=q_p^{1/2}(l)(Wt_a)^{-s(p)}(1+O(l^2/W))
\end{align}
where a partial isometric operator $\mathcal{E}^{(l)U}_{ij}$ 
  (see (\ref{E^l_ij})  for the  definition), exponent $s(p)\ge1$, and polynomial $q_p$ do not depend on $t$, $W$ and depend
  only on $p(U)$.
 If $p$ is of zero type, then the corresponding $\mathcal{E}^{(l)U}_{ii}$  is  an orthogonal projection  and the corresponding $q_p(l)$ is quadratic
 \begin{align}\label{type0}
&\lambda_{ij}^{(l)}(p)=1-q_p(l)(Wt_a)^{-1}(1+O(l^2/W)),\\
& K_{U}^{(l)}=\mathcal{E}^{(l)U}_{00}(1-l(l+1)(t_aW)^{-1} +O(l^4W^{-2})),\,\,\,
 \frac{\partial K_{U}^{(l)}}{\partial t_a}=W((p_*)^{(l)}_U)^*(p_*)^{(l)}_U(1+O(l^2W^{-1}))
\notag\end{align}
where $p_*(U)=U_{11}U_{12}$. The same is true for  $p(S)$. In particular,
for any fixed $s=0,1,2,\dots$ 
\begin{align}\label{U,V.1}
&\big( |U_{12}|^{2s}\big)_U^{(l)}=\mathcal{E}^{(l)U}_{00}(s!(t_aW)^{-s}+O(l^2/W^{s+1}))\\
& \big( |S_{12}|^{2s}\big)_S^{(l)}=\mathcal{E}^{(l)S}_{00}(s!(t_bW)^{-s}+O(l^2/W^{s+1}))\notag
%&\lambda^{(l)}(K_{0U})=1-l^2(t_aW)^{-1} +O(l^4W^{-2}), 
%\lambda^{(l)}(K_{0S})=1-l^2/t_bW),\notag
\end{align}
%where $K_{0U}$, $K_{0S}$ are defined by (\ref{K_U}) for $t_a=t_b=t_*:=\beta(a_+-a_-)^2$.
In addition, if  $r=B\big(p\big)_U$ or $r=B\big(p\big)_S$ with $p$ of the type at least 1 and with some bounded operator $B$, then for any $K_\alpha$ 
corresponding to the product of the type 0
\begin{align}\label{b_rr}
r^*r\le CW^{-1}(1+C/W-K_\alpha)
\end{align}
\end{lemma}

\textit{Proof of Lemma \ref{l:dec_C}}
The result of the application of $\widehat{C} $ to the product of the Grassmann variables has the form
\begin{align*}
&\prod_{j=1}^s \rho_{\alpha_j\beta_j}\tau_{\beta'_j\alpha'_j}\to\prod (U\rho S^{-1})_{\alpha_j\beta_j}(S\tau U^*)_{\beta'_j\alpha'_j}\\=&
\prod_{j=1}^s \Big(\sum_{\mu_j,\nu_j=1,2}
\sum_{\mu_j',\nu_j'=1,2}U_{\alpha_j\mu_j}
U^*_{\mu'_j\alpha'_j}(S^{-1})_{\nu_j\beta_j}
S_{\beta_j'\nu_j'}\rho_{\mu_j\nu_j}\tau_{\nu_j'\mu'_j}\Big)
\end{align*}
According to Lemma \ref{l:U,V}, the bound for the operator which is the product of matrix entries of $U$  depends on the number
of $U_{12}$ or $U_{21}$ in this product and the same for the product of matrix entries of $S$. In other words, the bound for the entry
of $\widehat{C} $ which correspond to transformation $\prod \rho_{\alpha_j\beta_j}\tau_{\alpha'_j\beta'_j}$ to
$\prod \rho_{\mu_j\nu_j}\tau_{\nu'_j\mu'_j}$ depends on the number of index changes ($1\to 2$ or $2\to 1$) which we need to transform
\[
\cup\{(\alpha_j,\beta_j)\}\to\cup \{(\mu_j,\nu_j)\}\quad\mathrm{and}\quad \cup\{(\beta_j',\alpha_j')\}\to\cup \{(\nu_j',\mu_j')\}.
\]
For all non-diagonal entries of $\widehat C$ we have at least one transformation of indexes which in view of Lemma \ref{l:U,V}   means at least
$O_*(\mathfrak{r}^{(1)})$operator in  the corresponding entry. 

If we  transform "good product" into another "good" one, then
we need at least two transformations and the resulting product of the entries of $U$ and $S$ can be written as a functions
of $|U_{12}|^2$ and $|S_{12}|^2$. Thus Lemma \ref{l:U,V}  yields $O(W^{-1})$ for the non-diagonal entries inside the block $\widehat C^{(11)}$,
and the diagonal entries from this block have the form $1+O(W^{-1})$.
 
 To obtain "semi-good" vectors from a "good" one, we need at least two transformations with at least 1 non-diagonal $U_{\alpha\beta}$
 and  at least 1 non-diagonal $S_{\alpha\beta}$ and corresponding operators will have the form at least $O_*(\mathfrak{r}^{(1)})O_*(\mathfrak{r}^{(1)})$.
 
 Relations (\ref{41})  are simple corollary from the equalities:
\begin{align*}
&\widehat C^{(11)}_0(e_0^+e_0^-)=\widehat C^{(11)}_0(e_0^+)\widehat C^{(11)}_0(e_0^-)+O(W^{-2}),\\
&\widehat C^{(11)}_0(n_{\alpha\beta}e_0^+e_0^-)=\widehat C^{(11)}_0(n_{\alpha\beta})\widehat C^{(11)}_0(e_0^+)\widehat C^{(11)}_0(e_0^-)+O(W^{-2}),\\
&\widehat C^{(11)}_0(n_{\alpha\beta})=n_{\alpha\beta}+O(W^{-1}),\quad \widehat C^{(11)}_0(e_0^\pm)=e_0^{\pm}\mp (t_*W)^{-1}c_0^{\pm}(n_{12}-n_{21})+O(W^{-2}). 
\end{align*}
Here, to obtain the last equality, we take into account that 
\[\big(U_{12}U^{-1}_{21}\big)_U\sim(t_aW)^{-1},\quad \big(S_{12}S^{-1}_{21}\big)_S\sim -(t_bW)^{-1}\]
$\square$

Now let us choose an appropriate  basis for $\mathcal{K}$.
 Denote
 \begin{align}\label{psi_ab}
 \Psi_{\bar k}(\bar a,\bar b)=\psi_{k_1}^{+}(\widetilde a_1)\psi_{k_2}^{-}(\widetilde a_2)\psi_{k_3}^{+}(\widetilde b_1)\psi_{k_4}^{-}(\widetilde b_2),
 \,\,\bar{k}=(k_1,k_2,k_3,k_4).
 \end{align}
 where $\psi_{k}^{+}(\widetilde a)$  $\psi_{k}^-(\widetilde b)$ are  defined in (\ref{psi_k+}).
Set
 \begin{align}\label{M_k,k'}
& \langle \mathcal{K}\rangle_{\bar k\bar k'}:= 
 \int \mathcal{K}\Psi_{\bar k'}(\bar a',\bar b')\Psi_{\bar k}(\bar a,\bar b)d\bar a'd\bar b'd\bar ad\bar b
 \end{align}
 Then the block operator $\mathcal{K}$  becomes a semi-infinite block matrix whose blocks $\langle \mathcal{K}\rangle_{\bar k\bar k'}$ are $70\times 70$
 matrices. The entries of each block  are integral operators in $L_2[\mathring{U}_2]\otimes L_2[\mathring{U}(1,1)]$.
\begin{lemma}\label{l:b_C} We have
  \begin{align}\label{b_C.1}
&\langle \widehat C\widehat F\rangle_{\bar k\bar k'}=\delta_{\bar k,\bar k'}\big(C_{d}F_0+\mathcal{C}_0F_0+O(W^{-1})\big)+(1-\delta_{\bar k,\bar k'})W^{-d(\bar k-\bar k')/2}O_*(\mathcal{C}),\\
&\langle \widehat F\widehat Q A_{ab}\rangle_{\bar k\bar k'}=\delta_{\bar k,\bar k'}( \lambda_{\bar k} F_0\widehat \Pi_0
+O(W^{-1}))+(1-\delta_{\bar k,\bar k'})O(W^{-d(\bar k-\bar k')/2})
\notag \end{align}
where $A_{ab}$ is defined in (\ref{phi_0}), $F_0=\widehat F\Big|_{\tilde a_{1,2}=\tilde b_{1,2}=0}$, $C_d=\widehat C_d\Big|_{\tilde a_{1,2}=\tilde b_{1,2}=0}$,
$\mathcal{C}_0=\mathcal{C}\Big|_{\tilde a_{1,2}=\tilde b_{1,2}=0}$,  \\ $\widehat \Pi_0=\Pi_0\oplus\Pi'_0\oplus\widetilde \Pi_0$ (see 
(\ref{exp_Pi})),  and
$ \lambda_{\bar k'}=\lambda_{k'_1}^+\lambda_{k'_2}^-\lambda_{k'_3}^+\lambda_{k'_4}^-$ with $\lambda_{k}^\pm$  defined in (\ref{la_k}). 
We set here
\begin{align}\label{d(k)}
d(\bar k)=\left\{\begin{array}{ll}1,& k=\pm\ell_i,\, or\, k=\pm 3\ell_i,\\
2,&|k|-even\\
3, & otherwise\end{array}\right.
\end{align}
where $|k|=|k_1|+|k_2|+|k_3|+|k_4|$,
\begin{align}\label{ell}
 \ell_1:=(1,0,0,0),\quad \ell_2:=(0,1,0,0),\quad \ell_3:=(0,0,1,0),\quad \ell_4=(0,0,0,1), 
 \end{align}
 \end{lemma}
 \textit{Proof.} 
 To obtain the  assertion  for the matrix $\widehat F\widehat QA_{ab}$, observe that if we expand all integral kernels with respect to $\tilde a,\tilde b$, then
 for entries with $\bar k-\bar k'=\pm \ell_i$
 zero order terms disappear after integration, and the first order terms  give $O(W^{-1/2})$.  The entries with
 $\bar k-\bar k'=\pm 3\ell_i$ are also $O(W^{-1/2})$,  since we have $\tilde a_i^3W^{-1/2}$ and $\tilde b_i^3W^{-1/2}$ 
  in $A^{\pm}_a$ and $A^{\pm}_b$ (see (\ref{K_a*})). The  entries with even $|\bar k-\bar k'|$ can obtain non-zero contribution 
 only from the terms which are of even order (thus at least quadratic) with respect to $\tilde a,\tilde b$, hence they are at least $O(W^{-1})$. And for all other entries, 
 the non-zero contribution can be obtained only from the terms which are of the odd order $3$ or more, and which cannot  contain only 
 $\tilde a_i^3$ or $\tilde b_i^3$.
 
 The assertion of the lemma for the matrix $ \widehat C$ follows   from Lemma \ref{l:U,V}. Indeed,  according to the lemma, the
 eigenvectors (or generalized  eigenvectors) of the entries of $\widehat C$ do not depend on $\widetilde a,\widetilde b$,   $O(1)$ terms  
  of eigenvalues also do not depend on $\widetilde a,\widetilde b$, and the first terms in the asymptotic expansion depends on $\widetilde a,\widetilde b$
 trough the coefficient $(Wt)^{-k}$ only. Hence the expansion of $t_a^{-1}$ or $t_b^{-1}$ with respect to $\widetilde a$ and  $\widetilde b$ will add additional $W^{-1/2}$ 
 in each order.
 
 $\square$
 
 The lemma    imply that at the same basis the operator $\mathcal{K}=\widehat F\widehat Q A_{ab} \widehat C\widehat F$ has the form
\begin{align}\notag
&\langle \mathcal{K}\rangle_{\bar k\bar k}=\lambda_{\bar k}(\Pi^{(F)}_0+F_0\widehat \Pi_0\mathcal{C}_0F_0)+O(W^{-1}),\\ \label{K_kk'}
&\langle \mathcal{K}\rangle_{\bar k\bar k'}=\langle\widehat F\widehat \Pi A_{ab} \widehat C_d\widehat F\rangle_{\bar k\bar k'}+W^{-d(\bar k-\bar k')/2}O_*(\mathcal{C})+O(W^{-3/2})\\
&\qquad\,\,\,\, =W^{-d(\bar k-\bar k')/2}
\big(O(1)+O_*(\mathcal{C})\big)+O(W^{-3/2}),\quad \bar k\ne \bar k',\notag\\ \notag
 &\Pi^{(F)}_0:=F_0\widehat \Pi_0C_{d}F_0.\quad
\end{align}
Consider now the decomposition of the Grassman space $\mathfrak{Q}_{70}$ into 3 subspaces 
\begin{align}\label{dec*}
\mathfrak{Q}_{70}=\mathfrak{L}+\big(\mathfrak{L}_1+\mathfrak{L}_2+\mathfrak{L}_3\big)+\mathfrak{L}'
\end{align} 
where $\mathfrak{L}$, is defined in (\ref{frak_L})   subspaces $\mathfrak{L}_1,\mathfrak{L}_2,\mathfrak{L}_3$ are defined similarly to (\ref{frak_L}) but  with $e^+_0e^-_0$ replaced by 
$e^+_0e^-_1$, $e^+_1e^-_0$, and $e^+_1e^-_1$ respectively. and $\mathfrak{L}'$ is a space of "non good" vectors. Then consider
each $70\times 70$ matrix $\langle \mathcal{K}\rangle_{\bar k\bar k'}$ as a block matrix, corresponding to the decomposition (\ref{dec*}) with
blocks $\langle \mathcal{K}\rangle_{\bar k\bar k'}^{(\mu\nu)}$, ($\mu,\nu=1,2,3$). Let $\langle \mathcal{K}\rangle_{\bar k\bar k'}^{(11)}$ correspond to
$\mathfrak{L}$ and $\langle \mathcal{K}\rangle_{\bar k\bar k'}^{(33)}$ correspond to $\mathfrak{L}'$. Then we use Proposition \ref{p:res}  with $H^{(11)}=\langle \mathcal{K}\rangle_{\bar 0\bar 0}^{(11)}$. 
Denote by $\mathcal{B}(z)=(H^{(22)}-z)^{-1}$. Now $\mathcal{B}(z)$ is also a block matrix with blocks $\mathcal{B}_{\bar k\bar k'}$ (they are quadratic $70\times70$
matrices for $\bar k\not=0,\bar k'\not=0$ or rectangular matrices if $\bar k=0,\bar k'\not =0$ or $\bar k'=0,\bar k\not=0$)
It follows from (\ref{K_kk'}) that
\begin{align}\label{b_B}
\|\mathcal{B}_{\bar k\bar k'}\|\le O(1)W^{-d(\bar k-\bar k')/2}
\end{align}
Indeed,  denoting $G_{\bar k\bar k}=(\langle\mathcal{K}\rangle_{\bar k\bar k}-z)^{-1}$ (remark that $G_{\bar 0\bar 0}
=(\langle\mathcal{K}\rangle_{\bar 0 \bar 0}'-z)^{-1}$ where $\langle\mathcal{K}\rangle_{\bar 0 \bar 0}'$ is $2\times 2$ block matrix constructed
from $\langle\mathcal{K}\rangle_{\bar 0 \bar 0}^{(\mu\nu)}$ with $\mu,\nu=2,3$), we have by (\ref{K_kk'})
\begin{align}\notag
&|\lambda_{0}(\langle\mathcal{K}\rangle_{\bar k\bar k})|\le 1-\delta,\quad\|\langle\mathcal{K}\rangle_{\bar k\bar k'}\|\le CW^{-1/2},\quad \bar k\ne \bar k' \\
& \mathcal{B}_{\bar k\bar k'}=\delta_{\bar k\bar k'}G_{\bar k\bar k}+
\sum G_{\bar k\bar k}\langle \mathcal{K}\rangle_{\bar k\bar k'}G_{\bar k'\bar k'}+\sum 
G_{\bar k\bar k}\langle\mathcal{K}\rangle_{\bar k\bar k_1}
G_{\bar k_1\bar k_1}\langle \mathcal{K}\rangle_{\bar k_1\bar k'}G_{\bar k'\bar k'}+
O(W^{-3/2}).
\label{expr_res}\end{align}
It is easy to check that
\[d(\bar k-\bar k_1)+d(\bar k_1-\bar k')\ge d(\bar k-\bar k').
\]
Hence (\ref{K_kk'}) implies (\ref{b_B}). Similarly, note that
\[d(\bar k)+d(\bar k-\bar k')+d(\bar k')\ge 4, 
\]
for all $\bar k,\bar k'$ except $\bar k=\bar k'=\ell_i$ or $3\ell_i$,  $\bar k=\bar k'=0$, and $\bar k=0$, $\bar k'=\ell_i, 3\ell_i$ or
$\bar k'=0$, $\bar k=\ell_i, 3\ell_i$. Thus we obtain
\begin{align*}
H^{(12)}(H^{(22)}-z)^{-1}H^{(21)}=&
%\sum\langle\mathcal{K}\rangle_{0\bar k}^{(1\mu)}\mathcal{B}_{\bar k\bar k'}^{(\mu\nu)}
%\langle\mathcal{K}\rangle_{\bar k'0}^{(\nu1)}
P_\mathfrak{L}\Big(\sum \langle\mathcal{K}\rangle_{\bar 0\bar k}G_{\bar k\bar k}\langle \mathcal{K}\rangle_{\bar k\bar0}\Big)P_\mathfrak{L}+
\sum \langle\mathcal{K}\rangle_{\bar 0\bar 0}^{(1\mu)}G_{\bar 0\bar 0}^{(\mu\nu)}\langle \mathcal{K}\rangle_{\bar 0\bar0}^{(\nu1)}\\
&+\sum \langle\mathcal{K}\rangle_{\bar 0\bar 0}^{(1\mu)}\mathcal{B}_{\bar 0\bar k}^{(\mu\nu)}\langle \mathcal{K}\rangle_{\bar k\bar0}^{(\nu1)}
+\sum \langle\mathcal{K}\rangle_{\bar 0\bar k}^{(1\mu)}\mathcal{B}_{\bar k\bar 0}^{(\mu\nu)}\langle \mathcal{K}\rangle_{\bar 0\bar0}^{(\nu1)}+O(W^{-2})
%=\Sigma_1+\Sigma_2+\Sigma_3+\Sigma_4+O(W^{-2}),
\end{align*}
where summations are with respect to $\bar k=\ell_i,3\ell_i$ and $\mu,\nu=2,3$. But denoting $\Sigma_1$ and $\Sigma_2$ the last two sums above,
we have
\begin{align}\label{b_sigma}
\Sigma_1+\Sigma_2=O(W^{-2})+W^{-1}O_*(\mathfrak{r}^{(1)})O_*(\mathfrak{r}^{(1)})
\end{align}
(see Definition \ref{d:type}). Indeed, consider $\Sigma_1$.
By  (\ref{K_kk'}) and (\ref{cal_R}) entries of $\langle\mathcal{K}\rangle_{\bar 0\bar0}^{(1\mu)}$ can be $O(W^{-1})$,   $O_*(\mathfrak{r}^{(2)})$ for $\mu=2$, or
 $O_*(\mathfrak{r}^{(1)})$ for $\mu=3$. In addition, (\ref{b_B}) implies
 \[
 \|\mathcal{B}_{\bar 0\bar k}^{(\mu\nu)}\|\le CW^{-1/2},%\quad \|\langle \mathcal{K}\rangle_{\bar k\bar0}^{(\nu1)}\|\le CW^{-1/2},
 \]
 and according to (\ref{K_kk'}) $\langle \mathcal{K}\rangle_{\bar k\bar0}^{(\nu1)}$ is $O(W^{-3/2})$ or at least $W^{-1/2}O_*(\mathfrak{r}^{(1)})$.
 All together this gives  (\ref{b_sigma}) for $\Sigma_1$. The estimate for $\Sigma_2$ can be obtained by the same argument.
 
%  for the case $\mu=2$.  If $\mu=\nu=3$ we get $O_*(\mathfrak{r}^{(1)})$ from $\langle\mathcal{K}\rangle_{\bar 0\bar0}^{(13)}$,
%$W^{-1/2}$ from $\mathcal{B}_{\bar 0\bar k}^{(33)}$, and $W^{-3/2}$ or $W^{-1/2}O_*(\mathfrak{r}^{(1)})$ from $\langle \mathcal{K}\rangle_{\bar k\bar0}^{(31)}$,
%which also implies (\ref{b_sigma}). For $\mu=3$, $\nu=2$ we get $O_*(\mathfrak{r}^{(1)})$ from $\langle\mathcal{K}\rangle_{\bar 0\bar0}^{(13)}$,
%$W^{-1/2}$ from $\mathcal{B}_{\bar 0\bar k}^{(32)}$, and $W^{-3/2}$ or $W^{-1/2}$
%we first expand $G_{\bar k\bar k}$ with respect to $\mathcal{C}_0$ (see (\ref{K_kk'}))
%\[
%G_{kk}=(\lambda_k\widehat\Pi_0^{(F)}-z)^{-1}
%\sum\limits_{m}(-1)^m\Big((\lambda_kF_0\widehat \Pi_0\mathcal{C}_0F_0+O(W^{-1}))(\lambda_k\widehat\Pi_0^{(F)}-z)^{-1}\Big)^m.
%\]
%This yields that $(G_{kk})^{\mu\nu}$ for $\mu\ne \nu$ gives $O(W^{-1})$ or at least $O_*(\mathfrak{r}^{(1)})$. Thus according to 
%(\ref{expr_res})  and (\ref{K_kk'})  $\mathcal{B}_{\bar 0\bar k}^{(32)}$ gives $W^{-3/2}$
%
%combined with  (\ref{K_kk'}) gives us at least $W^{-1/2}O_*(\mathfrak{r}^{(1)})$ for 
%   $\mathcal{B}_{\bar 0\bar k}^{(32)}$ and 
%and   at least $O(W^{-1/2})$ for  $\langle\mathcal{K}\rangle_{\bar k\bar0}^{(21)}$.

Now take  $T$ of (\ref{T}) and set
\begin{align}\label{fr_M}
\mathfrak{M}'=T(H^{(11)}-\widetilde{H})T, \quad 
\widetilde H:=H^{(12)}(H^{(22)}-z)^{-1}H^{(21)}.
\end{align}
By the above consideration
\begin{align}\label{ti_H}
\widetilde H=
P_\mathfrak{L}\Big(\sum \langle\mathcal{K}\rangle_{\bar 0\bar k}G_{\bar k\bar k}\langle \mathcal{K}\rangle_{\bar k\bar0}\Big)P_\mathfrak{L}+
\sum \langle\mathcal{K}\rangle_{\bar 0\bar 0}^{(1\mu)}G_{00}^{(\mu\nu)}\langle \mathcal{K}\rangle_{\bar 0\bar0}^{(\nu1)}+O(W^{-2})
+W^{-1}O_*(\mathfrak{r}^{(2)}).
\end{align}
Denote by $M$ an upper left $4\times 4$ block of $(H^{(11)}-\widetilde{H})$. Notice that, after above transformation with $T$, only the diagonal entries, 
 the entries $M_{i1}$ and $M_{4i}$ $(i=2,3)$ multiplied by $W$ and $M_{41}$ multiplied by $W^2$ may stay  $O(1)$ or more. 
It is easy to see that both $\langle\mathcal{K}\rangle_{\bar 0\bar 0}^{(1\mu)}$ and $\langle \mathcal{K}\rangle_{\bar 0\bar0}^{(\nu1)}$ give
at most $O_*(\mathfrak{r}^{(1)})$, and so the contribution of the second term of (\ref{ti_H}) to $WM_{i1}$ and $WM_{4i}$ $(i=2,3)$ is at most
$WO_*(\mathfrak{r}^{(2)})$ which we include in $\widetilde K_i$ (see (\ref{r_tiK})). Now we are going to compute the contribution of the 
second term of (\ref{ti_H}) to $W^2 M_{41}$. To this end we write (see (\ref{K_kk'}))
\[
\langle\mathcal{K}\rangle_{\bar 0\bar 0}^{(1\mu)}=F_0\widehat \Pi_0^{(11)}\mathcal{C}_0^{(1\mu)}F_0+O(W^{-1}), \quad 
\langle\mathcal{K}\rangle_{\bar 0\bar 0}^{(\nu 1)}=F_0\widehat \Pi_0^{(\nu\nu)}\mathcal{C}_0^{(\nu1)}F_0+O(W^{-1})
\]
 and expand $(\langle\mathcal{K}\rangle_{\bar 0\bar 0}-z)^{-1}$ with respect to $\mathcal{C}_0$:  %(see (\ref{P_ij}) for the definition of $P_0$)
%\begin{align}\label{err}
%P_o\Big(\Big(F_0\widehat \Pi_0(\mathcal{C}_0+W^{-1}\mathcal{X}_0)F_0(\Pi^{(F)}_0-z)^{-1}\Big)^m
%F_0\widehat \Pi_0(\mathcal{C}_0+W^{-1}\mathcal{X}_0)F_0\Big)P_0,
%\end{align}
\begin{align}\label{err}
(\widehat\Pi_0^{(F)}-z)^{-1}
\sum\limits_{m}(-1)^m\Big((F_0\widehat \Pi_0\mathcal{C}_0F_0+O(W^{-1}))(\widehat\Pi_0^{(F)}-z)^{-1}\Big)^m.
\end{align}
The entry $M_{41}$ is the result of the application of  the matrix $(H^{(11)}-\widetilde{H})$ to the vector $e_0^+e_0^-$ projected onto
$n_{12}n_{21}e_0^+ e_0^-$. But  matrices $\widehat \Pi_0$, $ F_0$ and $\big(\widehat\Pi_0^{(F)}-z\big)^{-1}$ do not increase the total number of 
 $\rho_{12},\rho_{21},\tau_{12},\tau_{21}$ in the Grassman vector to which they are applied. Hence, if the term in (\ref{err})  does not contain
 $O(W^{-1})$ we need to take the entries which add into the result   $\rho_{12},\rho_{21},\tau_{12},\tau_{21}$ from
 $\mathcal{C}_0$, which gives   $O_*(\mathfrak{r}^{(4)})$ 
 (recall that $\|O_*(\mathfrak{r}^{(4)})\|\le CW^{-2}$ by Lemma \ref{l:U,V}). 
 If the term in (\ref{err}) contains two or more $O(W^{-1})$, then it is $O(W^{-2})$, and if the term contain only one  $O(W^{-1})$, then we have at most $W^{-1}O_*(\mathfrak{r}^{(2)})$, since  both $\langle\mathcal{K}\rangle_{\bar 0\bar 0}^{(1\mu)}$ and $\langle \mathcal{K}\rangle_{\bar 0\bar0}^{(\nu1)}$ give
 $O_*(\mathfrak{r}^{(1)})$. Thus contribution of the second term of (\ref{ti_H}) to $W^2M_{41}$ can be included in $\widetilde K_5$ (see (\ref{r_tiK})).

Now rewrite the contribution of the first sum of (\ref{ti_H}) to $M$ as sum of the terms
\begin{align}\label{T_k}
P_0\mathcal{T}_{\bar k}P_0,\quad \mathcal{T}_{\bar k}=P_g\langle\mathcal{K}\rangle_{\bar 0\bar k}
G_{\bar k\bar k}\langle \mathcal{K}\rangle_{\bar k\bar0} P_g=\big(\langle\mathcal{K}\rangle_{\bar 0\bar k}
G_{\bar k\bar k}\langle \mathcal{K}\rangle_{\bar k\bar0})^{(11)},
\end{align}
where $P_0$ is defined in (\ref{P_ij}), and $P_g$ is the projection on the "good" vectors in $\mathfrak{Q}_{70}$ (evidently $P_0P_g=P_0$) and
the upper index $(11)$ corresponds to the upper left block of the matrix in the decomposition (\ref{dec}).
Since all matrices $\hat \Pi$, $\hat F$, $F_0$, $\hat C_d$, $\widehat\Pi_0$, $\widehat\Pi_0^{(F)}$ are block diagonal in decomposition (\ref{dec}),
the expression for $\mathcal{T}_{\bar k}$
%\[
%P_g\langle\mathcal{K}\rangle_{\bar 0\bar k}G_{\bar k\bar k}\langle \mathcal{K}\rangle_{\bar k\bar0}P_g
%\]
may include non-diagonal blocks of matrices $O_*(\mathcal{C})$ only. But then we should have at least two of such blocks which gives at
most $O_*(\mathfrak{r}^{(2)})$. In addition, according to (\ref{K_kk'}) $\langle\mathcal{K}\rangle_{\bar 0\bar k}$ and $\langle \mathcal{K}\rangle_{\bar k\bar0}$
give $O(W^{-1/2})$ each. Thus all such terms are of order $W^{-1}O_*(\mathfrak{r}^{(2)})$, and we have to take only terms that does not contain any $O_*(\mathcal{C})$
(remark that diagonal block of $O_*(\mathcal{C})$ is $W^{-1}$ so it can give the contribution only to the terms of order $O(W^{-2})$).
Hence to check that $(T_0\otimes T_0)M(T_0\otimes T_0)$ has the form (\ref{cal_M}), we need  to study the structure of two matrices
\begin{align}\label{M.1}
M_1=&P_0\langle \hat F\hat \Pi (1+ W^{-1}\mathcal{X})A_{ab}(\widehat C_d+\mathcal{C})\hat F\rangle_ {\bar 0\bar 0}P_0+err\\
M_2=&P_0\Big(\sum _{\bar k=\ell_i,3\ell i}\langle\hat F\hat \Pi  A_{ab}K_{US}\hat F\rangle_{\bar 0\bar k}(\lambda_{\bar k} \Pi^{(F)}_0-z)^{-1}
\langle \hat F\hat \Pi   A_{ab}K_{US}\hat F\rangle_{\bar k\bar0}\Big)P_0+err
\notag\end{align}
where $err$ means the error terms which we describe above, and $P_0$ is defined in (\ref{P_ij}).

 Let us check that up to $err$ terms the matrix $(T_0\otimes T_0)M_1(T_0\otimes T_0)$ has the same form as (\ref{cal_M}), but with
 functions $u$ and $y_1,y_2$ (see (\ref{M_0.1})), replaced by some constants $\tilde u,\tilde y_1,\tilde y_2$.
Notice that  the block structure of $Q$ (see (\ref{dec}))
and the fact that only even degrees of  $\widetilde a_1, \widetilde a_2,\widetilde a_1',\widetilde a_2',
\widetilde b_1, \widetilde b_2,\widetilde b_1',\widetilde b_2'$ give non-zero contribution yield
\begin{align}\label{M_12}
M_1=&P_0P_g\Big(\langle \hat F\Pi K_{US}\hat F\rangle_ {\bar 0\bar 0}+ W^{-1}F_0\Pi_0\mathcal{X}_0^{(11)}K_0F_0+ F_0\Pi_0\mathcal{C}^{(11)}_0F_0\Big)P_gP_0
+err.
\end{align}
Indeed, using that $\widehat\Pi$ in the decomposition (\ref{dec}) is block diagonal and $\mathcal{X}^{(12)}=0$ (see (\ref{X,Y})), we have
\begin{align}\label{M_13}
(\widehat F\widehat \Pi (1+ W^{-1}\mathcal{X})\widehat C\widehat F)^{(11)}=\widehat F\Big(\widehat \Pi^{(11)}\widehat C^{(11)}+
W^{-1}\widehat \Pi^{(11)}\mathcal{X}^{(13)}\mathcal{C}^{(31)}\Big)\widehat F.
\end{align}
By (\ref{dec}) $\mathcal{X}^{(13)}$ corresponds to the transformation of "non good"  vectors into "good" ones. But by (\ref{X_0}) there are only two "non good vectors" that after application of $\mathcal{X}$ become "good": $\rho_{11}\tau_{12}\rho_{22}\tau_{21}$ and
$\rho_{12}\tau_{22}\rho_{21}\tau_{11}$. To obtain these vectors from "good" ones, we need at least two "transformations", hence corresponding
entries of $\mathcal{C}^{(31)}$ should be  at least $O_*(\mathfrak{r}^{(2)})$, which gives $W^{-1}O_*(\mathfrak{r}^{(2)})$
and so can be included to an error term.

Let us now consider the contribution the  second and the third matrices in the r.h.s. of (\ref{M_12}). According to (\ref{X_0}),  $\mathcal{X}_0^{(11)}$
depends on $n_{21}-n_{12}$ and does not contain terms with $n_{12}n_{21}$, and $\Pi_0$ and $F_0$ do not increase the total number of 
 $\rho_{12},\rho_{21},\tau_{12},\tau_{21}$ in the Grassman vector to which they are applied. Thus the application of $F_0\Pi_0\mathcal{X}_0^{(11)}F_0$
 to $e_0^+e_0^-$ gives zero contribution to $M_{41}$, and gives  the same constant contribution to $M_{21}$ and $M_{31}$ but with the opposite sign
 (this contribution thus can be included in $\tilde u$). Similarly with $M_{42}$ and $M_{43}$. Same statement for $F_0\Pi_0\mathcal{C}^{(11)}_0F_0$
 follows from (\ref{41}).

 To study the first term of the r.h.s. of (\ref{M_12}), notice that since only even degrees of  $\widetilde a_1, \widetilde a_2,\widetilde a_1',\widetilde a_2',
\widetilde b_1, \widetilde b_2,\widetilde b_1',\widetilde b_2'$ give non-zero contribution, we need to study only the contribution of order $W^{-1}$
(the next order will be $W^{-2}$ and may give contribution only to $W^2M_{41}$, thus will be included in $u_0$ (see (\ref{cal_M}))).
 Observe that to obtain the non-zero contribution of order $W^{-1}$ we need to consider
the term of this order from
\begin{multline}\label{expr_0}
\Big(Q(c_{12})\otimes Q(c_{21})\otimes Q(c_{11})\otimes Q(c_{22}) \Big)\\
\times K_{US}(t_{a'},t_{b'}) A_a^+(\widetilde a_1,\widetilde a_1')A_b^+(\widetilde b_1,\widetilde b_1')A_a^-(\widetilde a_2,\widetilde a_2')A_b^-(\widetilde b_2,\widetilde b_2')
\hat F(\bar a,\bar b)\hat F(\bar a',\bar b')
\end{multline}
 (see (\ref{K_a*}), (\ref{F_a,b}), (\ref{Q(x)}) and (\ref{ch_ab})). Notice first that by (\ref{type0}) the terms containing the derivatives of
 $K_{US}$ with respect to $\tilde a,\tilde b$ will give us $O_*(\mathfrak{r}^{(2)})$ so after multiplication by $W$ they contribute to $\widetilde K_i$. 
 Thus we can change $K_{US}$ to $K_0$. Also
  it is easy to see that  we need to take at least
 one $\widetilde a_1',\widetilde a_2',\widetilde b_1',\widetilde b_2'$ from $Q(c_{12})\otimes Q(c_{21})$, since otherwise the respective term will
 contain $Q_0\otimes Q_0$ and will not contribute to  the entries which are important for us. The second observation is that if we take, e.g., the term
 containing $\widetilde a_1'$  from the first  matrix, then we need to complete it by $\widetilde a_1'$ or $(\widetilde a_1')^3$,
 or $\widetilde a_1$ or $(\widetilde a_1)^3$, otherwise the contribution will be $0$. Hence, in this case the the second matrix
should stay untouched. The next observation is that $Q'(c)$ has the form:
 \begin{align}\label{Q'}
 Q'(c)=\left(\begin{array}{cc}
-\beta^{-1} &0\\-1&0\end{array}\right)
 \end{align}
hence $Q''(c)=0$, and if we take $\widetilde a_1'$  from the first matrix, the respective term will be of the form
\[
F_0^2\cdot (Q'\otimes Q_0\otimes \widetilde Q)+F_0\frac{\partial F}{\partial a_1}\Big|_{\bar a=\bar b=(a_+,a_-)} (Q'\otimes Q_0\otimes \widetilde Q_1) 
\]
where $\widetilde Q$ and  $\widetilde Q_1$ are some $4\times 4$ matrix, whose coefficients depend on 
\[\langle \widetilde a_1'\cdot \widetilde a_1'\rangle_{\bar 0\bar 0},\,\,\langle  \widetilde a_1'\cdot (\widetilde a_1')^3\rangle_{\bar 0\bar 0},\,\,
\langle \widetilde a_1'\cdot \widetilde a_1\rangle_{\bar 0\bar 0},\,\,\langle \widetilde a_1'\cdot (\widetilde a_1)^3\rangle_{\bar 0\bar 0}\]
If we consider the contribution of the terms, containing $\widetilde b_1'$ taken from the second matrix in (\ref{expr_0}), then
it will have the form
\[
-F_0^2(Q_0\otimes Q'\otimes \widetilde Q)-F_0\frac{\partial F}{\partial b_1}\Big|_{\bar a=\bar b=(a_+,a_-)} (Q_0\otimes Q'\otimes \widetilde Q_1)
\]
with the same $\widetilde Q$ and  $\widetilde Q_1$, since the coefficients  at  $\widetilde a_1'$ differs from the respective coefficient at $\widetilde b_1'$ by the multiple $i$, the same for the coefficients at  $\widetilde a_1$ and $\widetilde b_1 $, $(\widetilde a_1)^3$ and 
$(\widetilde b_1)^3 $, and $(\widetilde a_1')^3$ and 
$(\widetilde b_1')^3$. Repeating the same argument for the terms containing $\widetilde a_2$  taken from $Q(c_{21})$ 
in (\ref{expr_0}) and
$\widetilde b_2$ taken from $Q(c_{12})$ we obtain the form (\ref{cal_M}) for $TM_1T$.

% Using that $\Pi_0(1+W^{-1}\mathcal{X}_0)e_0^+e_0^-$ does not contain terms with $n_{12}n_{21}$
%(see (\ref{X_0})),   we get
%\begin{align*}
%&\Big(P_0\Pi^{(F)}_0(1+W^{-1}\mathcal{X}_0)P_0\Big)_{41}=0, \\
%&\Big(P_0\Pi^{(F)}_0(1+W^{-1}\mathcal{X}_0)P_0\Big)_{4i} =O(W^{-1}),\quad
%\Big(P_0\Pi^{(F)}_0(1+W^{-1}\mathcal{X}_0)P_0\Big)_{i1}=O(W^{-1})
%\end{align*}

Now let us study the matrix $M_2$ of (\ref{M.1}). Define
\begin{align}\label{HGH}
&M_2'=
K_F^2\sum_{\bar k=\ell_i,3\ell_i}P_g\langle r\rangle_{\bar 0\bar k}P_g(\lambda_{\bar k} P_g\Pi^{(F)}_0P_g-z)^{-1}P_g\langle r\rangle_{\bar k\bar 0}P_g+O(W^{-2})
 \\
=&K_F^2z^{-1}\sum_{\bar k=\ell_i,3\ell_i}\sum_m(\lambda_{\bar k}/z)^m(K_F)^{m}P_g\langle r\rangle_{\bar 0\bar k} \Pi_0^m
\langle r\rangle_{\bar k\bar 0}P_g+O(W^{-2}),
\notag\end{align}
where $P_g$ is the projection on the "good" vectors in $\mathfrak{Q}_{70}$, and $K_F=F_0K_0F_0$ (see the formulation of Theorem \ref{t:M_0}).
Here $\ell_i$ are defined in (\ref{ell})
and $r$ collects the terms of order $W^{-1/2}$ which appear in (\ref{expr_0}). It is easy to see that $M_2=P_0M_2'P_0$.
Recall that we are interested in $(M_2)_{1i}$ and $(M_2)_{i4}$ ($i=2,3$).
For $\bar k=3\ell_i$  the only terms  in $r$ which can give non-zero contribution  are ones, containing  $\widetilde a_j^3$ or
$\widetilde b_j^3$. But then in both terms $P_g\langle r\rangle_{\bar 0\bar k}$ and $P_g\langle r\rangle_{\bar k\bar 0}$ the
first two matrices remain untouched and taking into account that
\[
 \Pi_0^m=Q_0^m\otimes Q_0^m \otimes (1+q_+^mP_{10} +q_-^mP_{01}+ (q_+q_-)^mP_{11}),\quad Q_0^m=\left(\begin{array}{cc}
1&m\beta^{-1} \\0&1\end{array}\right)
\]
we conclude that the terms with $\bar k=3\ell_i$ do not contribute in the entries of $P_0M_2'P_0$, which are interesting for us.

For $\bar k=\ell_1,\ell_3$, repeating the argument used above for $M'_1$ and taking into account that $\hat Q$ 
 depend on $\widetilde a_j',\widetilde b_j'$, but does not depend on $\widetilde a_j,\widetilde b_j$, we obtain that up to the term which 
 do not contribute to the "important" entries
 \begin{align*}
&P_g\langle r\rangle_{\bar 0\ell_1} \Pi_0^m\langle r\rangle_{\ell_1\bar 0}P_g=
K_F^2W^{-1}\Big(\nu_+^{(1)}F_0^{-1}\frac{\partial \widehat F}{\partial  a_1}+\nu_+^{(2)}\Big)\Big(\big(Q'Q_0^{m-1}\big)\otimes Q_0^m
\otimes Q_+^m\otimes Q_-^m\Big)\\
&P_g\langle r\rangle_{\bar 0\ell_3} \Pi_0^m\langle r\rangle_{\ell_3\bar 0}P_g=-K_F^2W^{-1}
\Big(\nu_+^{(1)}F_0^{-1}\frac{\partial \widehat  F}{\partial  b_1}+\nu_+^{(2)}\Big)\Big(Q_0^m\otimes \big(Q'Q_0^{m-1}\big)\otimes Q_+^m\otimes Q_-^m\Big)
\end{align*}
 where the coefficients $\nu_+^{(1)}$ and $\nu_+^{(2)}$ are not important for us.
  Similar relations hold for $\langle r \rangle_{\bar 0\ell_2}$ and $\langle  r \rangle_{\bar 0\ell_4}$ with some $\nu_-^{(1)}$ and $\nu_-^{(2)}$.

 Then after multiplication by  $(\lambda_{\bar k}/z)^m(K_F)^{m}$, summation over $m$
and then transformation (\ref{fr_M}) we obtain representation  (\ref{P_0GP_0}) for $\mathcal{M}$.

The analysis of 
\begin{align*}
&\mathcal{M'}=T_0\otimes T_0M', \quad \mathcal{M}''=M''T_0\otimes T_0,\quad \mathcal{M}'''=M'''\\
&M'=P_0(H^{(11)}-\widetilde H)(1-P_0), \quad M''=(1-P_0)(H^{(11)}-\widetilde H)P_0,\\
& M'''=(1-P_0)(H^{(11)}-\widetilde H)(1-P_0)
\end{align*}
 is simpler than that for $M$, since we need only to show that the entries of $M'$ and $M''$ are $O_*(\mathfrak{r}^{(1)})O_*(\mathfrak{r}^{(1)})$
 or $O(W^{-2})$.
Introduce $P_{sg}$ - projection onto the subspace of "semi good" vectors in $\mathfrak{Q}_{70}$. Similarly to (\ref{M_13}), using the decomposition (\ref{dec}),
we obtain
\begin{align*}
&P_0H^{(11)}(1-P_0)=P_{\mathfrak{L}}P_g\langle\mathcal{K}\rangle_{\bar 0\bar 0}P_{sg}P_{\mathfrak{L}}=P_{\mathfrak{L}}
\langle\mathcal{K}^{(12)}\rangle_{\bar 0\bar 0}P_{\mathfrak{L}}\\
&\mathcal{K}^{(12)}=(\widehat F\widehat \Pi (1+ W^{-1}\mathcal{X})\widehat C\widehat F)^{(12)}=\widehat F\Big(\widehat \Pi^{(11)}
(1+W^{-1}\mathcal{X}^{(11)})\mathcal{C}^{(12)}+
W^{-1}\widehat \Pi^{(11)}\mathcal{X}^{(13)}\mathcal{C}^{(32)}\Big)\widehat F\\
&=O_*(\mathfrak{r}^{(1)})O_*(\mathfrak{r}^{(1)})+W^{-1}O_*(\mathfrak{r}^{(2)})
\end{align*}
where we have used that by (\ref{cal_R}) $\mathcal{C}^{(12)}=O_*(\mathfrak{r}^{(1)})O_*(\mathfrak{r}^{(1)})$ and the argument given after (\ref{M_13}).
And for $P_0\widetilde H(1-P_0)$ we have similarly to  (\ref{T_k})
\begin{align*}
P_0\widetilde H(1-P_0)=P_{\mathfrak{L}}P_g\widetilde H P_{sg}P_{\mathfrak{L}},\quad
P_g\widetilde H P_{sg}=\sum \big(\langle\mathcal{K}\rangle_{\bar 0\bar k}G_{\bar k\bar k}
\langle \mathcal{K}\rangle_{\bar k\bar0}\big)^{(12)}+err
\notag\end{align*}
Repeating the argument given after (\ref{T_k}) we obtain, that the r.h.s. above either contains $\mathcal{C}^{(12)}$, or contains at least
two non-diagonal blocks of  $\mathcal{C}$. In both cases the contribution of the corresponding terms into the r.h.s. above is 
$W^{-1}O_*(\mathfrak{r}^{(1)})O_*(\mathfrak{r}^{(1)})$.
  As for $M'''$, since $\widetilde H=O(W^{-1})$,  (\ref{cal_M'}) follows from (\ref{K_kk'}) and (\ref{cal_R}).
 
 Now let us find $f^{(1)}$ and $g^{(1)}$  from Proposition \ref{p:res}.
Remark  that  since $v$ of the (\ref{repr}) contains the odd degrees of $\tilde a,\tilde b$ only with the coefficients of order $W^{-k/2}$ ($k\ge 1$),
 we have (see (\ref{d(k)})):
\begin{align*}
f_0=&(v,\Psi_{\bar 0})P_{\mathfrak{L}}e^{(0)}=(1+O(W^{-1}))
C_{0} F_0P_{\mathfrak{L}}e^{(0)},\,\,\\
f_{1}=&(1+O(W^{-1}))\Big( C_{0} (1-P_{\mathfrak{L}})e^{(0)},    \{f_{1\bar k}e^{(0)}\}_{\bar k\not=0}\Big)F_0,\\
f_{1\bar k}:=& F_0^{-1}(v,\Psi_{\bar k})\quad |f_{1\bar k}|\le CW^{-d(k)/2}
%=\prod f_{k_i}^{(i)}, \quad f_{k_i}^{(i)}=O(1)(1+(-1)^{k_i})+O(W^{-1/2})(1-(-1)^{k_i})
\end{align*}
where the first component of $f_1$ corresponds to $((1-P_{\mathfrak{L}})ve^{(0)},\bar \Psi_{\bar 0})$.
Then, similarly to (\ref{ti_H}) 
\begin{multline}
f^{(1)}=f_0-\sum_{\bar k=\ell_i,3\ell_i} P_{\mathfrak{L}}\langle\mathcal{K}\rangle_{\bar 0\bar k}G_{\bar k\bar k}f_{1\bar k}F_0\\-
\sum_{\mu=2,3} \langle\mathcal{K}\rangle_{\bar 0\bar 0}^{(1\mu)}G_{00}^{(\mu1)}f_{1\bar 0}F_0+O(W^{-2})F_0
+W^{-1}O_*(\mathfrak{r}^{(2)})F_0
\label{f'}
\end{multline}
Denoting $\Sigma_1'$ the first sum above, one can see similarly to (\ref{err}) 
\begin{align}\notag
\Sigma_1'=&P_{\mathfrak{L}}\Big(\sum _{\bar k=\ell_i,3\ell_i}P_g\langle \hat F\hat\Pi\hat C_d\hat F\rangle_{\bar 0\bar k}(\lambda_{\bar k} \widehat\Pi_0^{(F)}-z)^{-1}f_{1\bar k}\Big)F_0+\Big(O(W^{-2})+W^{-1}O_*(\mathfrak{r}^{(2)})\Big)F_0\\
=&P_{\mathfrak{L}}\Big(\sum _{\bar k=\ell_i,3\ell_i}P_g\langle r \rangle_{\bar 0\bar k}(\lambda_{\bar k} \widehat\Pi_0^{(F)}-z)^{-1}P_gf_{1\bar k}\Big)F_0
 +\Big(O(W^{-2})+W^{-1}O_*(\mathfrak{r}^{(2)})\Big)F_0
\label{sigma_1}\end{align}
Here we  replaced $P_{\mathfrak{L}}$ by $P_{\mathfrak{L}}P_g$, since $f_{1\bar k}=P_gf_{1\bar k}$ and $\hat\Pi$ and $\widehat \Pi^{(F)}_0$ have a block structure
and use $r$ defined in (\ref{HGH}). 

Repeating the argument which we used for $M$, we get that
all entries of $\langle\mathcal{K}\rangle_{\bar 0\bar 0}^{(1\mu)}$ and $G_{00}^{(\mu1)}$ have an order $O(W^{-1})$ or $O_*(\mathfrak{r}^{(1)})$
Then denoting $\Sigma_2'$ the second sum in (\ref{f'}) we have
\begin{align*}
\Sigma_2'=\Big(O(W^{-2})+O_*(\mathfrak{r}^{(1)})O_*(\mathfrak{r}^{(1)})\Big)F_0
\end{align*}
Since $f_{1\bar 0}\sim p(n_{11},n_{22})$, where $p$ is some polynomial, applying the argument given after (\ref{err}), we obtain 
for the forth component $(f^{(1)})_4$  of $f^{(1)}$ 
\[
(f^{(1)})_4=\Big(O(W^{-2})+W^{-1}O_*(\mathfrak{r}^{(2)})+O_*(\mathfrak{r}^{(4)})\Big)F_0
\]
 Additional terms which come into $(f^{(1)})_1$ from $\Sigma_1'$ and $\Sigma_2'$ are $O(W^{-1})+O_*(\mathfrak{r}^{(2)})$, hence
 \[(f^{(1)})_1=\Big(d_{1*}+O(W^{-1})+O_*(\mathfrak{r}^{(2)})\Big)F_0, \quad d_{1*}=C_0 (P_{\mathfrak{L}}e^{(0)})_1
 \]
Moreover,  repeating for the sum in the r.h.s. of (\ref{sigma_1})  the  argument given after (\ref{HGH}), we obtain
\begin{align*}
&W(f^{(1)})_2=\Big(\phi(K_F/z)+O(W^{-1})+O_*(\mathfrak{r}^{(2)})\Big)F_0,\quad\\
&W(f^{(1)})_3=\Big(-\phi(K_F/z)+O(W^{-1})+O_*(\mathfrak{r}^{(2)})\Big)F_0
\end{align*}
with some $\phi(\zeta)$ analytic in $\mathbb{B}_{1+\delta}$. Finally, since
\begin{align*}
Te_1=W^{-1}e_4,\quad Te_2=e_3,\quad Te_3=e_2,\quad Te_4=We_1,
%\\
%T^*e_1=We_4,\quad T^*e_2=e_3,\quad T^*e_3=e_2,\quad T^*e_4=W^{-1}e_1
\end{align*}
we obtain that $\widehat f=WTf^{(1)}$ satisfies conditions of (\ref{repr.fin}). The relations for $\widehat g=WT^*g^{(1)}$ can be obtained similarly.

\smallskip

To finish the proof of Theorem \ref{t:M_0}, we are left to prove (\ref{g=0}).

To this end consider the case when $n\sim \log W$. Since  by (\ref{cal_R})   $\|\mathfrak{r}\|\le CW^{-1/2}$, 
we have $\|\tilde K_i\|\le CW^{-1/2}$ and using the formula (\ref{A^-1}) for $\mathfrak{M}_1(z)$ (which is $\mathfrak{M}(z)$
with $\widetilde K_i=0$)
we evidently get the trivial bound
\begin{align*}
\|(\mathfrak{M}_1(z)-z)^{-1}\|\le &C\|G(z)\|^3,\quad G(z)=(K_F-z)^{-1},\\
\Rightarrow\|\mathfrak{G}(z)\|=&\|(\mathfrak{M}_1(z)-z+O(W^{-1/2}))^{-1}\| \\&\le
\|(\mathfrak{M}_1(z)-z)^{-1}\|\,\|1+O(W^{-1/2})(\mathfrak{M}_1(z)-z)^{-1}\|\le Cn^3.
\end{align*}
Here we used that $\|G(z)\|\le Cn$ for $z\in\omega_A$.
Thus, if we define $\mathfrak{M}_0(z)$ as $\mathfrak{M}_1(z)$ with $K_F$ replaced by $F_0^2$ in the upper left block
and $\mathfrak{G}_0(z)=(\mathfrak{M}_0(z)-z)^{-1}$, then, by (\ref{res_id}),
\begin{align*}
&|(\mathfrak{G}(z)-\mathfrak{G}_0(z))\widehat f,\widehat g)|\le
\|\mathfrak{G}(z)\|\Big(O(W^{-1/2})\|\mathfrak{G}_0(z)\widehat f\|\cdot \|\widehat g\|+\|(1-K_F)\mathfrak{G}_0(z)\widehat f\|\cdot \|\widehat g\|\Big)\\
&\le Cn^3\Big(O(W^{-1/2})n+\|(1-K_F)\mathcal{P}_L\|\,\|\mathfrak{G}_0(z)\widehat f\|\,\|\widehat g\|+\|1-K_F\|\,
\|(1-\mathcal{P}_L)\mathfrak{G}_0(z)\widehat f\|\,\|\widehat g\|\Big)\to 0
\end{align*}
Here we used the projection $\mathcal{P}_L$ defined in (\ref{P_L}), bound $\|(1-K_F)\mathcal{P}_L\|\le L^2/W$ which follows from
(\ref{U,V.1}),  the bound (\ref{f-Pf})  for the last term here, and $n\sim \log W$.
Now,  the above argument, bound  (\ref{b_I_+}) for $I_+$, and a similar bound for $I_-$, yield
\begin{align*}
\mathcal{R}_{nW}^{+-}(E,\varepsilon,\xi)=I_{\pm}+I_++I_-=&-\frac{1}{2\pi i}\oint_{\omega_A} z^{n-1}(\mathfrak{G}_0(z)\widehat f,\widehat g)dz+o(1),
\end{align*}
and thus we can rewrite $\mathcal{R}_{nW}^{+-}(E,\varepsilon,\xi)$ according to (\ref{int_z}).

Observe that by definition of $P_{\mathfrak{L}}$ (see (\ref{P_frak_L})) one can see that the constants $d_{1*}$ and $d_{2*}$ 
of (\ref{repr.fin}) are not zero. Indeed,
\begin{align*}
d_{1*}= &\big( e_1\otimes e_1\otimes s^+\otimes s^-,
e_1\otimes e_1\otimes e_1\otimes e_1\big)
=c^+_1c^-_1(c_0^+-c_1^+)^{-1}(c_0^--c_1^-)^{-1}\not=0\\
d_{2*}=&\Big( e_2\otimes e_2\otimes e_+\otimes e_-,
D^{(t)}_0(e_2\otimes e_2\otimes e_2\otimes e_2)\Big)\\=&\Big(D_0( e_2\otimes e_2\otimes e_+\otimes e_-),
e_2\otimes e_2\otimes e_2\otimes e_2\Big)
=(c_0^+-c_+)(c_0^--c_-)\not=0,
\end{align*}
 Hence  taking the integral with respect to $z$ (see (\ref{int_z})) and using that for  $u(1)\not=0$
\begin{align*}
-\frac{1}{2\pi i}\oint_{\omega_A}dz z^{n}G_F^3(z)u^2(F_0^2/z)=&F_0^{2(n-2)}\Big(\frac{1}{2}n(n-1)u^2(1)+4nu(1)u'(1)+O(1)\Big)\\
&\sim Cn^2u^2(1)F_0^{2(n-2)}+O(n)
\end{align*}
and the other terms of (\ref{int_z}) are maximum $O(n)$ (recall that $F_1$ and $F_2$ of (\ref{M_0.1}) contain the multiplier $n^{-1}$), we obtain
 \[\mathcal{R}_{nW}^{+-}(E,\varepsilon,\xi)=C n^2u^2(1)d_{1*}d_{2*}\int F_0^{2(n-2)}dUdS+O(n),\quad C_*\not=0.\]
On the other hand, we know that $\mathcal{R}_{nW}^{+-}(E,\varepsilon,0)=1$. Thus we conclude that $u(1)=0$. But using this 
in (\ref{int_z}), we obtain that all
the terms in (\ref{int_z}) are bounded except the one which contain $u_0$. Repeating the above argument we obtain that $u_0(1)=0$.

$\square$

\section{Proof of Lemmas \ref{l:bG} and \ref{l:bGP} }\label{s:bG}
\textit{Proof of Lemmas \ref{l:bG}}. 
To obtain the bound for $\|\mathfrak{G}\|$, we start from  the analysis of
\[\mathcal{M}_1=\mathcal{M}-\mathcal{M}'(\mathcal{M}'''-z)^{-1}\mathcal{M}'',\quad \widehat G_1:=(\mathcal{M}_1-z)^{-1}\]
and prove that
\begin{align}\label{b_G1}
\|\widehat G_1\|\le C/|z-1|,
\end{align}
or, equivalently,
\[
 \|(\widehat G_{1})_{ij}\|\le C/|z-1|.
\]
Since by (\ref{r_tiK}) $\widetilde K_i=WF_0r_1Br_2F_0$, where $r_1$ and $r_2$ are of the type at least 1, we have by  (\ref{b_rr}) 
with $K_\alpha=K_0$
\begin{align}\label{b_tiK}
|(\widetilde K_if,g)|^2\le &C
 W^2(F_0(r_2)^*r_2F_0f,f )(F_0r_1BB^*r_1^*F_0 g,g)\\&\le
C(F_0(1+C/W-K_0)F_0f,f)(F_0(1+C/W-K_0)F_0g,g)
\notag\end{align} 
Let us check that the operators which comes from $\mathcal{M}'(\mathcal{M}'''-z)^{-1}\mathcal{M}''$ satisfy (\ref{b_tiK}).
Since $\widetilde K_i=WF_0r_1Br_2F_0$, we have
\begin{align}\notag
%\label{b_tiK.1}
 &|((\mathcal{M}'(\mathcal{M}'''-z)^{-1}\mathcal{M}'')_{ij}\,f,g)|^2
\le W^2(F_0r_1Br_2F_0G_{\sigma*}F_0r_2^*B^*r_1^*F_0 g,g)\\
&\hskip5cm\cdot W^2(F_0(r_2')^*(B')^*(r_1')^*F_0 G_{\sigma*}F_0(r_1')B'(r_2')F_0f,f )\notag\\
&G_{\sigma*}:=(1+A/n-F_0K_{\sigma}F_0)^{-1},\quad\sigma=1,2
\label{G_a}\end{align}
Assume that $r_2$ contains $U_{12}$ or $U_{21}$. Then   by  (\ref{b_rr}) 
%$(r_2^*r_2)^{(l)U}$ is 
%$O(l^2/W^2)\mathcal{E}_{q,q}$ with some $q\in\mathbb{Z}$, and by (\ref{K_q}) $\big(K_\alpha\big)_U^{(l)}=(1-O(l^2/W)\mathcal{E}_{q',q'}^{(l)}$)
%If $q=q'$, then
\begin{align}\label{r_2r_2.1}
\|Br_2F_0G_{\sigma*}F_0r_2^*B^*\|&\le\|B\|^2\|G_{\sigma*}^{1/2}F_0r_2^*r_2F_0G_{\sigma*}^{1/2}\|\\
&\le CW^{-1}\|G_{\sigma*}^{1/2}F_0(1+C/W-K_{\sigma})F_0G_{\sigma*}^{1/2}\|\notag\\
&\le C W^{-1}\|G_{\sigma*}^{1/2}(1+C/W-F_0K_{\sigma}F_0)G_{\sigma*}^{1/2}\|\le C'W^{-1}.
\notag\end{align}
Here we used that $W\gg n$. 
%And if $q\not=q'$, then, since $F_0r_2^*r_2F_0$ is not $0$ only on the subspace of the functions of the form $\varphi(|U_{11}|^2)U_{11}^{2q}$,
%and $K_\alpha$ is not $0$ only on the subspace of the functions of the form $\varphi(|U_{11}|^2)U_{11}^{2q'}$, in this case
%\begin{align}\label{r_2r_2.2}
%\|G_{\alpha*}^{1/2}F_0r_2^*r_2F_0G_{1*}^{1/2}\|=(1+A/n)^{-1}F_0r_2^*r_2F_0\le CW^{-1}
%\end{align}
Hence
\begin{align*}
& W^{2}(F_0r_1Br_2F_0G_{\sigma*}F_0r_2^*B^*r_1^*F_0 g,g)\le C'W(F_0r_1r^*_1F_0\,g,g)\le C''(F_0(1+C/W-K_{0})F_0\,g,g)
\end{align*}
This bound combined with a similar bound for $(F_0(r_2')^*(B')^*(r_1')^*F_0 G_{\sigma*}F_0(r_1')B'(r_2')F_0f,f )$  yield (\ref{b_tiK}).

It is easy to see that that if some operator $\widehat A$ has the form similar to (\ref{cal_M}), i.e. its diagonal entries are the same (equal to $A$) and
other non-zero entries are only in the first line and the last column,  then the resolvent $\widehat G=(\widehat A-z)^{-1}$ has the same form and
\begin{align}\label{A^-1}
& \widehat G_{12}=-GA_{12}G,\quad \widehat G_{13}=-GA_{13}G,\quad \widehat G_{24}=-GA_{24}G,\quad \widehat G_{34}=-GA_{34}G,\\
&\widehat G_{14}=-GA_{14}G+GA_{12}GA_{24}G+GA_{13}GA_{34}G,\quad where\,\,G:=(A-z)^{-1}=\widehat G_{ii},
\notag\end{align}
Applying (\ref{A^-1}) to   $\mathcal{M}_1$  we conclude that it is sufficient to prove the corresponding bounds
for the operators
\begin{align}\label{L}
&G,\quad G\widetilde K_\alpha G,\quad G\widetilde K_\alpha G\widetilde K_\beta  G,\quad G F_\alpha  G,\quad
 G F_\alpha  G F_\beta G, \quad G\widetilde K_\alpha  G  F_\beta G,\quad G F_\alpha G \widetilde K_\beta  G,\\
&Gu_0\Big(\frac{K_F}{z}\Big) G,\,\,Gu\Big(\frac{K_F}{z}\Big) G,\,\, Gu\Big(\frac{K_F}{z}\Big) Gu\Big(\frac{K_F}{z}\Big)  G, \,\, Gu\Big(\frac{K_F}{z}\Big) G\widetilde K_\alpha G,\,\, Gu\Big(\frac{K_F}{z}\Big) GF_\alpha G,
\notag
\end{align}
with
\begin{align}\label{G_11}
G(z):=(K_F-z)^{-1}.
\end{align}
Observe, that  by (\ref{g=0}) $\|Gu(K_F/z)\|\le C$ and $\|Gu_0(K_F/z)\|\le C$, hence the bounds for the operators of the second line of
(\ref{L}) follow from the that of the first line.

Since the spectrum of  $K_F$ belongs to  $[0,1]$, it is evident that for $z\in\omega_A$
\begin{equation}\label{Gii}
\|G(z)\|\le C/|z-1|.
\end{equation}
To estimate other entries, we
set
\begin{align}\label{G_*}
G_*:=G(z)\Big|_{z=1+A/n},
\end{align}
and prove  the bounds
\begin{align}\label{b_GKG}
&\|G^{1/2}_*\widetilde K_\alpha G_*^{1/2}\| 
\le C,\quad
 \|G_*^{-1/2}G^{1/2}\|\le C,\\
 &\|G^{1/2}_*(\varphi_U/n)K_F\,\,\tilde x(K_F/z)G_*^{1/2}\|\le C\quad
 \|G^{1/2}_*(\varphi_S/n)K_F\,\,\tilde y(K_F/z)G_*^{1/2}\|\le C,
\notag 
\end{align}
where $\tilde x(\zeta)$ and $\tilde y(\zeta)$ are some analytic in $\mathbb{B}_{1+\delta}$ functions. Notice also that $G$ and $G_*$ commute.

It is easy to see  that operators from  (\ref{L})  can be represented as a sum of
 $G^{1/2}\prod G^{1/2}$, where $\prod$ is some product of the operators from  (\ref{b_GKG}) .

The definition of the operator norm, (\ref{b_tiK}), and the bound $F\le 1$ yield
\begin{align}\label{est}
&\|G^{1/2}_*\widetilde K_\alpha G^{1/2}_*\|^2=\sup_{|f|=|g|=1}|(\widetilde K_\alpha G^{1/2}_*f,G^{1/2}_*g)|^2\\
&\le
\sup_{|f|=1}|(F_0(1+\dfrac{C}{W}-K_0)F_0G^{1/2}_*f,G^{1/2}_*f)\sup_{|g|=1}|(F_0(1+\dfrac{C}{W}-K_0)F_0G^{1/2}_*g,G^{1/2}_*g)\notag\\
&=
\|G^{1/2}_*F_0(1+\dfrac{C}{W}-K_0)F_0G^{1/2}_*\|^2\le\|G^{1/2}_*(1+\dfrac{C}{W}-K_F)G^{1/2}_*\|^2\le1.\notag
\end{align}
Moreover, since  $G$ and $G_*$ commute  we have
\begin{align*}
%& \|G^{1/2}(z)|\widetilde K_\alpha|G^{1/2}(\bar z)\|= \|G^{1/2}(z)G^{-1/2}_*G^{1/2}_*F|\widetilde K_\alpha|FG^{1/2}_*G^{-1/2}_*G^{1/2}(\bar z)\|\\
& \|G^{1/2}(z)G^{-1/2}_*\|^2= \|G(z)G^{-1}_*\|\le\max_{|z|=1+A/n,0\le\lambda\le 1}\frac{1+A/n-\lambda}{|z-\lambda|}\le C,
\end{align*}
which gives the second inequality of (\ref{b_GKG}). 

To prove the inequalities of the second line of  (\ref{b_GKG}), 
observe that since $G_*$ and $\tilde x(K_F/z)$ commute, we have
\begin{align*}
&\|G^{1/2}_*(\varphi_U/n)K_F\tilde x(K_F/z)G_*^{1/2}\|\le C\|G^{1/2}_*(\varphi_U/n)K_FG_*^{1/2}\|\\
&\le
C\|G^{1/2}_*(\varphi_U/n)F_0^2G_*^{1/2}\|+C\|G^{1/2}_*(\varphi_U/n)F_0(1-K_0)F_0G_*^{1/2}\|\\
&\le C\|G^{1/2}_*(\varphi_U/n)F_0^2G_*^{1/2}\|+C\|G^{1/2}_*(\varphi_U/n)^2F_0^2G_*^{1/2}\|^{1/2}\|G_*^{1/2}F_0(1-K_0)^2F_0G_*^{1/2}\|^{1/2}
\end{align*}
Moreover, since
\begin{align*}
(\varphi_U/n)F_0^2\le C(1+A/n-F_0^2),
%\quad (\varphi_U/n)^2F_0^2\le C(1+A/n-F_0^2),\\
\quad (1-K_{0})^2\le(1-K_0),
\end{align*}
 using the last line of (\ref{est}) and the bound
\begin{align*}
&\|G_*^{1/2}(\varphi_U/n)F_0^2 G_*^{1/2}\|\le C \|G_*^{1/2}(1+A/n-F_0^2) G_*^{1/2}\|\le C \|G_*^{1/2}(1+A/n-K_F)G_*^{1/2}\|= C
%\\
%&\|G_*^{1/2}(\varphi_U/n)^2F_0^2 G_*^{1/2}\|\le C\\
%&\|G_*^{1/2}F_0(1-K_0)^2F_0 G_*^{1/2}\|\le \|G_*^{1/2}F_0(1-K_0)F_0 G_*^{1/2}\|\le\|G_*^{1/2}(1-K_F)G_*^{1/2}\|\le C
\end{align*}
we obtain the third bound of (\ref{b_GKG}).
%\begin{align*}
%&\|G_*^{1/2}(\varphi_U/n) K_{F}G_*^{1/2}\|
%\le\|G_*^{1/2}(\varphi_U/n)F_0^2 G_*^{1/2}\|+\|G_*^{1/2}(\varphi_U/n) F_0(1-K_{0})F_0G_*^{1/2}\|\\
%&\le C+\|G_*^{1/2}(\varphi_U/n)^2 F_0^2G_*^{1/2}\|^{1/2}\|G_*^{1/2} F_0(1-K_{0})^2F_0G_*^{1/2}\|\le C
%\end{align*}
The last bound (\ref{b_GKG}) one can prove similarly. Thus, we get (\ref{b_G1}).

Now let us prove (\ref{b_1})  for $\|\widehat G_1\mathcal{M}'(\mathcal{M}'''-z)^{-1}\|$.
%\begin{align}\label{G_21}
%\|\hat G_1\mathcal{M}'(\mathcal{M}'''-z)^{-1}\|\le C |z-1|^{-1}
%\end{align}
The structure of $\widehat G_1$ (see (\ref{A^-1}), (\ref{L})), the bounds (\ref{b_GKG}) combined with the structure  of $\mathcal{M}''$ 
 and $\mathcal{M}'''$ (see (\ref{r_tiK})) imply that it suffices find  the bound for
\begin{align*}
W\|G_{*}^{1/2}F_0r_1Br_2F_0G_{*\sigma}^{1/2}\|\le  \|B\|\|G_{*}^{1/2}F_0r_1\|\|r_2F_0G_{*\sigma}^{1/2}\|
\end{align*}
with $r_{1},r_{2}$ of  non-zero type (see  (\ref{G_*}) and (\ref{G_a}) for the definition of $G_{*}$ and $G_{*\sigma}$). But  in view of  (\ref{r_tiK})
\begin{align*}
\|G_{*}^{1/2}F_0r_1\|^2\le \|G_{*}^{1/2}F_0r_1r_1^*F_0G_{*}^{1/2}\|\le 
CW^{-1}\|G_{*}^{1/2}F_0(1+C/W-K_0)F_0G_{*}^{1/2}\|\le CW^{-1}
\end{align*}
The same bound for $\|r_2F_0G_{\alpha*}^{1/2}\|^2$ was obtained in (\ref{r_2r_2.1}).
%\begin{align*}
%\|r_2F_0G_{\alpha*}^{1/2}\|^2=\|G_{\alpha*}^{1/2}F_0r_2^*r_2F_0G_{\alpha*}^{1/2}\|\le C/W
%\end{align*}
Hence $\|\widehat G_1\mathcal{M}'(\mathcal{M}'''-z)^{-1}\|$ satisfies (\ref{b_1}). 

The bounds for $\|(\mathcal{M}'''-z)^{-1}\mathcal{M}''\widehat G_1\|$ and
$\|(\mathcal{M}'''-z)^{-1}\mathcal{M}''\widehat G_1\mathcal{M}'(\mathcal{M}'''-z)^{-1}\|$ can be obtained similarly. 
%Moreover, repeating the above argument for $(\mathcal{M}'''-z)^{-1}\mathcal{M}''\hat G_1$, we conclude that
%\begin{align*}
%&\|(\mathcal{M}'''-z)^{-1}\mathcal{M}''\hat G(\mathcal{M}''(\mathcal{M}'''-z)^{-1}\|\\&\le 
%C\|(\mathcal{M}'''-z)^{-1}\| \,\|G_{1*}^{1/2}F_0r_2Br_1F_0G_*^{1/2}\|\,\|G_*^{1/2}F_0r_1'F_0r_2'G_{1*}\|\le C\|G_{1*}\|\le C|z-1|^{-1}.
%\end{align*}
Then, using Proposition \ref{p:res} we obtain that the block matrix $\mathfrak{M}$ satisfies the bound (\ref{b_1})
%\[\|(\mathfrak{M}-z)^{-1}\|\le C/|z-1|\]

Since by (\ref{P_0GP_0}) $\mathfrak{M}'=\mathfrak{M}+O(W^{-1})$, we have
\[(\mathfrak{M}'-z)^{-1}=(1+(\mathfrak{M}-z)^{-1}O(W^{-1}))^{-1}(\mathfrak{M}-z)^{-1}=(\mathfrak{M}-z)^{-1}\big(1+O(n/W)\big).\]

$\square$

\textit{Proof of Lemma \ref{l:bGP}}

We start from the proof of (\ref{PM}) with an operators of a bit more general form  than (\ref{r_tiK}), since these operators appear in (\ref{repr.fin}).  Consider  
\[\widetilde K=W\widetilde K',\quad
\widetilde K'= F_0\big(q_1\big)_U\big(p_1\big)_SF_0\varphi_1\big(F_0K_{\bar m'}F_0\big)F_0\big(q_2\big)_U\big(p_2\big)_SF_0
\varphi_2\big(K_F)F_0,
\]
where
$p_{1,2}$ are of the type $1$ and $q_{1,2}$ are of the type 0,
$\varphi_{1,2}(\zeta)$ are  analytic in $\mathbb{B}_{1+\delta}$ function and  $K_{\bar m}$ is some operator of  the type $0$, which acts in $\mathcal{E}_{\bar m}(\mathcal{L}_2)$.
Here and below $\mathcal{E}_{\bar m}$ is an orthogonal projection on the space of the functions
\[
\phi(|S_{12}|^2,|U_{12}|^2)S_{11}^{m_1-k_1}S_{12}^{k_1}U_{11}^{m_2-k_2}U_{12}^{k_2},\quad \bar m=(m_1,m_2),\quad k_1,k_2\in\mathbb{Z}
\]
and if some of the above exponents is negative, then we replace the respective matrix entry with its conjugate. 
 Remark that the operators
$\big(q_1\big)_U\big(p_1\big)_S,K_{\bar m},\big(q_2\big)_U\big(p_2\big)_S, K_0$ should  feet to each other (otherwise their product is zero)
hence one could insert   $\mathcal{E}_{\bar m}$ after $\varphi_1$ and $\mathcal{E}_{0}$ after $\varphi_2$.
%
%\begin{align}\notag
%&\|\widetilde K'\mathcal{P}_L\|^2=\|\mathcal{P}_L(\widetilde K')*\widetilde K'\mathcal{P}_L\|
%\le CW^{-1}\|\mathcal{P}_L\mathcal{E}_0\overline{\varphi_2}\big(K_F)F_0\big(p_2\big)_S^*
%\big(p_2\big)_SF_0\varphi_2\big(K_F)\mathcal{E}_0\mathcal{P}_L\|\\
%&=CW^{-1}\|\big(p_2\big)_SF_0\varphi_2\big(K_F)\mathcal{E}_0\mathcal{P}_L\|=CW^{-1}\|\widetilde K''\|^2
%\label{K''}\end{align}
%where we used that the operator  at the left of $\big(p_2\big)_S$, written  as $B_1\big(p_1\big)_SB_2$ with bounded $B_1,B_2$, satisfies the bound 
%\begin{align*}
%\|B_2^*\big(p_1\big)_S^*B_1^*B_1\big(p_1\big)_SB_2\|\le C\|B_2^*\big(p_1\big)_S^*\big(p_1\big)_SB_2\|\le CW^{-1}\|B_2^*B_2\|\le C'W^{-1}
%\end{align*}

 Notice also that the cases when $p_1$ and $p_2$ both depend on   $U$ (or one of them depends on $U$, and another one -- on $S$), as well 
 the cases when the joint type of $\widetilde K'$ is more than 2,
are also possible, but their analysis is similar. For $W^2O_*(\mathfrak{r}^{(4)})$ the proof is similar also.

Assume that we proved the following relations
\begin{align}
&F_0\big(\varphi\big(F_0K_{\bar m}F_0\big)-\varphi(F_0^2)\big)F_0\mathcal{E}_{\bar m}\mathcal{P}_L=O(L^2W^{-1}), \quad
\|(K_{\bar m}-1)\mathcal{E}_{\bar m}\mathcal{P}_L\|\le CL^2/W\label{comm}\\
&[p_2,\varphi (F_0)]=\varphi'(F_0)\mathcal{D}_S+O(W^{-3/2}),\quad \mathcal{D}_S=\Phi_1(S_1)\big(p_2p_*\big)_S+
\bar\Phi_1(S_1)\big(p_2\bar p_*\big)_S\label{comm'}\\
&\|[\big(p^{(s)}\big)_S,\varphi(F_0)]\|\le CW^{-(1+s)/2},\quad \|[\big(p^{(s)}\big)_S,\Phi_1]\|\le CW^{-(s+1)/2},\label{comm0}\\
&\|[\varphi_1(F_0K_mF_0),\varphi(F_0)]\|\le C/\sqrt{W },\quad\|[\varphi_1(F_0K_mF_0),\Phi_1]\|\le C/\sqrt{W }
% \|(K_m-1)\mathcal{E}_m\|\,
 % \|\big(\varphi\big(F_0K_{\bar m}F_0\big)-\varphi\big(F_0^2\big)\mathcal{E}_m\|\le C/\sqrt{W},
  \label{comm1} \\
& \|\big(p_{1,2}\big)_S\|\le C/\sqrt{W},\quad
\|\big(p_{1,2}\big)_S\mathcal{P}_L\| \le CL/W,\quad\|\mathcal{D}_S\|\le CW^{-1},
\label{comm2} \end{align}
where $\varphi$ is any analytic in $\mathbb{B}_{1+\delta}$ function, $p^{(s)}$ is any product of the type $s$,
$\Phi_1$ is the  operator of multiplication by $\Phi_1(S)=cn^{-1}S_{11}\bar S_{12}F_{0}$ ($c=\alpha_2(a_+-a_-)$ see (\ref{F_a,b})) and $p_*(S)=\bar S_{11}S_{12}$.

Then, using the first inequality of (\ref{comm}), the first bound of (\ref{comm0}), and the first bound of (\ref{comm2}) we get
\begin{align*}
\widetilde K'\mathcal{P}_L= B_1\big(p_1\big)_SF_0\varphi_1\big(F_0K_{\bar m}F_0\big)F_0\big(q_2\big)_U\big(p_2\big)_SF_0^2\varphi_2(F_0^2)\mathcal{P}_L
+O(L^2W^{-2})
\end{align*}
where $B_1$ is a bounded operator to the left of $\big(p_1(S)\big)_S$ in $\widetilde K'$.
Then   (\ref{comm'}) with $\widetilde\varphi_2(\zeta)=\zeta^2\varphi_2(\zeta^2)$  combined with the first bound of (\ref{comm2}),
(\ref{comm}) for $\big(q_2\big)_U$, and (\ref{comm2}) yield
\begin{align}\notag
\widetilde K'\mathcal{P}_L=&B_1\big(p_1\big)_SF_0\varphi_1\big(F_0K_{\bar{m}}F_0\big)F_0\big(q_2\big)_U
\Big(\widetilde\varphi_2(F_0)\big(p_2\big)_S+\widetilde\varphi'_2(F_0)\mathcal{D}_S\Big)\mathcal{P}_L
+O(L^2W^{-2})\\
=&B_1\big(p_1\big)_SF_0\varphi_1\big(F_0K_{\bar{m}}F_0\big)F_0
\Big(\widetilde\varphi_2(F_0)\big(p_2\big)_S+\widetilde\varphi'_2(F_0)\mathcal{D}_S\Big)\big(q_2\big)_U\mathcal{P}_L
+O(L^2W^{-2}).
\label{comm3}\end{align}
Using consequently the first bound of (\ref{comm1})  for $\varphi(\zeta)=\widetilde\varphi_2(\zeta)$ combined with the first bound of (\ref{comm2}) and the fact
that $\big(p_2\big)_S\big(q_2\big)_U$ commute with $\mathcal{P}_L$, then the first relation of (\ref{comm}), then the first bound of (\ref{comm0}) with 
$\varphi(\zeta)=\zeta^2\varphi_1(\zeta^2)\widetilde\varphi_2(\zeta)$ and finally the second bound of (\ref{comm2}),
 we obtain for the term in the r.h.s. of (\ref{comm3}) which contains $\big(p_2\big)_S$:
\begin{align*}
B_1& F_0\big(p_1\big)_SF_0\varphi_1\big(F_0K_{\bar{m}}F_0\big)F_0
\widetilde\varphi_2(F_0)\big(p_2\big)_S\big(q_2\big)_U\mathcal{P}_L\\=&B_1 F_0\big(p_1\big)_S
\widetilde\varphi_2(F_0)F_0\varphi_1\big(F_0K_mF_0\big)F_0\mathcal{P}_L\big(q_2\big)_U\big(p_2\big)_S\mathcal{P}_L
%+\widetilde\varphi'_2(F_0)\mathcal{D}_S\Big)\big(q_2\big)_U\mathcal{P}_L
+O(L^2W^{-2})\\
=&B_1 F_0\big(p_1\big)_SF_0^2
\varphi_1\big(F_0^2\big)\widetilde\varphi_2(F_0)\mathcal{P}_L\big(q_2\big)_U\big(p_2\big)_S\mathcal{P}_L+O(L^2W^{-2})\\
=&B_1 F_0^3
\varphi_1\big(F_0^2\big)\widetilde\varphi_2(F_0)\big(p_1\big)_S\big(q_2\big)_U\big(p_2\big)_S\mathcal{P}_L+O(L^2W^{-2})
=O(L^2W^{-2}).
\end{align*}
Similarly for the first term of $\mathcal{D}_S$ (see (\ref{comm1})) we get
\begin{align*}
B_1& \big(p_1\big)_SF_0\varphi_1\big(F_0K_{\bar{m}}F_0\big)F_0\widetilde\varphi'_2(F_0)\Phi_1(p_2p_*)_S\big(q_2\big)_U\mathcal{P}_L\\
&=B_1F_0^2\widetilde\varphi'_2(F_0)\Phi_1\varphi_1\big(F_0^2\big)\big(p_1\big)_S\big(p_1\big)_S(p_2p_*)_S\big(q_2\big)_U\mathcal{P}_L
%&\widetilde K'\mathcal{P}_L=B_1\varphi_1\big(F_0^2\big)F_0^2\big(p_1\big)_S
%\Big(\widetilde\varphi_2(F_0)\big(p_2\big)_S+\widetilde\varphi'_2(F_0)\mathcal{D}_S\Big)
+O(L^2W^{-2})=O(L^2W^{-2})
%&= B_2\Big(\widetilde\varphi_2(F_0)\big(p_1\big)_S\big(p_2\big)_S+\widetilde\varphi'_2(F_0)\Big(\Phi_1\big(p_1\big)_S\big(p_2p_*\big)_S+
%\bar\Phi_1\big(p_1\big)_S\big(p_2\bar p_*\big)_S\Big)\Big)\mathcal{P}_L
%+O(L^2W^{-2})\\
%=O(L^2W^{-2})
\end{align*}
%where $B_2$ is a bounded operators at the left of $\big(p_1\big)_S$ at the first line.
The second term of $\mathcal{D}_S$  can be analysed similarly.

Thus we are left to prove  relations (\ref{comm})-(\ref{comm2}).
Remark that (\ref{comm2}) and the first bound of (\ref{comm1}) are  direct corollaries of Lemma \ref{l:U,V}, where we need only to take into account
that $\big(p_2p_*\big)_S$ is also reduced by $L^{(l)S}$ and it is of the second type, hence its norm is bounded by $O(W^{-1})$. 
%The second relation 
%of (\ref{comm1}), if we write $\big(\varphi\big(F_0K_{\bar m}F_0\big)-\varphi\big(F_0^2\big)$ as a difference of the Taylor series and use
%the first relation of (\ref{comm1}) (recall that $F_0$ and $K_{\bar m}$ commute with $\mathcal{E}_m$).

To prove  the first  relation of (\ref{comm}),   we use first the Cauchy formula and (\ref{res_id})
\begin{align}\notag
&\varphi\big(F_0K_{0}F_0\big)-\varphi(F_0^2)=\frac{1}{2\pi i}\oint_{|z|=1+\delta}\varphi(z)G_1(z)F_0(K_0-1)F_0G_0(z)dz\\&=
\dfrac{1}{2\pi i}\oint_{|z|=1+\delta}\varphi(z)G_1(z)F_0\Big(F_0G_0(z)(K_0-1)+[K_0,F_0G_0(z)]\Big)\\
&G_1=(F_0K_{0}F_0-z)^{-1},\quad G_0=(F_0^2-z)^{-1},
\label{comm4}\end{align}
%Here $O((K_{\bar m}-1)^2)$ denotes a sum of  terms  
%\[(F_0K_{\bar m}F_0)^{s_1}F_0(K_{\bar m}-1)F_0(F_0K_{\bar m}F_0)^{s_2}F_0(K_{\bar m}-1)F_0F_0^{2s_3}.
%\]
%Since $F_0$ and $K_{\bar m}$
%commute with $\mathcal{E}_{\bar m}$ these terms after multiplication by $\mathcal{E}_{\bar m}$ will give us $O(W^{-1})$ terms.
%Here  we consider $K_0$ instead of $K_m$, since it makes formulas shorter, but for $K_m$ the proof is the same.
To estimate the above commutator, we are going to expand
\[
F_0G_0(z)=-z^{-1}\sum\limits_{s=0}^\infty z^{-s}F_0^{2s+1}.
\]
Notice that for any $p>0$ $[K_{m},F_0^{p}]$ is an integral operator with a kernel\\
 $(F^p_0(S_1)-F^p_0(S_2))\widetilde p(\widetilde S)K_{0S}(\widetilde S)K_{0U}$,
where $\widetilde S=S_1S_2^{-1}$ and $K_m\sim (\widetilde p)_S$.  Now use the formula
\begin{align}\label{exp_diff}
e^{-x}-e^{-y}=-(x-y)e^{-x}-(x-y)^2\int_0^1te^{-tx-(1-t)y}dt 
\end{align}
for $x=c(2s+1)n^{-1}|(S_1)_{12}|^2$ and $y=c(2s+1)n^{-1}|(S_2)_{12}|^2$ (recall that $c=\alpha_2(a_+-a_-)$ by (\ref{F_a,b})).  Then
\begin{align}\notag
y-x=&(2s+1)cn^{-1}\Big(|\widetilde S_{12}|^2(2|(S_1)_{12}|^2+1)+
\big(\widetilde S^{-1}_{11}\overline{\widetilde S^{-1}}_{12}(S_1)_{12}(\overline{ S_1})_{22}+
\overline{\widetilde S^{-1}_{11}\overline{\widetilde S^{-1}}_{12}(S_1)_{12}(\overline{ S_1})_{22})}\Big)\\
=&(2s+1)cn^{-1}\Big(\bar p_*(S_1)p_*(\widetilde S)+p_*(S_1)\bar p_*(\widetilde S)\Big)
+O((2s+1)n^{-1}|\widetilde S_{12}|^2).
\label{exp_diff1}\end{align}
Multiplying this relation by $\widetilde p(\widetilde S)K_0(\widetilde U,\widetilde S)z^{-s-1}$, summing with respect to $s$, and using the second bound of (\ref{comm2}) for $(\widetilde pp_*)_S$, we obtain
the first relation of (\ref{comm}). The second relation of (\ref{comm}), as well as (\ref{comm2}), are the direct corollaries of Lemma \ref{l:U,V}.

The proof of (\ref{comm'}) is very similar. We expand $\varphi(F_0)$ into the Taylor series  $\sum \varphi_sF_0^s$. The commutator
$[\big(p_2\big)_S,F_0^s]$ is an integral operator with a kernel $(F^s_0(S_1)-F^s_0(S_2))K_{0}(\widetilde S,\widetilde U)p_2(\widetilde S)$. Hence, using
(\ref{exp_diff}) and (\ref{exp_diff1}) like before, we obtain after multiplication by $K_{0S}(\widetilde S)p_2(\widetilde S)$ that the remainder term
will give us $O(W^{-3/2})$ by (\ref{comm2}). Then multiplying the relation by $\varphi_s$ and summing with respect to $s$ we get (\ref{comm'}).
The proof of (\ref{comm0}) is very similar (and even simpler).

For the proof of (\ref{comm1}) we  use again the Cauchi formula and write
\begin{align*}
&[\varphi\big(F_0K_{0}F_0\big),\varphi(F_0)]=\dfrac{1}{(2\pi )^2}\oint_{|z|=1+\delta}\varphi(z_1)\varphi(z_2)[G_1(z_1),G_{01}(z_2)]dz_1dz_2\\
&\|[G_1(z_1),G_{01}(z_2)]\|=\|G_1(z_1)G_{01}(z_2)[F_0K_0F_0,F_0]G_{01}(z_2)G_1(z_1)\|\le C/\sqrt{W}
\end{align*}
where $G_1$, $G_{0}$ are defined in (\ref{comm4}) and $G_{01}$ can be obtained from $G_0$, if we replace $F_0^2$ by $F_0$. The last relation here
follows from the fist bound of (\ref{comm0}) with $s=0$ and bounds for $\|G_1\|$ and
$\|G_{01}\|$. The second bond of (\ref{comm1}) can be obtained similarly.

\medskip
To prove the bound for $(1-\mathcal{P}_{L})\widehat{G}_0\widehat f$ with $\widehat{G}_0$ of (\ref{M_0}),  it  suffices to prove similar bounds for $(1-\mathcal{P}_{LS})\widehat{G}_0\widehat f$
and $(1-\mathcal{P}_{LU})\widehat{G}_0\widehat f$. From the structure of $\widehat{G}_0$ (see (\ref{A^-1})) we conclude that we need to prove corresponding bounds
for
\begin{align}\label{f-Pf.S}
&(1-\mathcal{P}_{LS}){G}_0 f,\quad (1-\mathcal{P}_{LS}){G}_0 F_\alpha G_0 f,\quad (1-\mathcal{P}_{LS}){G}_0 F_\alpha G_0F_\beta G_0 f,\\
&f=e^{-(c_1|U_{12}|^2+c_2|S_{1,2}|^2)/n},\quad G_0=(e^{-(c_1|U_{12}|^2+c_2|S_{1,2}|^2)/n}-z)^{-1}
\notag\end{align}
with some $c_{1,2}$. By \cite{Vil:68} we have for any $\varphi(2|S_{12}|^2)$
\begin{align*}
\|(1-\mathcal{P}_{LS})\varphi\|^2_{S}=&\int_{\rho>L}|a(\rho)|^2\rho\tanh(\pi \rho)d\rho,\quad
a(\rho)=\int_{0}^\infty\varphi(x)\mathfrak{P}^{-1/2-i\rho}(x-1)dx\\
\|\Delta_{S}^m\varphi\|^2=&\int_{0}^\infty\Big(\rho^2+\frac{1}{4}\Big)^{2m}|a(\rho)|^2\rho\tanh(\pi \rho)d\rho,
\end{align*}
where $\mathfrak{P}^{l}_{00}$ is the Legendre function (see the proof of Lemma \ref{l:U,V} for the definition). Hence,
\begin{align*}
\|(1-\mathcal{P}_{LS})\varphi\|^2\le \Big(L^2+\frac{1}{4}\Big)^{-2m}
\|\Delta_{S}^m\varphi\|^2=\Big(L^2+\frac{1}{4}\Big)^{-2m}(\Delta_{S}^{2m}\varphi,\varphi)
\end{align*}
\begin{proposition}\label{p:Delta} For any smooth enough function $\varphi$
\[|\Delta_{S}^m\varphi(x)|\le  2^{2m}(m!)^2\sum_{k=0}^{2m}(x+1)^k|\varphi^{(k)}(x)|\]
\end{proposition}
 Proposition \ref{p:Delta} implies
\begin{align*}
(\Delta_{S}^{2m}&G_0f,G_0f)\le 2^{4m}((2m)!)^2\sum_{k=0}^{2m} I_k,\quad\\
 I_k:=&\int_0^\infty dx(x+1)^k\Big|\frac{\partial^k}{\partial x^k} 
\frac{e^{-c_1x/n}}{e^{-c_1x/n}-z}\Big| \Big| \frac{e^{-c_1x/n}}{e^{-c_1x/n}-z}\Big|,
\end{align*}
Expanding $(e^{-c_1x/n}-z)^{-1}$ into the series with respect to $e^{-jc_1x/n}$ and taking the derivative we get
\begin{align*}
I_k&\le Cn^2\sum_{j=2}^\infty |z|^{-j}j^{k}\int_0^\infty( \tilde x+c_1/n)^ke^{-j\tilde x}d\tilde x \\
&\le Cn^2\sum_{j=2}^\infty |z|^{-j}j^{k}\int_0^\infty \tilde x^ke^{-j(\tilde x-c_1/n)}d\tilde x
%\\
%&
\le Cn^2k!\sum_{j=2}^\infty j^{-1}|e^{c_1/n}/z|^{j} \le C(2m)!n^2\log n
\end{align*}
Here we used the change of variable $\tilde x=c_1x/n$ and take $|z|=1+A/n$ with sufficiently big $A$.  Thus,
\begin{align*}
&(\Delta_{S}^{2m}G_0f,G_0f)\le n^2\log n \cdot 2m ((2m)!)^32^{4m}(L^2+1/4)^{-2m}\\
&\Rightarrow \|(1-\mathcal{P}_{LS}){G}_0 f\|\le
2n^2 (\log n) m \exp\{4m\log 2+4m(\log(2m)^{3/2}-\log L-1)\}\le e^{-c\log^{4/3}n}
\end{align*}
if we take $m=\tilde cL^{2/3}=\tilde c\log^{4/3}n$ with sufficiently small $\tilde c$. The bounds for $(1-\mathcal{P}_{LS})f$, other functions from (\ref{f-Pf.S}) 
and similar  bounds with $(1-\mathcal{P}_{LU})f$ can be obtained similarly. For the last bound in (\ref{f-Pf}) we need to estimate 
$\|G_0f\|,\,\|G_0^2F_\alpha f\|, \,\|G_0^3F_\alpha F_\beta  f\|$. Using the change of variables $x=n\tilde x$ we get, e.g., for $\|G_0^3F_\alpha F_\beta  f\|$:
\begin{align*}
\|G_0^3F_\alpha F_\beta  f\|\le n\Big(\int_0^1+\int_{1}^\infty\Big)\frac{(\tilde x+C/n)^4e^{-c\tilde x}}{|z-e^{-c\tilde x}|^6} d\tilde x\le 
Cn\int_0^1 d\tilde x\frac{e^{-c\tilde x}}{|z-e^{-c\tilde x}|^2}+Cn \le \frac{nC}{|z-1|}
\end{align*}
The bounds (\ref{b_ti_f}) follow from the bound for $ \|\widetilde K\mathcal{P}_L \|$ obtained above and  (\ref{f-Pf}), since
\begin{align*}
\|\widetilde K F_0\|\le \|\widetilde K\mathcal{P}_L F_0\|+\|\widetilde K(1-\mathcal{P}_L )F_0\|\le  C(L^2/W)\|F_0\|+O(e^{-c\log^{4/3}n})
\end{align*}
$\square$
\section{Proof of Theorem \ref{t:2}}
Analysis of $\mathcal{K}^+$ of (\ref{K+_X,Y}) is much more simple then of $\mathcal{K}$, and can be done by similar argument. 

Define
\begin{align*}
\widetilde \Omega_{\pm}^+=&\{a_1, a_1', b_1, b_1',b_2,b_2'\in\Omega_+, a_2,a_2'\in\Omega_-\}, \\
\widetilde \Omega_{+}^+=&\{a_1,a_1', a_2, a_2', b_1,b_1', b_2,b_2'\in\Omega_+\},\\
\widetilde \Omega_{-}^+=&\{a_1, a_1', a_2,a_2'\in\Omega_- ,\,b_1, b_1', b_2, b_2'\in\Omega_+\}
\end{align*}
Repeating the argument of Section \ref{s:K} one can obtain
\begin{align}\label{repr2_+}
\mathcal{R}_{Wn}^{++}(E,\varepsilon,\bar\xi)=&-\frac{W^4}{2\pi i}\oint_{\omega_A} z^{n-1}(\mathcal{G}^+_{\pm}(z)f_{\pm},g_\pm)dz-\frac{W^4}{2\pi i}\oint_{\omega_A}  z^{n-1}(\mathcal{G}^+_{+}(z)f_{+},g_+)dz
\\&-\frac{W^4}{2\pi i}\oint_{\omega_A}  z^{n-1}(\mathcal{G}^+_{-}(z)f_{-},g_-)dz+o(1)
=I^+_{\pm}+I^+_{+}+I^+_{-}+o(1),\notag
\end{align}
where 
\begin{align*}
&\mathcal{K}_\pm^+=\mathbf{1}_{\Omega_{\pm}^+}\mathcal{K}^+\,\mathbf{1}_{\Omega_{\pm}^+},\quad \mathcal{K}_{+}^+=\mathbf{1}_{\widetilde\Omega_{+}^+}\mathcal{K}^+\,\mathbf{1}_{\widetilde\Omega_{+}^+},\quad
\mathcal{K}_-^+=\mathbf{1}_{\widetilde\Omega_{-}^+}\mathcal{K}^+\,\mathbf{1}_{\widetilde\Omega_{-}^+},\\
&\mathcal{G}^+_{\pm}=(\mathcal{K}_{\pm}^+-z)^{-1},\quad 
 \mathcal{G}_{+}(z)=(\mathcal{K}_{+}^+-z)^{-1},\quad
\mathcal{G}_{-}(z)=(\mathcal{K}_{-}^+-z)^{-1} 
\end{align*}
and $f^+_{\pm},f^+_+,f^+_-$ and $g^+_{\pm},g^+_+,g^+_-$ are projections of $f^+$ and $g^+$ onto the subspaces corresponding to
$\mathcal{K}_{\pm}^+,\,\mathcal{K}^+_{+},\,\mathcal{K}^+_{-}$. 

We will prove below that this time the main contribution to (\ref{repr2_+}) is given by $I_+^+$.

\subsection{Analysis of $I^+_+$}
Similarly to Section \ref{s:I_+}, in this case it is more convenient to consider $X$ like 
unitary matrix which is close to the $a_+I_2$, and $Y$ as a Hermitian matrix which is close to $a_+I_2$. Then  $X_i$ and $Y_i$ will be parametrized as (cf  (\ref{ch_ab}))
\begin{align}\notag
&X_{11}=a_+(1+i\theta_+\widetilde a_1/\sqrt{W}),\,X_{22}=a_+(1+i\theta_+\widetilde a_2/\sqrt{W}),\, X_{12}=\dfrac{ia_+\theta_+(x+iy)}{\sqrt{2W}}=-\overline{X_{12}},\\ \label{ch_xy}
&Y_{11}=a_+(1+\theta_+\widetilde b_1/\sqrt{W}),\,Y_{22}=a_+(1+\theta_+\widetilde b_1/\sqrt{W}),\,Y_{12}=\dfrac{a_+\theta_+(p+iq)}{\sqrt{2W}}=\overline{Y_{12}}.
\end{align}
This change transforms the measure $dX_i dY_i/(-\pi^2)$ from (\ref{int_rep_+}) to
\[
(a_+)^8(\theta_+)^8(2\pi)^{-4}d \bar a_i \,d \bar b_i  \,d \bar x_i \, d\bar y_i\,d\bar p_i d\bar q_i,
\]
and it "kills" the factor $W^4$ in front of $I_+^+$ in (\ref{repr2_+}). Set also
\begin{align*}
\hat F_+^+(X,Y)= \exp\Big\{\dfrac{i}{2n}\Tr Y(\varepsilon I+\hat\xi'/\rho(E)) - \dfrac{i}{2n}\Tr X(\varepsilon I+\hat\xi/\rho(E))\Big\}.
\end{align*}
It is easy to see also that with this parametrization the operator $\mathcal{K}_+^+$ has the form
\begin{align*}
\mathcal{K}_+^+(X,Y; X',Y') =&
 (-\lambda^+_0)^{-4}\hat F_+^+(X,Y)A_{ab}^+(\bar a,\bar a',x,x',y,y';\bar b,\bar b',p,p',q,q') \\
 &\times\hat F_+^+(X',Y')\widehat Q(\rho,\tau;\rho',\tau'; X',Y') (1+o(1))
\end{align*}
where
\begin{align*}
&A_{ab}^+(\bar a,\bar a',x,x',y,y';\bar b,\bar b')\\
&\qquad\qquad =A_{a}^+(\widetilde a_1,\widetilde a_1')A_{a}^+(\widetilde a_2,\widetilde a_2')
A_{b}^+(\widetilde b_1,\widetilde b_1')A_{b}^+(\widetilde b_2,\widetilde b_2')A_x^+(x,x')A_x^+(y,y')A_x^+(p,p')A_x^+(q,q')
\end{align*}
and $A_x$ is defined in (\ref{A_x}).

The main order of $A_x^+$ has  the form
$A^+_*$ (see(\ref{K_a*})), hence the largest eigenvalue of $A_{ab}^+$ is $1+O(W^{-1/2})$, and the next eigenvalue is smaller then $1-\delta$.

Moreover, if we consider $\widehat Q(\rho,\tau;\rho',\tau'; X',Y')$ in this parametrization, we get
\begin{align}\notag
  (-\lambda^+_0)^{-4}\widehat Q(\rho,\tau;\rho',\tau')=& (-\lambda^+_0)^{-4}\Big(\prod_{\mu,\nu=1,2}e^{\beta(\rho_{\mu\nu}-\rho'_{\mu\nu})(\tau_{\nu\mu}-\tau'_{\nu\mu})}
  e^{-c_+\rho'_{\mu\nu}\tau'_{\nu\mu}}\Big) (1+O(W^{-1/2}))\\
  =&\big(\Pi_{0,+}\oplus R_{0,+}\big)(1+O(W^{-1/2}))
  \label{hat_Q^+} 
   \end{align}
 where 
 \begin{align}\label{Q^+(x)}
& \Pi_{0,+}:=Q_+\otimes Q_+\otimes Q_+\otimes Q_+
\end{align}
with $Q_+$ of (\ref{Q_pm}) corresponds to "good" vectors, and $R_+$ corresponds to all other vectors. Similarly to Lemma \ref{l:R} one can prove
\begin{lemma}\label{l:R_+}
 Given $R_{0,+}$  of (\ref{hat_Q^+}), we have
 \begin{align*}
\lambda_0(  R_{0,+})<1-\delta.
 \end{align*}
\end{lemma}
\textit{Proof.}
Indeed, as was mentioned in the proof of Lemma \ref{l:R}, any "non-good" product can be represented as an "absolutely non-good" part $\tilde\eta$ and  a "good" part $p(n_{11},n_{22},n_{12},n_{21})$.  Here "absolutely non-good" are  the products which do not contain any $n_{\alpha,\beta}$. It is easy to see that  the exponential part of $\widehat Q$ (see (\ref{hat_Q^+}) ) transforms
$\tilde \eta\to \beta^m\tilde \eta$, where $m\ge 2$ is the degree $\tilde\eta$.  But $\beta/\lambda_0^+<1-\delta$, thus the eigenvalue is smaller than $1-\delta$.
$\square$

Now Lemma \ref{l:R_+} and the consideration above yield that first eigenvalue $\lambda_0(\mathcal{K}_+^+)$ of $\mathcal{K}_+^+$ is $1+O(W^{-1/2})$, and the next eigenvalue
is smaller than $1-\delta$. Since $\mathcal{K}_+^+$ is a compact operator, according to the spectral theorem we can rewrite its resolvent $\mathcal{G}^+_{+}(z)$ as
\begin{align}\label{G_++}
\mathcal{G}^+_{+}(z)=\dfrac{P_\mu}{\lambda_0(\mathcal{K}_+^+)-z}+R(z),\quad \|R(z)\|\le C_\delta,
\end{align}
where $R(z)$ is analytic operator-functions in $\{z: 1-\delta/2<|z|<1+\delta \}$,  and $P_\mu$ is a rank one operator of the form $P_\mu= \mu\otimes\mu^*$ with
vectors $\mu$, $\mu^*$ such that
\begin{align}\label{mu}
\mathcal{K}_+^+\mu=\lambda_0(\mathcal{K}_+^+)\mu,\quad (\mathcal{K}_+^+)^*\mu^*=\overline {\lambda_0(\mathcal{K}_+^+)}\mu^*.
\end{align}
Thus if we change the contour $\omega_A$ to $L_1\cup L_0$
\begin{equation}\label{cont}
L_1=\{z:|z|\le 1-\delta/2\},\quad L_0=\{z:|z-\lambda_0(\mathcal{K}_+^+)|\le \delta/4\},
\end{equation}
we get 
\begin{align*}
I_\pm&=-\dfrac{1}{2\pi i}\Big(\oint_{L_0}+\oint_{L_1}\Big)z^{n-1}\Big(\mathcal{G}^+_{+}(z) f_+, g_+\Big) dz= 
-\dfrac{1}{2\pi i}\oint_{L_0} z^{n-1}\Big(\mathcal{G}^+_{+}(z) f_+, g_+\Big) dz+O(e^{-cn})\\
&=
-\dfrac{1}{2\pi i}\oint_{L_0}z^{n-1}\Big(\Big(\dfrac{P_\mu}{\lambda_0(\mathcal{K}_+^+)-z}\Big) f_+, g_+\Big) dz=(\lambda_0(\mathcal{K}_+^+))^{n-1}(P_\mu f_+,g_+)+O(e^{-cn}).
\end{align*}
Since we will prove below that $I^+_-$ and $I^+_\pm$ are of order $O(e^{-cn})$, we have from (\ref{repr2_+})
\begin{align*}
\mathcal{R}_{Wn}^{++}(E,\varepsilon,\bar\xi)=e^{i(\xi_1+\xi_2-\xi_1'-\xi_2')/\rho(E)}\cdot (\lambda_0(\mathcal{K}_+^+))^{n-1}(P_\mu f_+,g_+)+O(e^{-cn}).
\end{align*}
Notice that according to consideration above
\[
\lambda_0(\mathcal{K}_+^+)=1+C_1/\sqrt{W}+O(W^{-1}),\quad (P_\mu f_+,g_+)=c_1+O(W^{-1/2}).
\]
In addition, due to the definition of $\mathcal{R}_{Wn}^{++}(E,\varepsilon,\bar\xi)$ (see (\ref{G_2})), we have
\begin{align*}
\mathcal{R}_{Wn}^{++}(E,\varepsilon,\bar\xi)\Big|_{\xi'=\xi}=1.
\end{align*}
Thus $C_1=0$, $c_1=1$, and
\[
I_\pm\to e^{i(\xi_1+\xi_2-\xi_1'-\xi_2')/\rho(E)},
\]
which gives Theorem \ref{t:2}.

\subsection{Analysis of $I^+_-$}
In this case we will consider $X$ like 
unitary matrix which is close to the $a_-I_2$, and $Y$ as a Hermitian matrix which is close to $a_+I_2$. Then  $X_i$ and $Y_i$ will be parametrized as (cf  (\ref{ch_xy}))
\begin{align*}
&X_{11}=a_-(1+i\theta_-\widetilde a_1/\sqrt{W}),\,X_{22}=a_-(1+i\theta_-\widetilde a_2/\sqrt{W}),\, X_{12}=\dfrac{i\theta_-a_-(x+iy)}{\sqrt{2W}}=-\overline{X_{12}},\\
&Y_{11}=a_+(1+\theta_+\widetilde b_1/\sqrt{W}),\,Y_{22}=a_+(1+\theta_+\widetilde b_1/\sqrt{W}),\,Y_{12}=\dfrac{\theta_+a_+(p+iq)}{\sqrt{2W}}=\overline{Y_{12}}.
\end{align*}
This change transforms the measure $dX_i dY_i/(-\pi^2)$ from (\ref{int_rep_+})  to
\[
(2\pi)^{-4}d \bar a_i \,d \bar b_i  \,d \bar x_i \, d\bar y_i\,d\bar p_i d\bar q_i,
\]
and it "kills" the factor $W^4$ in front of $I_-^+$ in (\ref{repr2_+}).

In this parametrization the operator $\mathcal{K}_+^-$ has the form
\begin{multline*}
\mathcal{K}_+^-(X,Y; X',Y') \\
= (-\lambda^+_0)^{-2}(-\lambda^-_0)^{-2}A_{ab}^-(\bar a,\bar a',x,x',y,y';\bar b,\bar b',p,p',q,q') \widehat Q(\rho,\tau;\rho',\tau'; X',Y') (1+o(1)),
\end{multline*}
where
\begin{align*}
&A_{ab}^-(\bar a,\bar a',x,x',y,y';\bar b,\bar b')\\
&\qquad\qquad =A_{a}^-(\widetilde a_1,\widetilde a_1')A_{a}^-(\widetilde a_2,\widetilde a_2')
A_{b}^+(\widetilde b_1,\widetilde b_1')A_{b}^+(\widetilde b_2,\widetilde b_2')A_x^-(x,x')A_x^-(y,y')A_p^+(p,p')A_p^+(q,q')
\end{align*}
where $A_p^+(x,x')$ is defined in (\ref{A_x}), and
\begin{align*}
&A_x^-(x,x')=(-\lambda^-_0)^{1/2}\frac{a_- \theta_-}{(2\pi)^{1/2}}\exp\Big\{(a_-\theta_-)^2\big(\beta (x-x')^2/2-\frac{c_-}{4}(x^2+(x')^2)\big)\Big\}(1+o(1)).
\end{align*}
Similarly to Section \ref{s:A} one can get that  the largest eigenvalue of  $A_{ab}^-$ is $1+O(W^{-1/2})$, and the next eigenvalue is smaller then $1-\delta$.
Considering $\widehat Q(\rho,\tau;\rho',\tau'; X',Y')$ in this parametrization, we get 
\begin{align}
  \widehat Q(\rho,\tau;\rho',\tau')=&\Big(\prod_{\mu,\nu=1,2}e^{\beta(\rho_{\mu\nu}-\rho'_{\mu\nu})(\tau_{\nu\mu}-\tau'_{\nu\mu})}\Big) (1+O(W^{-1/2})).
  \label{hat_Q^-} 
   \end{align}
Here we used $1+(a_+a_-)^{-1}=0$. In the same way as in Lemma \ref{l:R_+} it is easy to see that the largest eigenvalue of matrix in (\ref{hat_Q^-}) is
$\beta^4(1+O(W^{-1/2}))$, and thus
\[
\| (-\lambda^+_0)^{-2}(-\lambda^-_0)^{-2}\widehat Q\|\le \dfrac{\beta^4}{|\lambda_0^+|^4}(1+O(W^{-1/2}))<1-\delta.
\]
This implies
\[
\|\mathcal{K}_+^-\|<1-\delta,
\]
and so
\[
I^+_-=O(e^{-cn}).
\]

\subsection{Analysis of $I^+_\pm$}
In this case we diagonalize $X$ and use parametrization (\ref{ch_ab}) for $X$ and parametrization (\ref{ch_xy}) for $Y$.
Then repeating almost literally the consideration in Section \ref{s:I_+} (just changing $K_S$ to $K_U$) we get
\[
I^+_\pm=O(e^{-cn}).
\]
\section{Auxiliary results}
\subsection{Analysis of $A_{ab}$}\label{s:A}
Take $c_\pm$ of (\ref{c_pm}) and consider the operator $A_b$ (\ref{K_ab}) near the point $a_+$ with $b,b'$ defined as 
$b_{1i}$ in (\ref{ch_ab}), where
\begin{align}
&\theta_{\pm} =(|\kappa_{\pm}|/\kappa_{\pm})^{1/2},\quad \kappa_{\pm}=\big( c_{\pm}^2/4-\beta c_{\pm}\big)^{1/2}.
\label{kappa}\end{align}
It is easy to see that since $\phi''(a_+)=c_+$, after this change of variables and  normalization by $(-\lambda_0^+)^{1/2}$ of (\ref{la_pm}),
the kernel $A_b$ takes the form
\begin{align}\label{K_a*}
(-\lambda_0^+)^{1/2}A_b&\to  A^+_b=A^+_*(1+W^{-1/2}p_+(\widetilde b))(1+W^{-1/2}p_+(\widetilde b'))+O(e^{-c\log^2W}),\, \\
A^+_*&=\dfrac{(-\lambda_0^+)^{1/2}a_+\theta_+}{\sqrt{2\pi}}\exp\Big\{(a_+\theta_+)^2\big[\beta (\widetilde b-\widetilde b')^2/2-c_+\widetilde b^2/4-c_+(\widetilde b')^2/4\big]\Big\}\notag \\
p_+(\widetilde b)&=c_{3+}\widetilde b^3+c_{4+}\widetilde b^4W^{-1/2}+c_{5+}\widetilde b^5W^{-1}+\dots
\notag\end{align}
where the coefficients  $c_{3+},c_{4+},\dots $ are expressed in terms of the derivatives of $\phi_0$ at $a_+$.

Introduce the orthonormal basises
\begin{align}\label{psi_k+}
\psi_k^\pm(\widetilde b)=|\kappa_\pm|^{1/4}H_k(|\kappa_\pm|^{1/2}b) e^{-|\kappa_\pm|\widetilde b^2/2},
\end{align}
where $\{H_k(x)\}$ are   Hermit polynomials which are orthonormal with the weight $e^{-x^2}$.

Notice that if we make the change of variables for $a,a'$ as $a_{1i}$ in (\ref{ch_ab})
with  $\theta_+$ of (\ref{kappa}), then
\begin{align}\label{K_a*.1}
(-\lambda_0^+)^{1/2}A_a&\to  A^+_a=A^+_*(1+W^{-1/2}\hat p_+(\widetilde a))(1+W^{-1/2}\hat p_+(\widetilde a'))+O(e^{-c\log^2W}),\, \\
\hat p_+(\widetilde a)&=ic_{3+}\widetilde a^3-c_{4+}\widetilde a^4W^{-1/2}-ic_{5+}\widetilde a^5W^{-1}+\dots
\notag\end{align}
with the same $A^+_*$ and  $c_{3+},c_{4+}$ as in (\ref{K_a*}). 

%At the neighbourhood of the point $a_-$ we introduce similarly an operator $A_*^-$ and an orthonormal basis
%\begin{align}\label{psi_k-}
%\psi_k^-(\widetilde b)=|\kappa_-|^{1/4}H_k(|\kappa_-|^{1/2}b) e^{-|\kappa_-|b^2/2}
%\end{align}

\begin{proposition}\label{p:A_ab}
Let $\kappa_{+},\kappa_-$ be defined as in (\ref{kappa}). 
Then the matrices of the operators $A_*^+$ and $A_*^-$ are diagonal in the basis $\{\psi_k^+\}$ and $\{\psi_k^-\}$
and the corresponding eigenvalues have the form ( cf (\ref{la_pm})).
\begin{align}\label{la_k}
\lambda_k^\pm&=\lambda_k(A_*^\pm)=q_\pm^k,\quad k=0,1,2\dots,\quad q_\pm:=\frac{\beta}{\kappa_\pm+c_\pm/2-\beta},
\quad |q_\pm|<1
%\\
%\lambda_0(K_*^\pm)&=\lambda_0^{-1/2}(Q(c_\pm))
%\notag
\end{align}
The matrices of operators $A^+$ and $A^-$ have the same (up to the error $W^{-1}$) diagonals as $A_*^+$ and $A_*^-$ respectively, and
\[(A^\pm)_{k,k'}=O(W^{-1/2})(\delta_{|k-k'|,1}+\delta_{|k-k'|,3})+O(W^{-1})\delta_{|k-k'|,2}+O(W^{-(|k-k'|-3)/2}).\]
\end{proposition}

\textit{Proof.} To simplify formulas, we consider the kernel
\begin{align*}
M(x,y)=(2\pi)^{-1/2}e^{-(\mathcal{A}x,x)/2},\,\bar x=(x,y),\quad \mathcal{A}=\left(\begin{array}{cc}\mu&\nu\\ \nu&\mu \end{array}\right),\,\lambda_{\pm}=\mu\pm \nu,\,\Re\lambda_\pm>0.
\end{align*}
Then, taking $\kappa=\sqrt{\mu^2-\nu^2}=\sqrt{\lambda_+\lambda_-}$, we obtain that 
\begin{align*}
(2\pi)^{-1/2}\int e^{-(\mathcal{A}x,x)/2+\kappa y^2/2}(\frac{d}{dy})^ke^{-\kappa y^2}\ dy=q^k (\mu+\kappa)^{-1/2}e^{\kappa x^2/2}(\frac{d}{dy})^ke^{-\kappa x^2},
\quad q=\frac{\nu}{\mu+\kappa}
\end{align*}
Since the operator with the kernel $ e^{-(\mathcal{A}x,x)/2}$ is compact, we  have  $|q|< 1$. Notice also that
\[
(-\lambda_0^+)^{1/2} (\mu+\kappa)^{-1/2}=1.
\]
If we change the variables
 \[x_1=\theta x,\, y_1=\theta y,\quad \theta =e^{-i(\mathrm{arg}\lambda_++\mathrm{arg}\lambda_-)/4}=e^{-i\mathrm{arg}\kappa/2}, \]
 then
for the new matrix $\widetilde A=\theta^2A$ has eigenvalues $\theta^2\lambda_+, \theta^2\lambda_-$, whose real parts  are still positive,
 $\widetilde\kappa=|\kappa|$, and $\widetilde q=q$.

\subsection{Proof of Lemma \ref{l:b_K_ab}}\label{s:b_K_ab}
To simplify formulas, set
\begin{align*}
&\Lambda_1(t,s)=\Re \Lambda(e^{i\varphi_0}t,e^{i\varphi_0}s)=\frac{\cos 2\varphi_0}{2}(\beta(t-s)^2-\frac{t^2+s^2}{2})-\frac{E\sin\varphi_0}{2}( t+s)\\
&+\frac{\log t}{2}+\frac{\log s}{2}+\Re\phi_0(a_+)\\
&\Lambda_2(t,s)=\Re \Lambda(e^{i\varphi_0}+te^{i\psi},e^{i\varphi_0}+se^{i\psi})=\frac{\cos 2\psi}{2}(\beta(t-s)^2-\frac{t^2+s^2}{2})-\frac{E\sin\psi}{2}( t+s)\\
&-\dfrac{\cos (\psi+\varphi_0)}{2}(t+s)+\frac{\log\big( (t+\cos\phi)^2+\sin^2\phi\big)}{4}+\frac{\log \big((s+\cos\phi)^2+\sin^2\phi\big)}{4}
\end{align*}
with $\Lambda(x,y)$ defined in (\ref{K_ab}) and $\phi=\varphi_0-\psi$.

Then (\ref{b_K.1}) for $|\varphi_0|>\pi/4$ follows from the inequalities
\begin{align}\label{b_K.2}
&\Lambda_1(t,s)\le -c((t-1)^2+(s-1)^2),\quad  \Lambda_2(t,s)\le -c(t^2+s^2)\\
&\Re \Lambda(te^{i\varphi_0},e^{i\varphi_0}+se^{i\psi})\le \Lambda_1(t,1)+\Lambda_2(0,s)
\notag \end{align}
and for $|\varphi_0|< \pi/4$ (\ref{b_K.1})   follows from the first inequality in (\ref{b_K.2}), if it is valid for all $t,s\ge 0$.

The first inequality in (\ref{b_K.2}) follows from the relations
\begin{align*}
\Lambda_1(1,1)=0,\quad\frac{\partial \Lambda_1(1,1)}{\partial t}=\frac{\partial \Lambda_1(1,1)}{\partial s}=0,\quad D\Lambda_1=
\left(\begin{array}{cc}\frac{\partial^2 \Lambda_1(t,s)}{\partial t^2}&\frac{\partial^2 \Lambda_1(t,s)}{\partial t \partial s}\\
\frac{\partial^2 \Lambda_1(t,s)}{\partial t \partial s}&\frac{\partial^2 \Lambda_1(t,s)}{\partial s^2}\end{array}\right)<-cI
\end{align*}

The relation for $\Lambda_1$ and its first derivatives follow from the fact that $x=y=e^{i\varphi_0}$ is the stationary point of $\Lambda(x,y)$.
To prove the last bound, due to the symmetry $\Lambda_1$ it suffices to check that
\begin{align}\label{b_K.3}
&\max\Big\{\frac{\partial ^2\Lambda_1(t,s)}{\partial t^2},\frac{\partial ^2\Lambda_1(t,s)}{\partial s^2}\Big\}\pm \frac{\partial^2 \Lambda_1(t,s)}{\partial t \partial s}<-c\\
& \Leftrightarrow
\frac{\cos 2\varphi_0}{2}(2\beta\pm 2\beta-1)-\inf_{t<1}\frac{1}{2t^{2}}\le -c
\notag\end{align}
The last inequality is valid since the absolute value of the first term  above is less than $\frac{1}{2}$, while the second term is less than $-\frac{1}{2}$.
Notice, that for $|\varphi_0|< \pi/4$ the last inequality is valid also for all $t,s>0$, which implies (\ref{b_K.1}) in this case.

For $|\varphi_0|> \pi/4$, to prove the second bound of (\ref{b_K.2}), it suffices to use that  $s=t=0$ is the stationary point of $\Lambda_2(t,s)$
and the analogue of the first line of (\ref{b_K.3}) is valid. Hence we need to check that 
\begin{align*}
\frac{\cos 2\psi}{2}(2\beta \pm 2\beta -1)-\inf_{t>0}\frac{ (t+\cos\phi)^2-\sin^2\phi}{2( (t+\cos\phi)^2+\sin^2\phi\big)^2}\le -c
\end{align*}
Since the function  under $\inf$  (we call it $d(t)$) for $t\ge 0$ has only one stationary point (maximum), $d(t)$ can take its minimum  either at $t=0$ or at $t\to \infty$.
Since $0<\phi=\varphi_0-\psi<\pi/4$, we have $d(0)=\cos (2\phi)\ge 0$, and $d(t)\to 0$ as $t\to\infty$, $d(t)\to 0$, the  above
inequality takes the form
\[\cos 2\psi(2\beta \pm 2\beta -1)/2<-c\]
Since  $\cos 2\psi>0$  for $\pi/2>\varphi_0>\pi/4$ and $4\beta-1<0$, the last inequality is valid.
The last bound in (\ref{b_K.2}) follows from the relations
\begin{align*}
&\Re (e^{i\varphi_0}+se^{i\psi}-te^{i\varphi_0})^2= \Re e^{2i\varphi_0}(1-t)^2+\Re e^{2i\psi}s^2+2(1-t)s\cos (\varphi_0+\psi)\\
&\le \Re e^{2i\varphi_0}(1-t)^2+\Re e^{2i\psi}s^2,
\end{align*}
since $\psi+\varphi_0\ge \pi/2\Rightarrow\cos (\varphi_0+\psi)\le 0$.

For  $\varphi_0=\pi/4$ we have
\begin{align*}
\Re \Lambda(te^{i\pi/4},se^{i\pi/4})=(-t-s+\log t+\log s+2)/2
\end{align*}
which obviously yields (\ref{b_K.1}).

To prove (\ref{b_K.1'}), recall first that now the kernel of the operator has $-\Lambda(e^{i\varphi},e^{i\varphi'})$ in the exponent, and notice that
\begin{align*}
-\Re \Lambda(e^{i\varphi},e^{i\varphi'})=&-\beta(\cos\varphi-\cos\varphi')^2/2+\beta(\sin\varphi-\sin\varphi')^2/2-(\sin\varphi-\sin\varphi_0)^2/2\\
 &-(\sin\varphi'-\sin\varphi_0)^2/2\\
&\le\beta(\sin\varphi-\sin\varphi')^2/2-(\sin\varphi-\sin\varphi_0)^2/2-
(\sin\varphi'-\sin\varphi_0)^2/2\\
&\le -(1-2\beta)(\sin\varphi-\sin\varphi_0)^2/2 -(1-2\beta)(\sin\varphi-\sin\varphi_0)^2/2.
\end{align*}

$\square$

\subsection{Proof of Lemma \ref{l:U,V}}\label{s:U,V}
  
 It is known that
\begin{equation}\label{LU}
L_2(U)=\displaystyle\oplus_{l=0}^\infty L^{(l)U}, \quad L^{(l)U}=\mathrm{Lin}\,\{t^{(l)U}_{mk}\}_{m,k=-l}^l
\end{equation}
where $\{t^{(l)U}_{mk}(U)\}_{m,k=-l}^l$ are the coefficients of the irreducible representation of the shift operator $T_U\widetilde U=U\widetilde U$. 
It follows from the properties of the unitary representation that
\begin{align*}%\label{prop_t}
t^{(l)U}_{mk}(U^{-1})=\overline{t^{(l)U}_{km}(U)},\quad t^{(l)U}_{mk}(U_1U_2)=\sum t^{(l)U}_{mj}(U_1)t^{(l)U}_{jk}(U_2).
\end{align*}
According to \cite{Vil:68}, Chapter III,
\begin{align}\label{t,U}
&t^{(l)U}_{mk}(U)=e^{-i(m\phi+k\psi)/2}P^{(l)}_{mk}(\cos\theta),
&U=\left(\begin{array}{rr}\cos\frac{\theta}{2}e^{i(\phi+\psi)/2}&i\sin\frac{\theta}{2}e^{i(\phi-\psi)/2}\\
i\sin\frac{\theta}{2}e^{-i(\phi-\psi)/2}&\cos\frac{\theta}{2}e^{-i(\phi+\psi)/2}\end{array}\right).
\end{align}
where
\begin{align}\notag
 &P^{(l)}_{mk}(\cos\frac{\theta}{2})=
 %\frac{i^{m-k}}{2^m}\Big(\frac{(l-m)!(l+m)!}{(l-k)!(l+k)!}\Big)^{1/2}(1-z)^{(m-k)/2}(1+z)^{(m+k)/2}P_{l-m}^{(m+k,m-k)}(z)
 \frac{c_{mk}}{2\pi}\int_0^{2\pi}d\varphi(\cos\frac{\theta}{2}+i\sin\frac{\theta}{2}e^{i\varphi})^{l+k}
(\cos\frac{\theta}{2}+i\sin\frac{\theta}{2}e^{-i\varphi})^{l-k}e^{i(m-k)\varphi},\\
&c_{mk}=\Big(\frac{(l-m)!(l+m)!}{(l-k)!(l+k)!}\Big)^{1/2} .
\label{P_mk}\end{align}
It is known also
that $\{t^{(l)U}_{mk}(U)\}_{m,k=-l}^l$  make an orthonormal basis in  $L^{(l)U}$.

For any function $v(U)$ consider the matrix $ v^{(l)U}=\{v^{(l)U}_{mk}\}$ defined as
\begin{align}\label{v_km}
v^{(l)U}_{mk}:=\int v(U)\overline{ t^{(l)U}_{mk}(U)}dU.
\end{align}
It is easy to see that if we consider an integral operator $\widehat v$ with the kernel $v(U_1U^{-1}_2)$, then 
\begin{align*}
(\widehat vt_{mk})(U)=
\sum v^{(l)U}_{mj}t^{(l)U}_{jk}( U).
\end{align*}
Hence, we obtain that $L^{(l)U}$ reduces $\widehat v$ and  the reduced operator $\widehat v^{(l)U}$ is uniquely  defined by the matrix $v^{(l)U}$. Moreover, if $v$ is some product of the matrix entries, then due to the integration with respect to $\phi,\psi$ in (\ref{v_km}) 
there is only one $k$ and only one $m$ such that $v^{(l)U}_{mk}\not=0$, hence, if 
we denote $\mathcal{E}_{ij}^{(l)}$ an isometric operator such that
\begin{align}\label{E^l_ij}
\mathcal{E}_{ij}(t_{mk}^{(l)U})=\delta_{j,m}t_{ik}^{(l)U},
\end{align}
then $\widehat v^{(l)U}=v^{(l)U}_{km}\mathcal{E}_{km}$.
Let us find the matrices, corresponding to $\big( |U_{12}|^{2p}\big)_U$  in $L^{(l)U}$. Using  (\ref{P_mk}) it is easy to see that
\begin{align}\label{cal_E}
&\big( |U_{12}|^{2s}\big)_U^{(l)}=\lambda^{(l)U}_{00}\mathcal{E}_{00}\\
%&\int |U_{12}|^{2s}K_U(|U_{12}|^2)\overline{ t^{(l)U}_{mk}(U)}dU
%=itW\int_0^\pi\sin\theta d\theta \int_0^{2\pi}d\phi d\psi\sin^{2s}(\theta/2)\\
%&\times e^{-(tW)\sin^2(\theta/2)}e^{i\phi}
%e^{im\phi+in\psi}\overline{ P^{(l)}_{mk}(\cos\theta)}=\delta_{m,0}\delta_{k,0}\lambda^{(l)U}_{00},\notag\\
%&\Rightarrow \left(\big( |U_{12}|^{2p}\big)_U\right)^{(l)}=,}\\
\lambda^{(l)U}_{00}(s)=&tW\int_0^\pi  e^{-tW\sin^2(\theta/2)}\sin^{2s}(\theta/2)
P^{(l)}_{00}(\cos\theta)\sin\theta d\theta
\notag\end{align}
Expanding the function under the integral (\ref{P_mk}) for $\sin(\theta/2)\sim 0$, it is easy to obtain
\begin{align}\label{P(1)}
P^{(l)}_{mm}(1-x)=1-x(l+m)(l+m+1)/2+O(l^4x^2).
\end{align}
Using this asymptotic in the above integral representation of $\lambda^{(l)U}_{00}$ we get the first relation in (\ref{U,V.1})
%\begin{align}\label{la_00}
%\lambda^{(l)U}_{00}=p!(tW)^{-p}(1-l(l+1)/tW)
%\end{align}
Similarly for the  product  zero type $U$ ($K_{sqU}:=\big( |U_{11}|^{2s}U_{11}^{2q}\big)_U$, and here and below, if $s<0$, then we replace
$U_{11}$ with $U_{22}$), we get
\begin{align}\label{K_q}
&\big( K_{sqU}\big)^{(l)}=\lambda^{(l)U}_{-q,-q}\mathcal{E}_{-q,-q},\\
&\lambda^{(l)U}_{-q,-q}=1+O(W^{-1})-(l+q)(l+q+1)/(tW)+O(l^2/W^2)
\notag\end{align}
In particular, for $s=q=0$ we obtain the second relation  in (\ref{type0}) for $K_{0U}$. The norm of  any operator $(p)_U$ of the type $s$ is bounded because
of the inequality
\begin{align*}
\|(p)_U\|\le Wt\int |U_{11}|^{m}|U_{12}|^s e^{-Wt|U_{1,2}|^2}dU\le \delta_{s,0}+C/W^{s/2}.
\end{align*}

To analyse the  products of the first type, 
consider $p_{qs}^{(1)}=|U_{11}|^{2p}U_{12}U_{11}^{2q+1}$. 
 Then (\ref{t,U}) and (\ref{v_km}) yield
\begin{align}\label{rr.1}
&\big( p_{qs}^{(1)}\big)_U^{(l)}=\lambda^{(l)U}( p_{qs}^{(1)})\mathcal{E}_{-1-q,-q},\quad
%\\
%&
\big(\big( p_{qs}^{(1)}\big)_U^{(l)}\big)^*\big( p_{qs}^{(1)}\big)_U^{(l)}=|\lambda^{(l)U}( p_{qs}^{(1)})|^2\mathcal{E}_{-q,-q},\quad
%\big( p_{qs}^{(1)}\big)_U^{(l)}\big(\big(p_{qs}^{(1)}\big)_U^{(l)}\big)^*=|\lambda^{(l)U}(p_{qs}^{(1)})|^2\mathcal{E}_{-1-q,-1-q}
%\notag
\end{align}
where 
\begin{align}\label{la_-1,0}
\lambda^{(l)U}( p_{qs}^{(1)})=&{tW}\int_0^\pi  e^{-tW\sin^2(\theta/2)}\sin(\theta/2)\cos^{2(s+q)+1}(\theta/2)
P^{(l)}_{-1-q,-q}(\cos\theta)\sin\theta d\theta\\
&=(tW)^{-1}\sqrt{(l+q+1)(l-q)}\Big(1+O(l^2/tW)\Big)
\notag\end{align}
Comparing (\ref{K_q}) with $q=0$ with the last two formulas for $q=0$, we obtain (\ref{type0}).

To prove (\ref{b_rr}), notice that
\begin{align*}
r^*r\le  C\big(p\big)_U^*\big(p\big)_U
\end{align*}
By  (\ref{rr.1}) $(\big(p\big)_U^*\big(p\big)_U)^{(l)}=O(l^2/W^2)\mathcal{E}_{q,q}^{(l)U}$ with some $q\in\mathbb{Z}$, and by (\ref{K_q}) $\big(K_\alpha\big)_U^{(l)}=(1-O(l^2/W))\mathcal{E}_{q',q'}^{(l)}$).
If $q=q'$, then
\begin{align}\label{b_rr.1}
&\big(p\big)_U^*\big(p\big)_U\le CW^{-1} (1+C/W-K_\alpha)
\end{align}
And if $q\not=q'$, then $\big(\big(p\big)_U^*\big(p\big)_U\big)^{(l)}$ is not zero only on the  image of $\mathcal{E}_{q,q}^{(l)U}$
and $K_\alpha=0$ hence the r.h.s. here is $O(W^{-1})$  and (\ref{b_rr.1}) is still true. The case of $\big(p(S)\big)_S$ is similar.
The analysis of the difference operators in $L_2(S)$ is very similar. The difference is that for the hyperbolic group the irreducible representations
are labelled by the continuous parameter $l'=-\frac{1}{2}+i\rho$,  $\rho\in\mathbb{R}$,
\[
t^{(l')S}_{mk}=e^{i(m\phi+k\psi)}\mathfrak{P}^{(l)}_{mk}(\theta),\quad m,k\in\mathbb{Z},
\]
and $\mathfrak{P}^{(l)}_{mk}(\theta)$ has the form (\ref{P_mk}) with $\cos(\theta/2)$ replaced by $\cosh(\theta/2)$,
$i\sin(\theta/2)$ replaced by $\sinh(\theta/2)$ and $c_{mk}$ replaced by 1 (see \cite{Vil:68}, Chapter VI) .
Then the same argument yields the second line of (\ref{U,V.1}) and the last line for $K_{0S}$.

\subsection{Proof of Proposition \ref{p:Delta}}\label{s:Delta}
Consider first $\mathcal D=\dfrac{d}{dx}x^2\dfrac{d}{dx}$. It easy to see that for any sufficiently smooth $\varphi(x)$
\begin{equation}\label{c_k}
(\mathcal D)^m\varphi(x)=\sum\limits_{k=1}^{2m}c_{mk} x^k \varphi^{(k)}(x)
\end{equation}
with some integer coefficients $c_{mk}$.
Take $\varphi(x)=e^x$. Then we get
\begin{equation}\label{Q_m}
(\mathcal D)^m e^x=Q_{2m}(x) e^x,\quad Q_{2m}(x)=\sum\limits_{k=1}^{2m}c_{mk} x^k.
\end{equation}
It is easy to see also that
\begin{equation}\label{rec}
Q_{2(l+1)}(x)=(x^2+2x)Q_{2l}(x)+(2x^2+2x)Q'_{2l}(x)+x^2Q''_{2l}(x)
\end{equation}
Define $\Sigma_m=\sum\limits_{k=1}^{2m}|c_{mk}|$. Then considering both sides of (\ref{rec}) we obtain
\[
\Sigma_{l+1}\le (3+4\cdot (2l)+(2l)(2l-1))\Sigma_l\le 4(l+1)^2\Sigma_l.
\]
Thus, since $\Sigma_1=3<4$, 
\[
\Sigma_m\le 2^{2m} (m!)^2,
\]
and so the same bound holds for each individual $c_{km}$. Hence
\begin{equation}\label{d_in}
|(\mathcal D)^m\varphi(x)|=| \sum\limits_{k=1}^{2m}c_{mk} x^k \varphi^{(k)}(x)|\le 2^{2m} (m!)^2\sum\limits_{k=1}^{2m} x^k | \varphi^{(k)}(x)|.
\end{equation}
Recall now that $$\triangle_S=-\dfrac{d}{dx}x(x+1)\dfrac{d}{dx}, \quad x=|S_{12}|^2\ge 0.$$
Notice that 
\begin{align*}
&x(x+1)\le (x+1)^2;\\
&\frac{d}{dx} x(x+1)=2x+1<2x+2=\dfrac{d}{dx} (x+1)^2;\\
&\frac{d^2}{dx^2} x(x+1)=2=\dfrac{d^2}{dx^2} (x+1)^2.
\end{align*}
Therefore Proposition \ref{p:Delta} follows from (\ref{d_in}) where we put $x+1$ instead of $x$ into the r.h.s.

\end{document}